\documentclass[useAMS,usenatbib]{mn2e}
\usepackage{latexsym}
\usepackage{amsmath}
\usepackage{amssymb}
\usepackage{fnpos}
\usepackage{hyperref}
\usepackage{tabularx}
\usepackage{xspace}
\usepackage{enumerate}
\usepackage{graphicx}
\usepackage{caption}
\usepackage{gensymb}
\usepackage{subcaption}
\usepackage{longtable}
\usepackage{lscape}

\usepackage[normalem]{ulem} 
\usepackage[dvipsnames]{xcolor} 
\usepackage{wasysym} 
\definecolor{darkgreen}{rgb}{0.0,0.55,0.0}
\definecolor{darkblue}{rgb}{0.0,0.0,0.5}

\newcommand{\eal}[2]{\ifmmode{\mathrm{#1\,#2}}\else{#1\textsc{$\,$\lowercase{#2}}}\fi\xspace}
\newcommand{\feal}[2]{\ifmmode{\mathrm{#1\,#2}}\else{[#1\textsc{$\,$\lowercase{#2}}]}\fi\xspace}

\title[3D models of the environments of evolved stars]{3D models of the circumstellar environments of evolved stars: Formation of multiple spiral structures}

\author[Aydi et al.]{Elias Aydi$^{1,2,3}$\thanks{E-mail: aydielia@msu.edu} and Shazrene Mohamed$^{2,3,4,5}$\\
$^{1}$Center for Data Intensive and Time Domain Astronomy, Department of Physics and Astronomy, Michigan State University, East Lansing, MI 48824, USA\\
$^{2}$South African Astronomical Observatory, P.O. Box 9, 7935 Observatory, South Africa\\
$^{3}$Astronomy Department, University of Cape Town, 7701 Rondebosch, South Africa\\
$^{4}$NITheCS National Institute for Theoretical and Computational Sciences, South Africa\\
$^{5}$Department of Physics, University of Miami, Coral Gables, FL 33124, USA\\
}

\begin{document}

\date{Accepted ***. Received ***; in original form ***}
\pagerange{\pageref{firstpage}--\pageref{lastpage}} \pubyear{2015}
\maketitle

\label{firstpage}
\begin{abstract}
We present 3D hydrodynamic models of the interaction between the outflows of evolved, pulsating, Asymptotic Giant Branch (AGB) stars and nearby ($< 3$ stellar radii) sub-stellar companions ($M_{\mathrm{comp}} \lesssim 40$ M$_J$). Our models show that due to resonances between the orbital period of the companion and the pulsation period of the AGB star, multiple spiral structures can form; the shocks driven by the pulsations are enhanced periodically in different regions as they encounter the denser material created by the sub-stellar companion's wake. We discuss the properties of these spiral structures and the effect of the companion parameters on them. We also demonstrate that the gravitational potential of the nearby companion enhances the mass loss from the AGB star. For more massive ($M_{\mathrm{comp}} > 40$ M$_J$) and more distant companions ($> 4$ stellar radii), a single spiral arm forms. 
We discuss the possibility of observing these structures with the new generations of high-resolution, high-sensitivity instruments, and using them to `find' sub-stellar companions around bright, evolved stars.
Our results also highlight possible structures that could form in our solar system when the Sun turns into an AGB star. 
\end{abstract}


\newpage

\section{Introduction}
\label{Intro}

At the tip of the Asymptotic Giant Branch (AGB),
stars become unstable to large amplitude, radial pulsations which result in periodic variations in their optical and near-infrared (NIR)
light curves on timescales of the order of $\sim$ 1 to several years \citep{Wood_1979,Feast_etal_1989,Whitelock_etal_2008}. These pulsating stars are called Mira variables; they are long period variables that exhibit  high mass-loss rates, up to 10$^{-5} M_{\odot}$ yr$^{-1}$ \citep{Bowen_1988}, via slow (3 -- 30\,km\,s$^{-1}$), cool (100 -- 1000\,K) outflows that extend for hundreds of astronomical units. 

Detailed understanding of the pulsation mechanism is still lacking, however, it is suggested that they are excited in the convective zone of the star \citep{Freytag_Hofner_2008,Freytag_etal_2017,Trabucchi_etal_2021}. The pulsations are considered an important step in driving the slow, dense Mira winds; layers in the  atmosphere are pushed outwards, however, they fall back as they do not have sufficient acceleration to overcome the gravity of the star. The infalling material encounters outward moving material launched by another cycle of pulsation. The collision of these inward and outward moving layers creates shock waves that propagate through the outer atmosphere, lifting the gas far enough away from the star to produce density enhancements that are sufficiently cool for solid particles, dust grains, to condense 
\citep{Bowen_1988,Hofner_Olofsson_2018}.

\begin{figure}
\begin{center}
\includegraphics[scale =0.34]{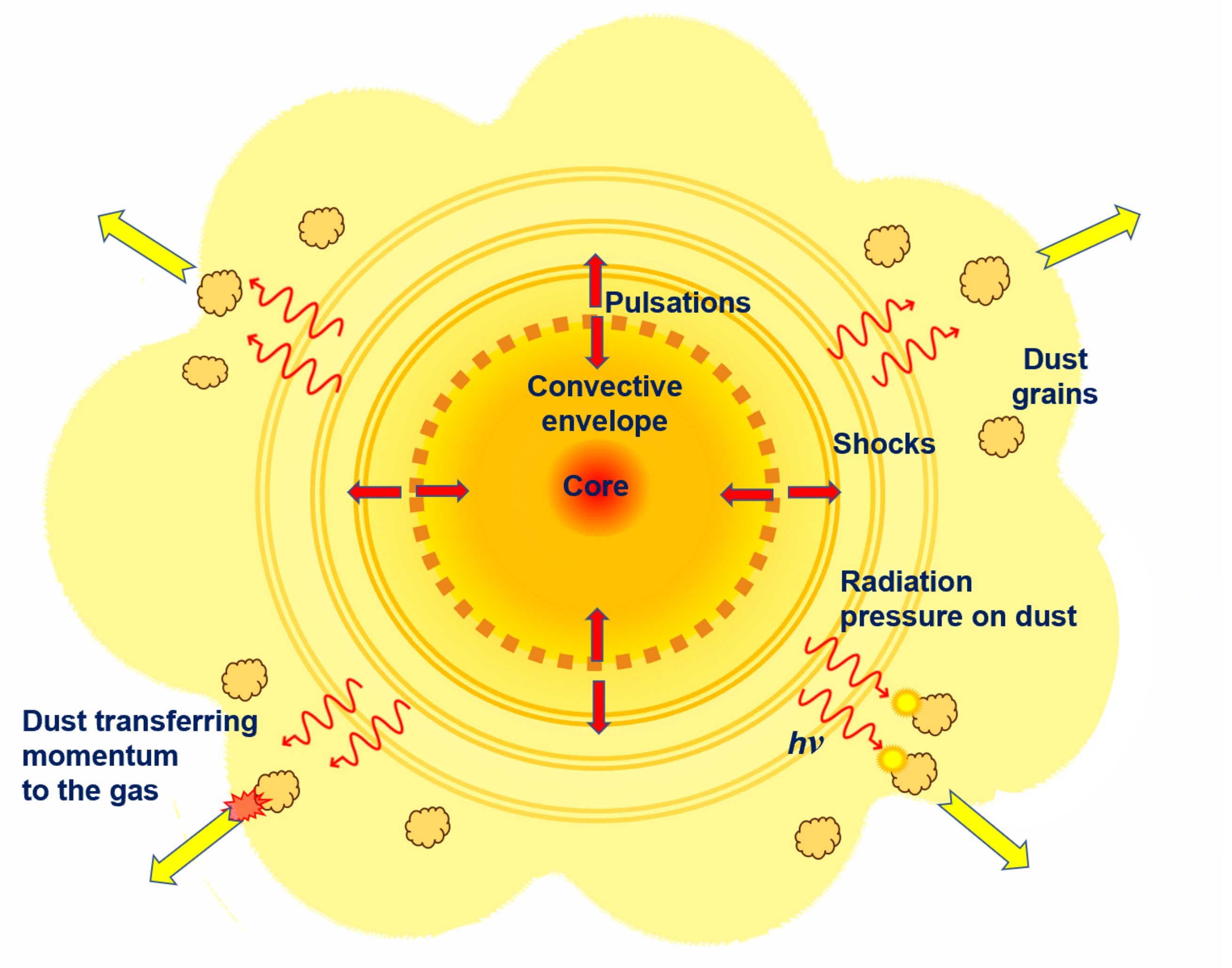}
\caption{A schematic illustration (not to scale) of the outflow in Mira variables. The pulsation creates shocks in the circumstellar environment driving the outflow \citep{Freytag_etal_2017}. While moving outward, the gas cools down and when its temperature drops below the dust condensation temperature and if the density of the medium is high enough, dust starts to form. The dust grains absorb the radiation from the star, gain momentum, and move outward, colliding with and transferring momentum to the gas resulting in a slow, dense stellar wind.}
\label{Fig:mira_wind}
\end{center}
\end{figure}

Mira variables are known to produce copious amounts of dust in their outflows, usually indicated by a high excess in infrared (IR) emission. The dust plays a role in driving the outflow by mediating between and coupling the stellar radiation field to the gas \citep{Gilman_1972}. The dust grains absorb the photon energy, $h\nu$, and the photon momentum, $h\nu/c$, they heat up and start moving outward. While moving outward, they collide with and transfer momentum to the gas resulting in a dust-driven wind \citep{Lamers_etal_1999}. An illustration scheme of both processes that drive the outflow is presented in Figure~\ref{Fig:mira_wind}.

State-of-the-art 1D simulations have been able to account for many of the observed properties of Mira outflows described above, such as, their spectra, colours, and mass-loss properties for C-rich Miras (e.g., \citealt{Nowotny_etal_2013,Eriksson_etal_2014}). These time-dependent models have also been used to study how changes in the stellar wind properties produce shell-like circumstellar structures and detached shells observed around AGB stars, as well as linking them to the spherical arcs observed in their progeny, planetary nebulae (e.g, \citealt{Schoenberner_1981,Kahn_etal_1990,Steffen_etal_1998,Myasnikov_etal_2000,Corradi_etal_2000,Simis_etal_2001,Schonberner_etal_2005,Mattsson_etal_2010}).

While asymmetric structures are well documented and ubiquitous in post-AGB systems and planetary nebulae \citep{Balick_etal_1987,Balick_Frank_2002,vanWinckel_2003,Corradi_etal_2004,Sabin_etal_2007,Corradi_etal_2011,Lykou_etal_2011,Ueta_etal_2014,Manick_etal_2015,Montez_etal_2017}, with more recent high resolution, high sensitivity observations, e.g., with the Atacama Large Millimeter/submillimeter Array (ALMA; \citealt{Wootten_Thompson_2009}) and the  Spectro-Polarimetric High-contrast Exoplanet REsearch (SPHERE; \citealt{Beuzit_etal_2019}), it has become increasingly clear that deviations from spherical symmetry happen much earlier than previously thought - indeed, while the stars are still on the AGB. Observations of asymmetric structures, e.g., spirals, bipolar outflows, arcs, disks and equatorial outflows in the circumstellar environments of AGB stars are now commonplace (e.g., \citealt{Mauron_etal_2006,Morris_etal_2006,Mauron_etal_2006,Olofsson_etal_2010, Lagadec_etal_2011,Lagadec_2011_II,Maercker_etal_2012,Abate_etal_2013,Ramstedt_etal_2014,Starkenburg_etal_2014,Decin_etal_2015,kim_etal_2015,Lagadec_Chesneau_2015,Doan_etal_2017,Ramstedt_etal_2017,khouri_etal_2018,lagadec_2018,Ramstedt_etal_2020,Doan_etal_2020}).

Many of these structures have been linked to shaping by a binary companion, as described by the references above as well as detailed theoretical studies (e.g., \citealt{Theuns_etal_1996,Soker_1998,Mastrodemos_Morris_1999,Balick_Frank_2002,Gawryszczak_etal_2002,Nordhaus_etal_Blackman_2006,Mohamed_Podsiadlowski_2007,Mohamed_Podsiadlowski_2012,Kim_etal_2012,Staff_etal_2016,Chen_etal_2017,Saladino_etal_2019,Saladino_Pols_2019,Malfait_etal_2021,Maes_etal_2021} and references therein). 
The C-rich AGB star, R Sculptoris, proved to be a good catalyst for both observational and theoretical studies of the shaping processes \citep{Maercker_etal_2012}; having recently undergone a thermal pulse, it is surrounded by a spherical, thin detached shell which was then revealed to harbour a spiral, created by an unseen  low-mass companion interacting with the dense AGB outflow. Modelling the high-resolution ALMA data not only improved our understanding of the dramatic changes to the mass-loss properties during a thermal pulse, but also helped constrain the properties of the orbiting companion.

Indeed, the high luminosity of AGB stars and the large variations in that brightness caused by their pulsations, make it difficult to detect low-mass companions using radial velocity or transit measurements. However, imaging of the circumstellar environment has been used to infer the presence of planets or sub-stellar companions (e.g., \citealt{Kervella_etal_2017}), and although they are difficult to detect, the number of giant planets detected around evolved stars has increased in the past few years, particularly using radial velocity techniques (e.g., \citealt{Sato_etal_2003,Niedzielski_etal_2007,Sato_etal_2007,Liu_etal_2008,
Sato_etal_2008,Niedzielski_etal_M2009,Niedzielski_etal_D2009,Gettel_etal_J2012,
Gettel_etal_S2012,Nowak_etal_2013,Brucalassi_etaL_2014,Niedzielski_etal_2015,Marshall_etal_2019,Quirrenbach_eatl_2019,Marshall_etal_2019,Jeong_etal_2021,Dollinger_etal_2021}); with the majority of these planets at  orbital separations ranging from close to intermediate (0.5 AU to 4.0 AU). 
Furthermore,  \citet{Soszynski_etal_2021} suggested that accreting planetary companions -- turned brown dwarfs -- are the cause of the long secondary periods observed in red giant stars, illustrating the influence of sub-stellar companions on the observed properties of giant stars. However, detailed 3D modelling and predictions of the effects of such close-in sub-stellar companions on the circumstellar envelopes of evolved stars have been lacking. 

\citet{Struck_etal_2002} carried out 2D simulations of planets and brown dwarfs in close orbit around Mira variables and they found substantial wakes behind the companion and episodic accretion phenomena, which may give rise to observable optical events and affect SiO maser emission. \citet{Kim_etal_2012} studied the formation of an Archimedian spiral structure around AGB stars due to the presence of a sub-stellar companion at intermediate distance from the star (10--20\,AU), making estimates for the density contrasts between the arm and inter-arm regions. 
Using models that included a more detailed treatment of the AGB stellar wind physics, e.g., a sinusoidal piston (at the inner boundary - near the stellar surface), radiative cooling, and silicate dust formation, \citet{Wang_Willson_2012} built several 1D models of a pulsating star with a planetary companion in close orbit (1.019 -- 1.35 $R_*$). Their results showed the formation of multiple spiral arms centered on the AGB star due to the interaction of the planet with the stellar pulsation. The number and period of the spiral arms vary depending on the mass of the companion and its orbital period.

To further aid observational studies, in this paper, we extend the theoretical work of \citet{Wang_Willson_2012} to 3D, allowing us to better predict the effect of a close-in companion on the morphology of the outflow at different viewing/inclination angles. Furthermore, unlike previous multi-dimensional studies where the flow at the inner boundary is prescribed and sets the wind mass-loss properties  (e.g., all the previous SPH and grid code models above), in this work we have an outer atmosphere within which the wind launching takes place, allowing us to better study the influence of the companion on the outflow properties as well. In addition, modelling the interaction of close-in substellar companions with a pulsating AGB star could give us insight into the future solar system, when our Sun turns into a giant, pulsating star. In Section~\ref{Num} we describe the model setup and numerical details, and the resulting simulations are presented in Section~\ref{Results}. The formation and shaping of the AGB outflow into multiple spirals is discussed in detail in Section~\ref{Disc}, and we touch briefly on the observability of these structures, given the sensitivity and resolution of current and future facilities, and the less certain understanding of planet survivability around evolved stars. We conclude with a short summary and highlight potential avenues for future studies in Section~\ref{Concl}.


\section{Numerical method and setup}
\label{Num}
\subsection{Smoothed particle hydrodynamics}
SPH \citep{Gingold_Monaghan_1977,Lucy_1977,Monaghan_1992} is a Lagrangian method where the fluid is discretized into a finite number of particles. Each particle has fluid properties, such as density, velocity, pressure, and internal energy. Starting from an estimate of the density obtained by summing and smoothing the masses of neighbouring particles, the other properties are derived using the Lagrangian formulation of the fluid equations. For the interpolation in this case we use a cubic spline smoothing kernel. The method conserves momentum, energy, entropy, and angular momentum by its Lagrangian construction \citep{Springel_Hernquist_2002}. 
Artificial viscosity is added in order to model shocks and density discontinuities, where the viscous forces are added to the equation of motion as an excess of pressure on the particles \citep{Monaghan_Gingold_1983,Balsara_1995}.\\
We use the SPH cosmological simulation code GADGET-2 \citep{Springel_2005}, which is a TreeSPH code that is capable of following a collisionless fluid with the N-body method and an ideal gas by means of SPH. The gravitational forces and self-gravity in GADGET-2 are derived by the use of a hierarchical Barnes-Hut (BH) oct-tree algorithm \citep{Barnes_Hut_1986}.

\subsection{Model setup}

\subsubsection{Modelling the pulsation and atmosphere}
The stellar pulsations are thought to originate in the convective zone of the AGB star (see, e.g., \citealt{Freytag_etal_2017}). These pulsations result in shock waves that travel through the extended atmosphere and begin to drive the outflow. Our models adopt the approximation from \citet{Bowen_1988} where the pulsation is simulated by an artificial piston (in this work, a shell of particles) placed at the edge of the photosphere ($R_{\mathrm{piston}}$ $\sim 0.92\,R_{\mathrm{photo}}$). The artificial piston, placed directly below the atmosphere of the star, oscillates sinusoidally starting from a maximum amplitude $U_{\mathrm{amp}}$. The piston acts like a boundary layer where the radius and velocity of the boundary vary with time as follows:
\begin{equation}
R = R_0 + U_{\mathrm{amp}}(P/2\pi) \sin (2\pi t/P_{\mathrm{puls}})\,,
\end{equation}
\begin{equation}
U =  U_{\mathrm{amp}} \cos (2\pi t/P_{\mathrm{puls}})\,,
\end{equation}
\\
where $R$ and $U$ are the piston radius and velocity at time $t$, respectively, $R_0$ is the initial radius of the piston, $U_{amp}$ is the piston velocity amplitude, and $P_{\mathrm{puls}}$ is the pulsation period. 

The fundamental pulsation mode, $P_{\mathrm{puls}}$ is derived from the period-mass-radius relationship of \citet{Ostlie_Cox_1986} which relates the $P_{\mathrm{puls}}$ to the mass ($M_*$) and radius ($R_*$) of the star through:
\begin{equation}
\log P_{\mathrm{puls}} = -1.92 - 0.73 \log M_* + 1.86 \log R_*.
\end{equation}

The atmosphere of the star is represented by another shell of gas particles, just above the artificial piston. We use equal mass particles ($\approx$ 7.476 $\times 10^{21}$\,g $\approx$ 3.75 $\times 10^{-12}$\,M$_{\odot}$), set in a 1/$r^2$ density distribution. The combined mass of the artificial piston and atmosphere is not included in the mass of the central star. A density distribution of 1/$r^2$ agrees well with the observations \citep{Struck_etal_2002}; using an exponential density profile similar to the one from \citep{Wang_Willson_2012} would be preferable, however, such a distribution is difficult to achieve in SPH models using equal mass particles. 
The inner regions of the star are not modelled and instead a point mass (particle) is used to simulate its gravity. The standard parameters of the central star used in all our models are listed in Table~\ref{table:star}.

\begin{table}
\centering
\caption{Standard model parameters.}
\begin{tabular}{lr}
\hline
$M_*$ & 1\,M$_{\odot}$\\
$R_*$ & 268\,R$_{\odot}$\\
$P_{\mathrm{puls}}$ & 394\,d\\
$U_{\mathrm{amp}}$ & 4 km\,s$^{-1}$\\
$T_{\mathrm{eff}}$ & 2966 K\\
\hline
\end{tabular}
\label{table:star}
\end{table}

\begin{figure}
\begin{center}
\includegraphics[width=\columnwidth]{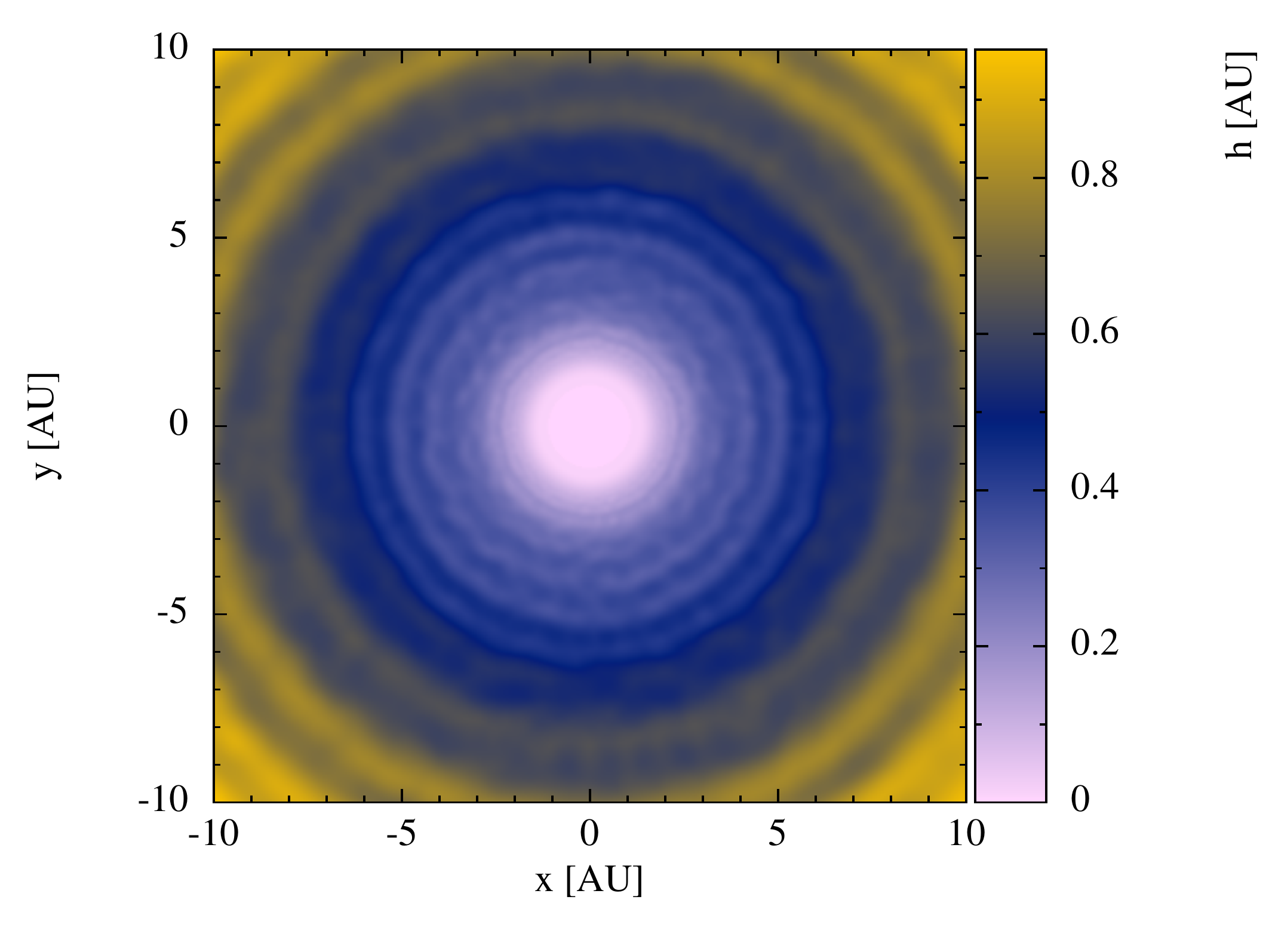}
\vspace{-0.2cm}
\caption{A rendered cross-section of the smoothing length, $h$, of each particle near the atmosphere of the star.}
\label{Fig:smoothing}
\end{center}
\end{figure}

\subsubsection{Numerical resolution and smoothing}

Our models consist of 4.2$\times 10^6$ gas particles; 9.2$\times 10^5$ particles for the piston and 3.3$\times 10^6$ particles in the atmosphere of the star. Each gas particle has a mass of $\approx$ 7.476 $\times 10^{21}$\,g $\approx$ 3.75 $\times 10^{-12}$\,M$_{\odot}$. The star and the companion are added to the model as Gadget-2 ``star'' (point-mass) particles with masses equal to $M_*$ (see Table~\ref{table:star}) and $M_{\mathrm{comp}}$ (see Table~\ref{table:models}), respectively. 

The smoothing length, $h$ in SPH simulations defines the scale over which the fluid properties of the particles are interpolated. Therefore, $h$ effectively determines the resolution of a simulation. To achieve the goals of the Lagrangian adaptivity in SPH, \textsc{gadget-2} allows the smoothing length of each particle to vary in both space and time. Hence, the resolution increases in high-density regions (e.g., where shocks are forming from the stellar pulsation or interaction of the stellar wind with the companion) and decreases in low density regions. In Figure~\ref{Fig:smoothing} we show a cross-section plot around the regions of the periodic shock-wave formation, showing that for the $\approx$ 4 million particles we use, the smoothing length is $\lesssim$ 0.8 AU.

\subsubsection{Modelling the radiation pressure on dust}
\label{sec_dust_calc}
The radiation pressure on dust results in a net outward acceleration of the dust grains that then transfer momentum to the gas by collision. Dust formation is typically modelled via two approaches, the so-called moment method \citep{Gail_Sedlmayr_1988,Gauger_etal_1990}, which effectively calculates the growth of dust grains, assuming the formation via primarily single \textit{monomers}, such as C, SiO and Fe. 
The other method is the chemical kinetic approach, which was introduced by \citet{Cherchneff_2006} and in which the chemistry of dust formation is described by using a chemical kinetic network of reactions. These approaches  work successfully for C-rich AGB stars where a dust-driven wind is achieved, however, for O-rich AGB stars the nucleating species are still uncertain and pre-existing grain seeds are used to initiate the process and trigger a dust-driven wind \citep{Hofner_2008,Bladh_Hofner_2012,Bladh_etal_2013}. There are also uncertainties regarding the grain composition \citep{Bladh_etal_2013} and the Fe-enrichment needed to reproduce the observable features \citep{Hofner_etal_2021}. 

Given these uncertainties, and the significant numerical cost to follow this in detail, similarly to \citet{Wang_Willson_2012} we adopt the simplified method from \citet{Bowen_1988} to model dust formation. The approach consists of comparing the radiative equilibrium temperature ($T_{\mathrm{eq}}$) of the gas to the condensation temperature ($T_{\mathrm{cond}}$) of the dust. If $T_{\mathrm{eq}}$ is less then $T_{\mathrm{cond}}$ an acceleration is added to the gas particles. As detailed radiative transfer is not included in our models (due to computational cost), we assume radiative equilibrium \citep{Lamers_Cassinelli_1999}:

\begin{equation}
\label{Teq}
T_{\mathrm{eq}} = T_{\mathrm{eff}} \left(\frac{R_*}{2r} \right)^{\frac{2}{4+p}} \,,
\end{equation}
where $r$ is the radius of the gas from the central star and $p$ is a power-law coefficient adopted from \citet{Bladh_Hofner_2012} for different types of dust grains. We experiment with both silicates and carbonaceous grains. For the silicate (Mg$_2$SiO$_4$) grains, we use a  dust condensation temperature $T_{\mathrm{cond}}$ = 1100\,K and $p$ = -0.9. For amorphous carbon (amC) grains, we adopt $T_{\mathrm{cond}}$ = 1500\,K and $p$ = 2.1. A more realistic approach to calculate the optical depth in the cirumstellar medium and the equilibrium temperature of the gas will be presented in a future work.

The dust cross-section for radiation pressure, in cm$^2$\,g$^{-1}$ of gas is derived using:
\begin{equation}
k_D = k_{\mathrm{max}} \frac{1}{1 + \mathrm{exp}[(T_{\mathrm{eq}} - T_{\mathrm{cond}})/\delta]},
\end{equation}
where $\delta$ is the parameter for the condensation temperature range and the maximum cross-section per unit mass $k_{\mathrm{max}}$ is specified as an input parameter. A value of $k_{\mathrm{max}}$ = 1.5 cm$^2$\,g$^{-1},$ is chosen to maintain the ratio between the acceleration due to the radiation pressure on dust and the gravitational acceleration, $\frac{dv}{dt}$(rad)/$\frac{dv}{dt}$(g), less than or equal to 0.95 \citep{Bowen_1988}.

The acceleration of the gas particles due to the radiation pressure on dust, which is added to the particles acceleration is defined as follows:
\begin{equation}
\frac{dv}{dt}(\mathrm{rad}) =\frac{k_D L }{4\pi r^2 c},
\end{equation}
where $L$ is the luminosity of the star, $r$ is the distance between the gas particle and the star, and $c$ is the speed of light.

Our models do not take dust destruction into account for several reasons: (1) The formation of dust occurs beyond $\sim$ 2--3$R_{*}$ ($\sim$ 2--4 AU), (2) The hot shocks that lead to dust destruction occur below $\sim$ 2$R_{*}$, (3) Our assumption of adding acceleration to the gas particles considers instantaneous dust formation depending on $T_{\mathrm{eq}}$, and the shock waves generated by the pulsation every pulsation period do not reduce significantly the dust abundance \citep{Wang_Willson_2012}.

\subsubsection{Cooling}
Cooling or thermal relaxation by radiative transfer is not included in this work. The models are all adiabatic such that the net radiative energy per unit mass gained/lost, $Q$, is always equal to zero. The results from these models produces realistic mass-loss rates on the order of $\sim$ 10$^{-7}$ M$_{\odot}$ yr$^{-1}$ without including dust formation. Note that the \citet{Bowen_1988} models with adiabatic processes resulted in high mass-loss rate, but his results were less catastrophic than the models of \citet{Wood_1979}. More realistic cooling, such as relaxation towards $T_{eq}$ at a finite rate will be implemented in future versions of the code.

\begin{figure*}
\begin{center}
\includegraphics[width=0.49\textwidth]{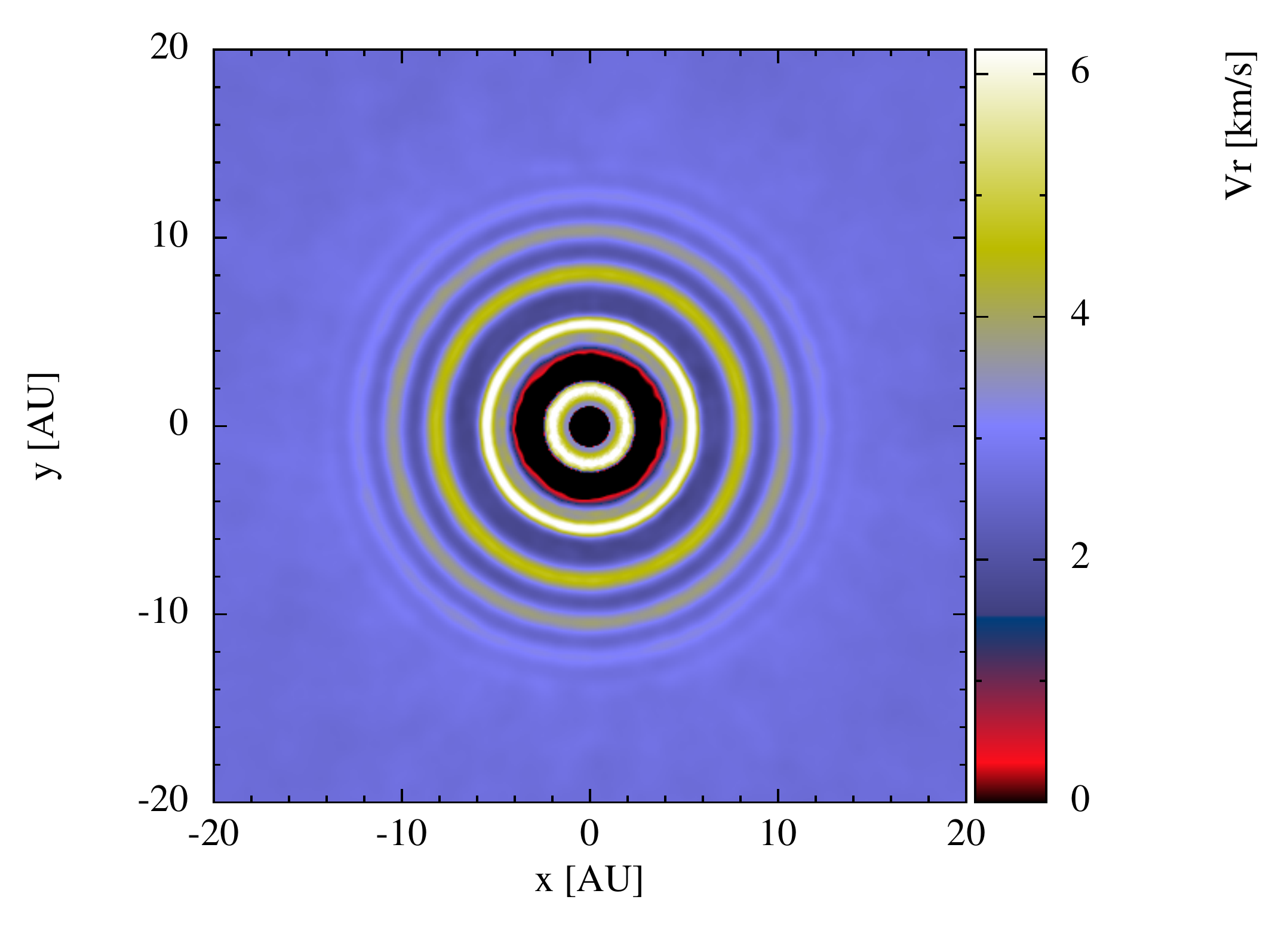}
\includegraphics[width=0.49\textwidth]{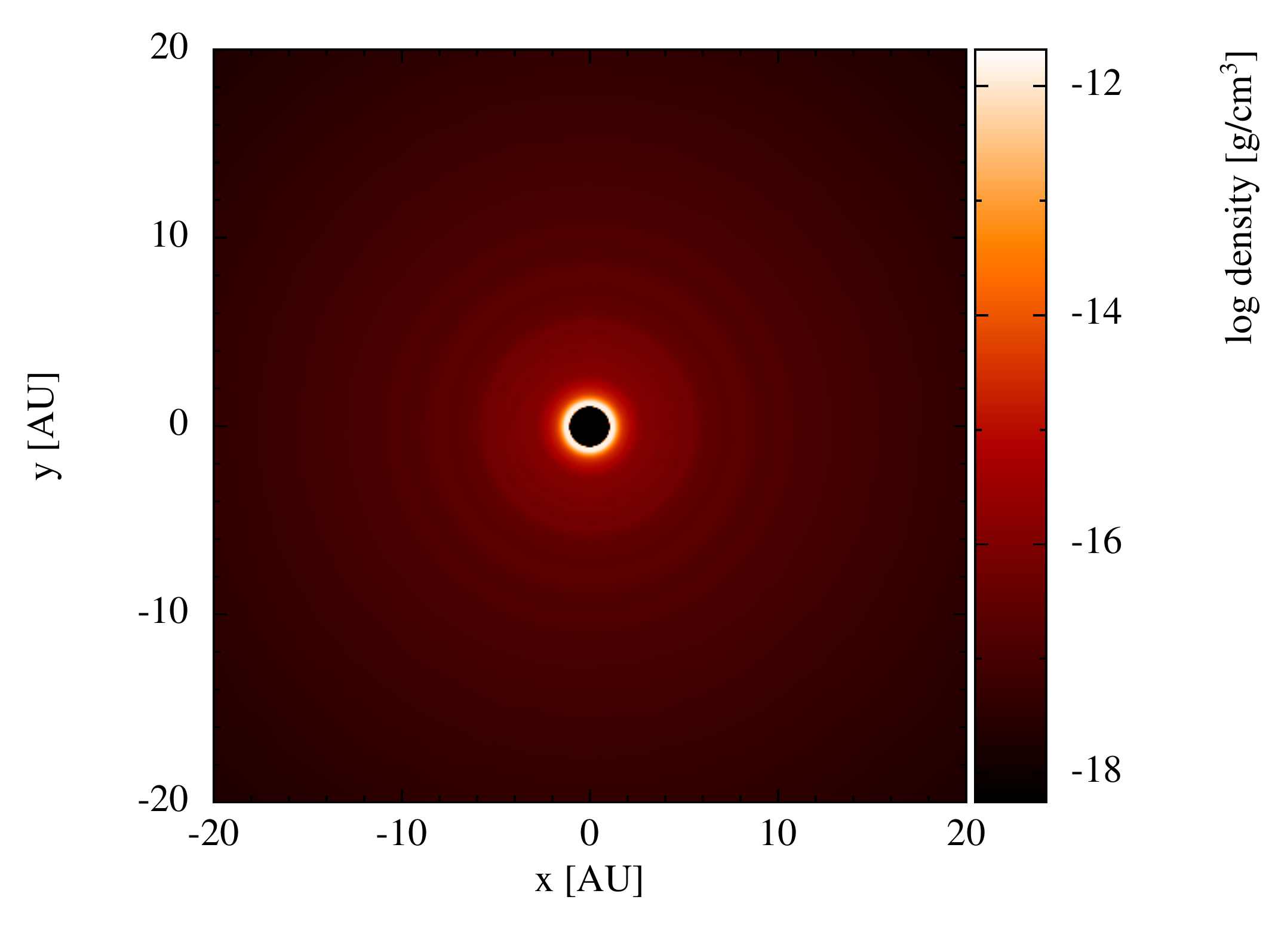}
\includegraphics[width=0.49\textwidth]{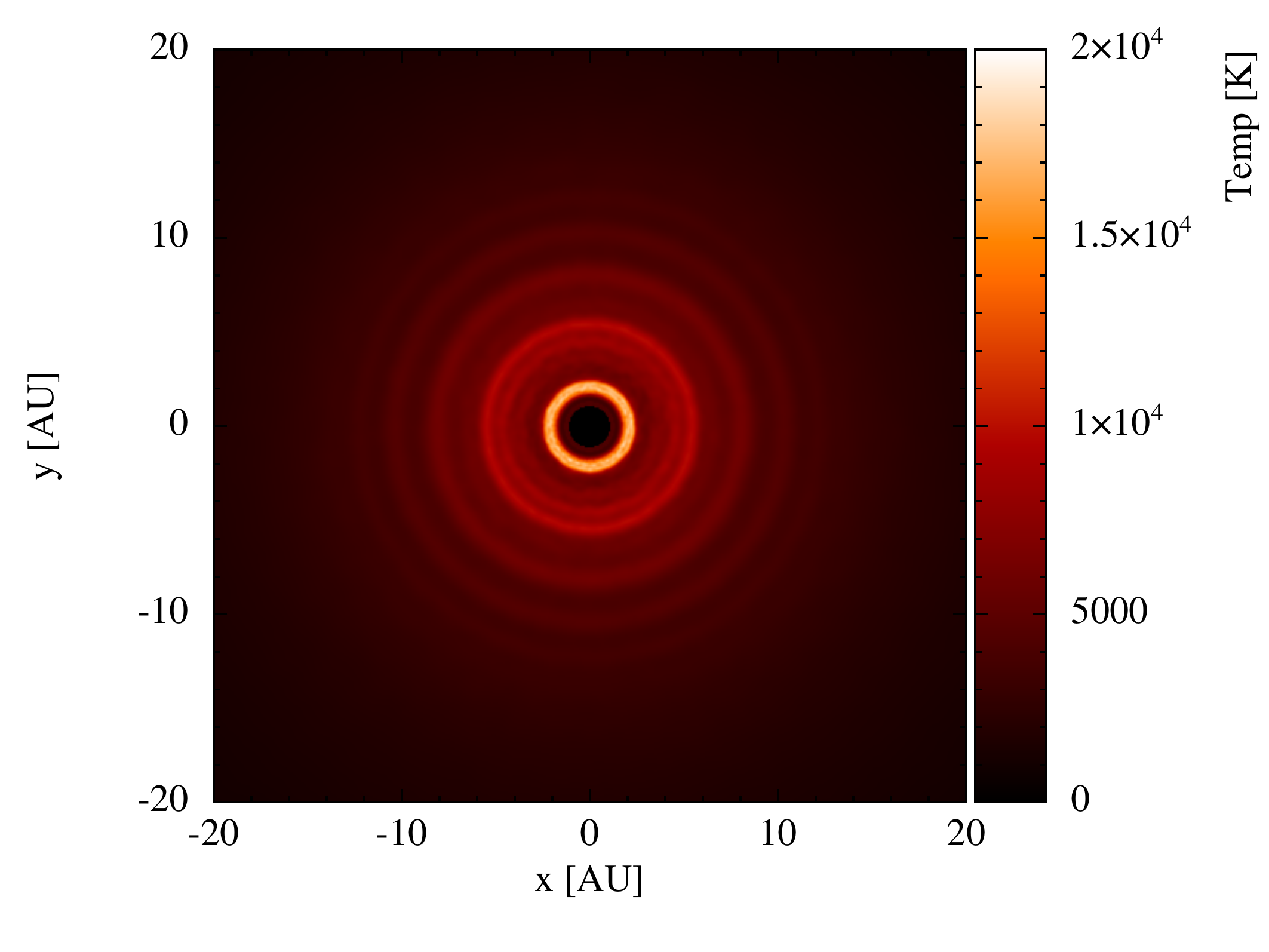}
\includegraphics[width=0.49\textwidth]{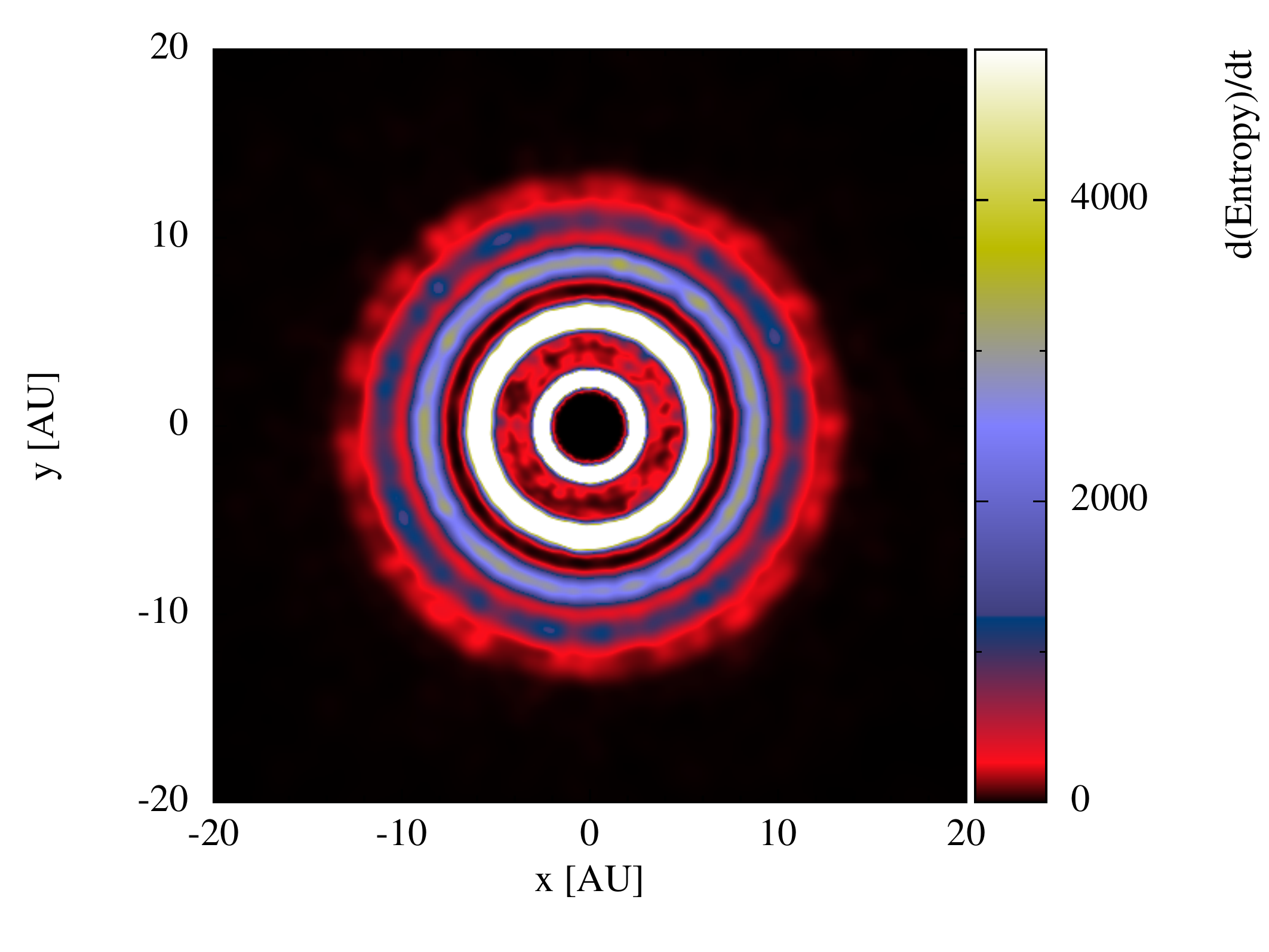}
\vspace{-0.4cm}
\caption{Cross-section slices through the orbital plane x-y of a single star, showing radial velocity, density, temperature, and entropy, from top left to bottom right, respectively. The plots show the formation of circular shock waves in 2D (spherical shells in 3D) due to the stellar pulsation.
}
\label{Fig:shock-waves}
\end{center}
\end{figure*}

\begin{figure*}
\begin{center}
\includegraphics[width=0.48\textwidth]{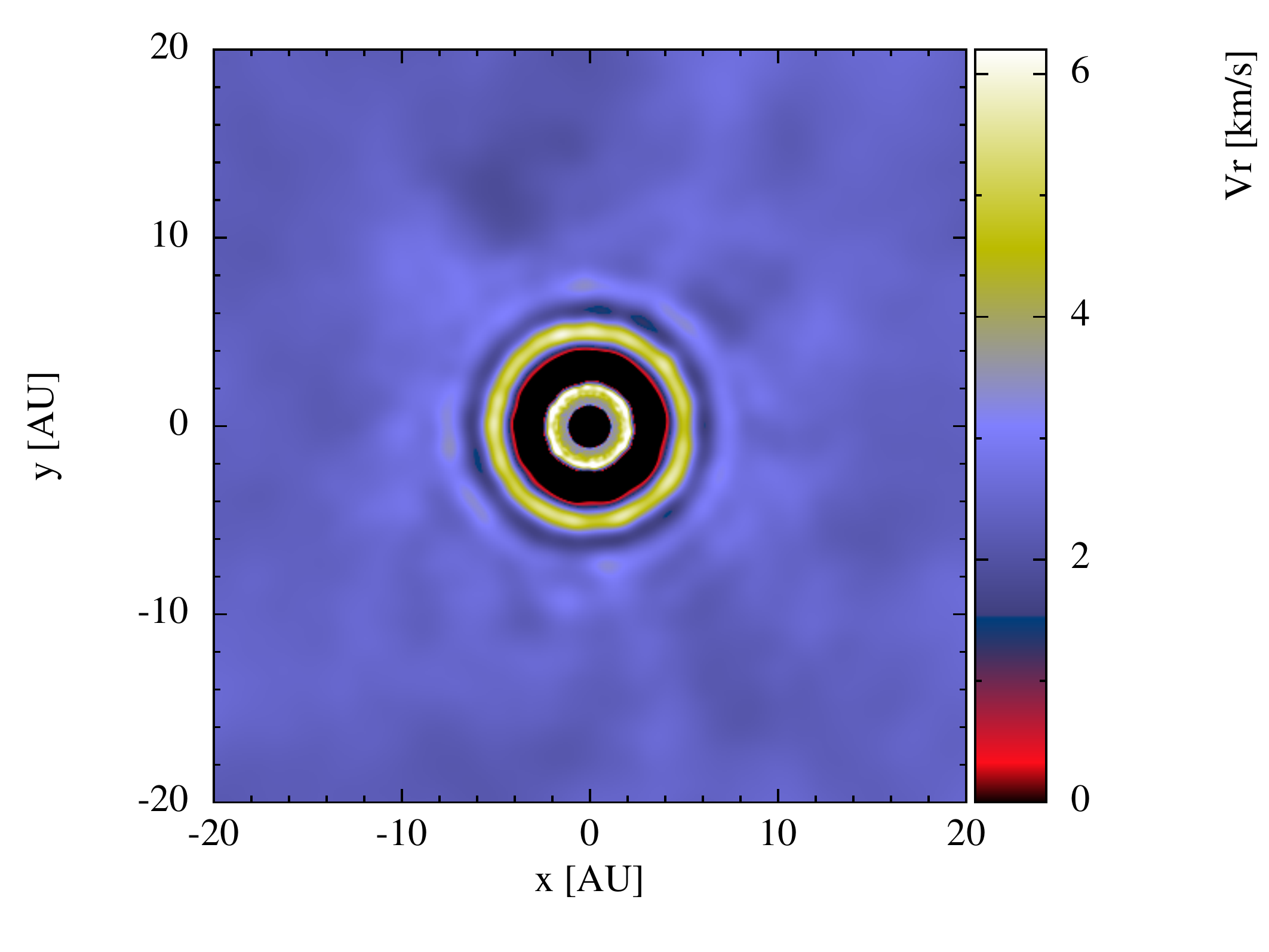}
\includegraphics[width=0.48\textwidth]{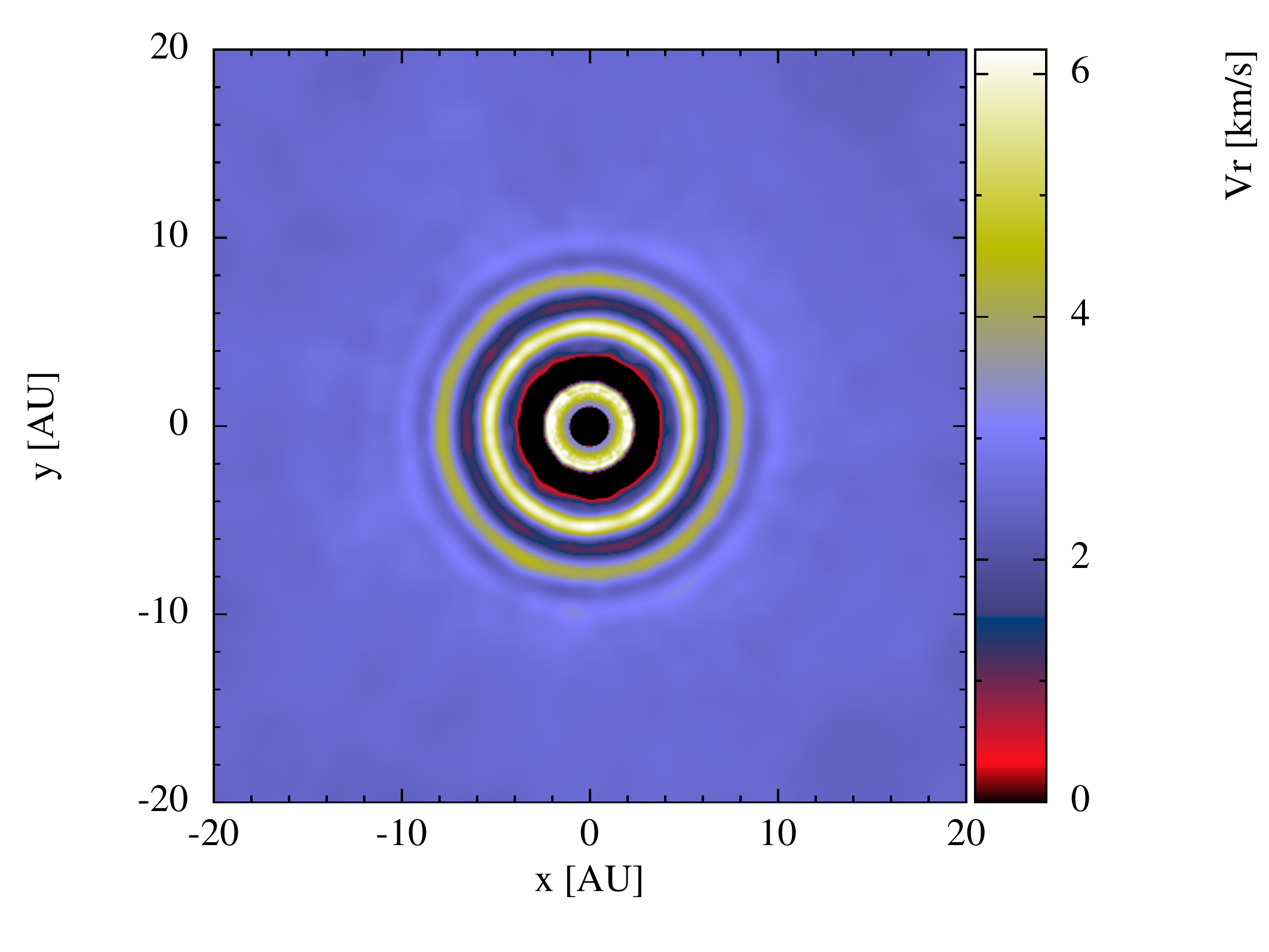}
\vspace{-0.1cm}
\caption{Radial velocity cross-sections through the orbital plane x-y of a single star, using $2.5 \times 10^5$ particles (\textit{left}) and $1.1 \times 10^6$ particles (\textit{right}).}
\label{Fig:single_star_resolution}
\end{center}
\end{figure*}

\subsubsection{The orbit of the companion:}

The sub-stellar companion is placed in a circular orbit around the pulsating star. The companion orbits in the equatorial/orbital plane (the x-y plane) and is added as a point-mass particle. The orbit of the companion is not affected by tidal drag or mass loss in this version of the model. The companion is assumed not to accrete any material due to: (1) the relatively small mass of the companion compared to the mass of the star, (2) the short timescale of the simulation compared to the accretion timescale in case of a sub-stellar companion (see e.g.,  \citealt{Villaver_Livio_2009}).

For models where the companion is at an orbital distance of $d_{\mathrm{comp}}$\,$<$3$R_*$ from the center of the AGB star, it is considered in circular orbit around the center. The star is then kept stationary, particularly since the barycenter of the system is very close to the stellar center (separated by\,$\leq$\,0.06$R_*$). In models where $d_{\mathrm{comp}}$\,$>$3$R_*$, the star and the companion are both orbiting the barycenter of the system which lies within the radius of the star for most of the models. For such models, the stellar center is separated by $\sim 0.1 - 1.0 R_*$ from the barycenter of the system. $d_{\mathrm{comp}}$ is always the distance between the companion and the center of the star. In Appendix~\ref{appA} we elaborate on how the properties of the companion, such as, the relative velocity ($v_{\mathrm{orb}}$), orbital period ($P_{\mathrm{orb}}$), angular momentum $L$, and $d_{\mathrm{comp}}$ were derived. In some models we intentionally fix some of the properties of the companion to test the effect of changing other properties.

\begin{figure*}
\begin{center}
\includegraphics[width=0.48\textwidth]{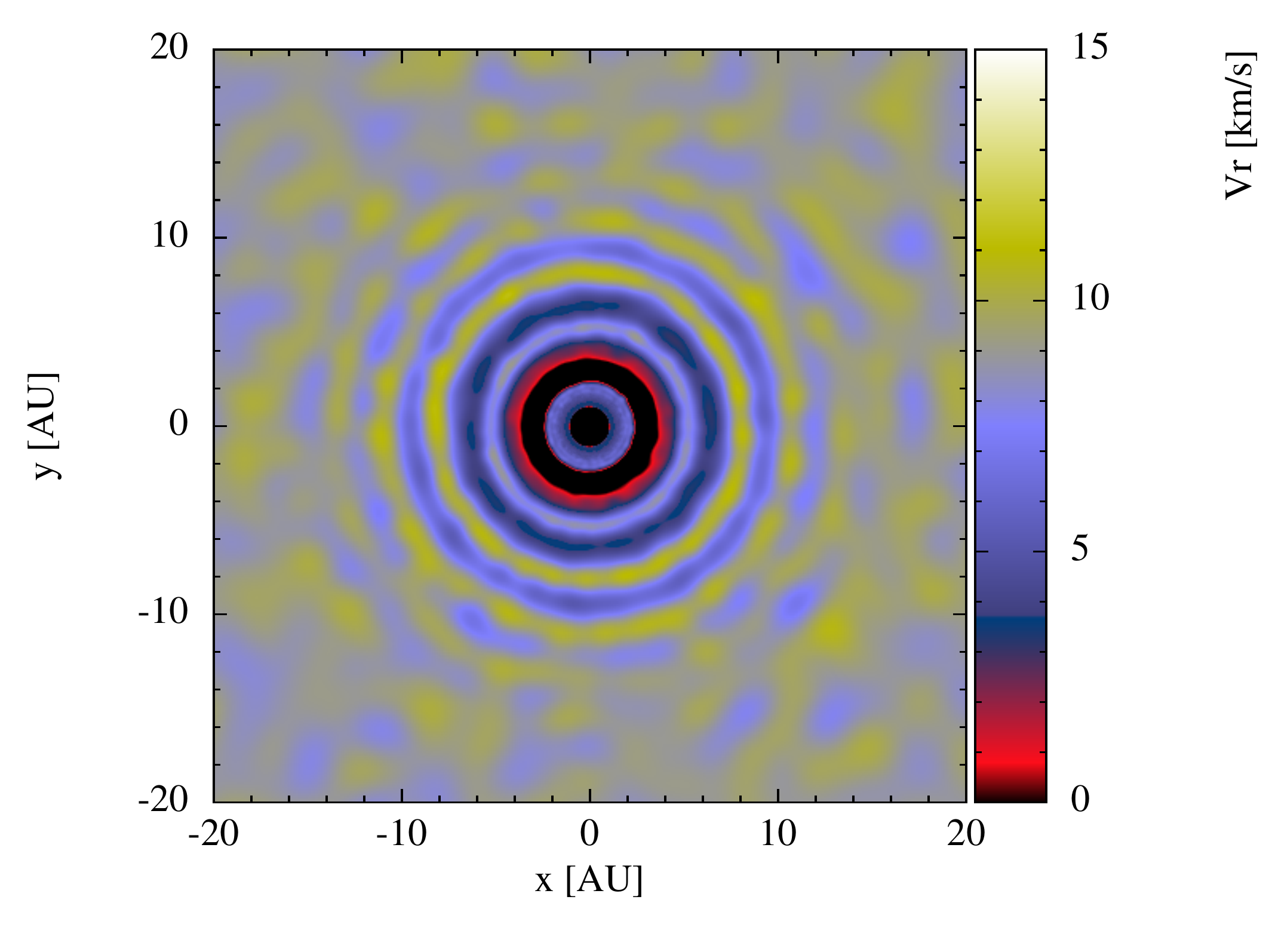}
\includegraphics[width=0.48\textwidth]{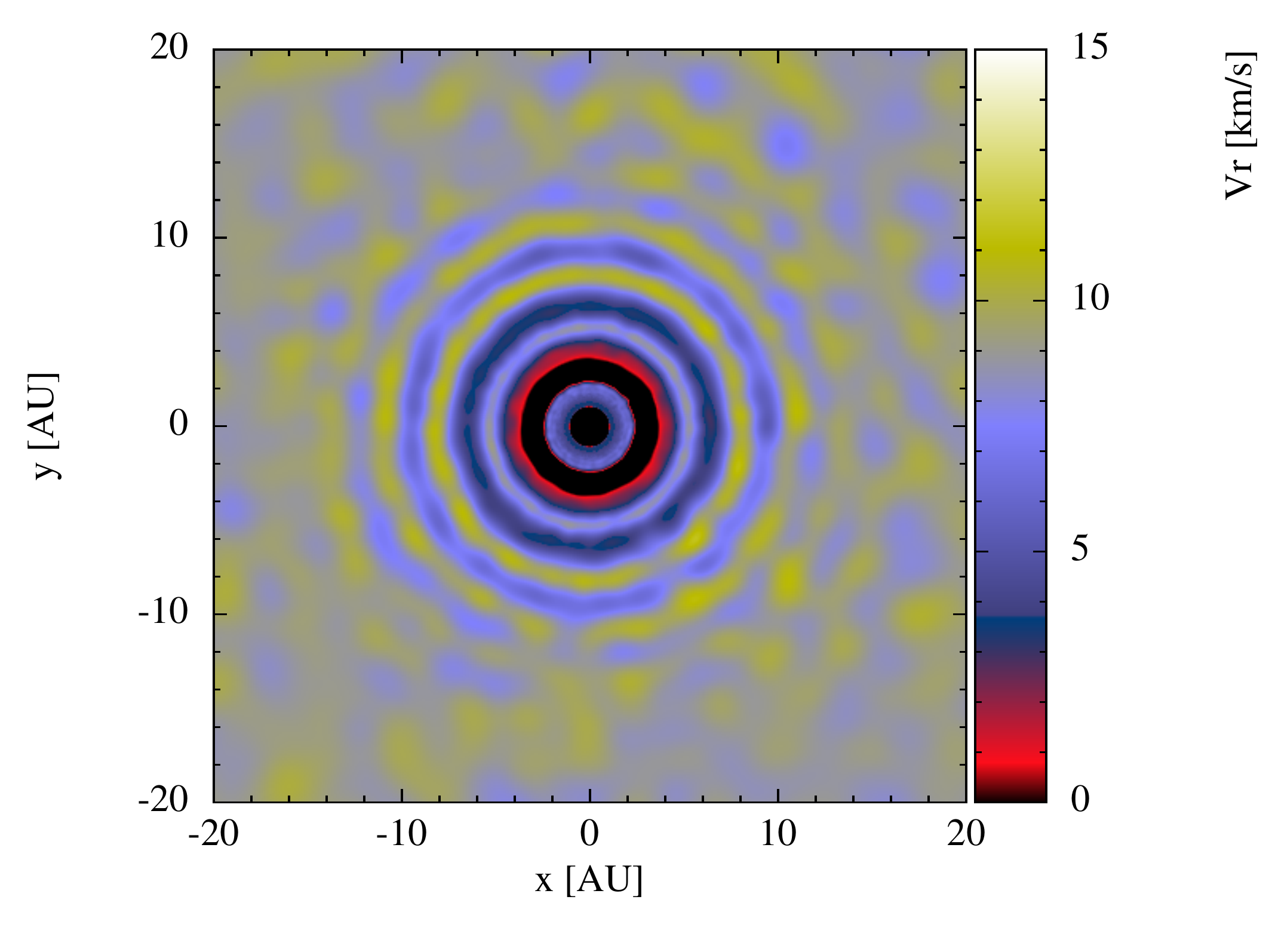}
\includegraphics[width=0.48\textwidth]{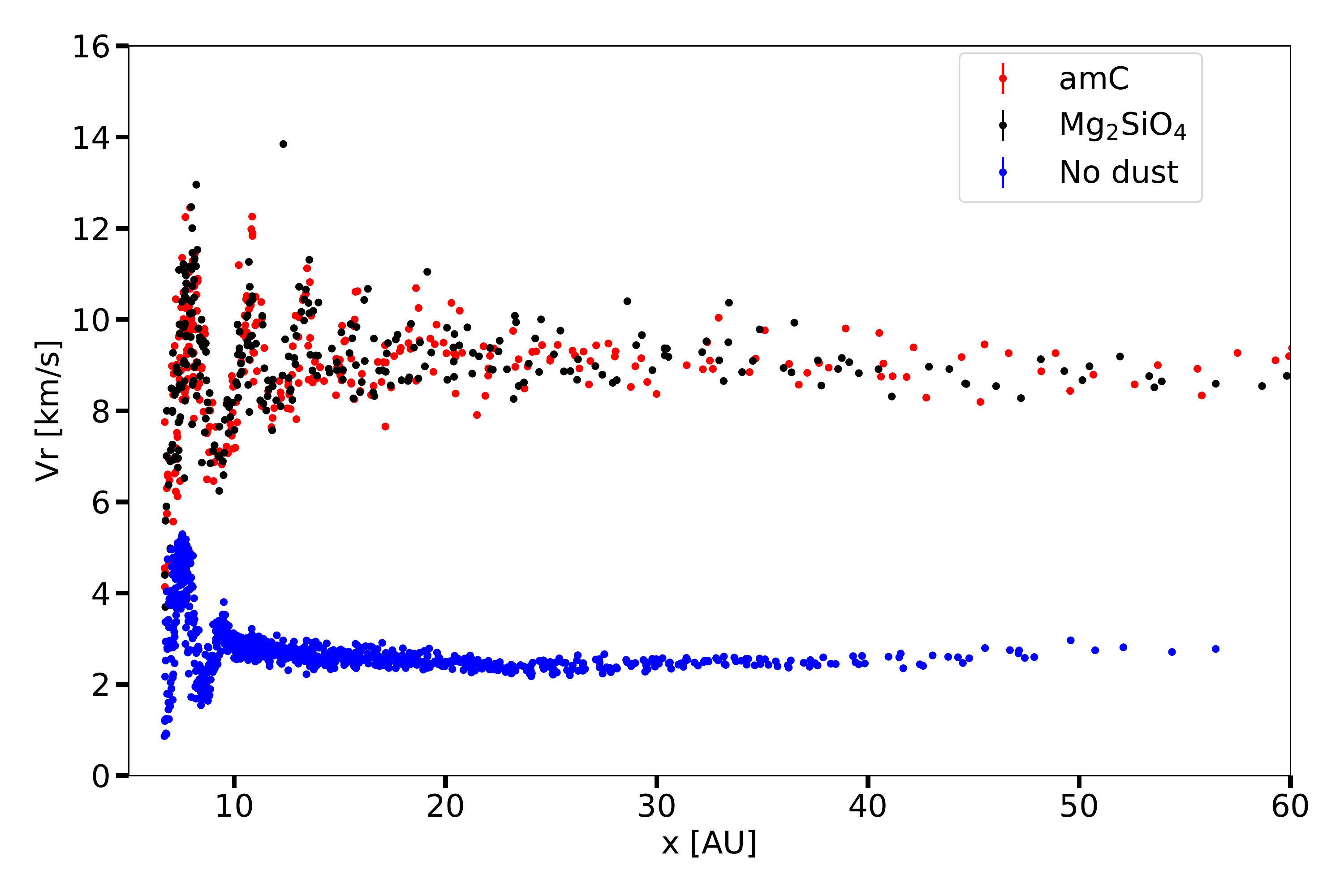}
\includegraphics[width=0.48\textwidth]{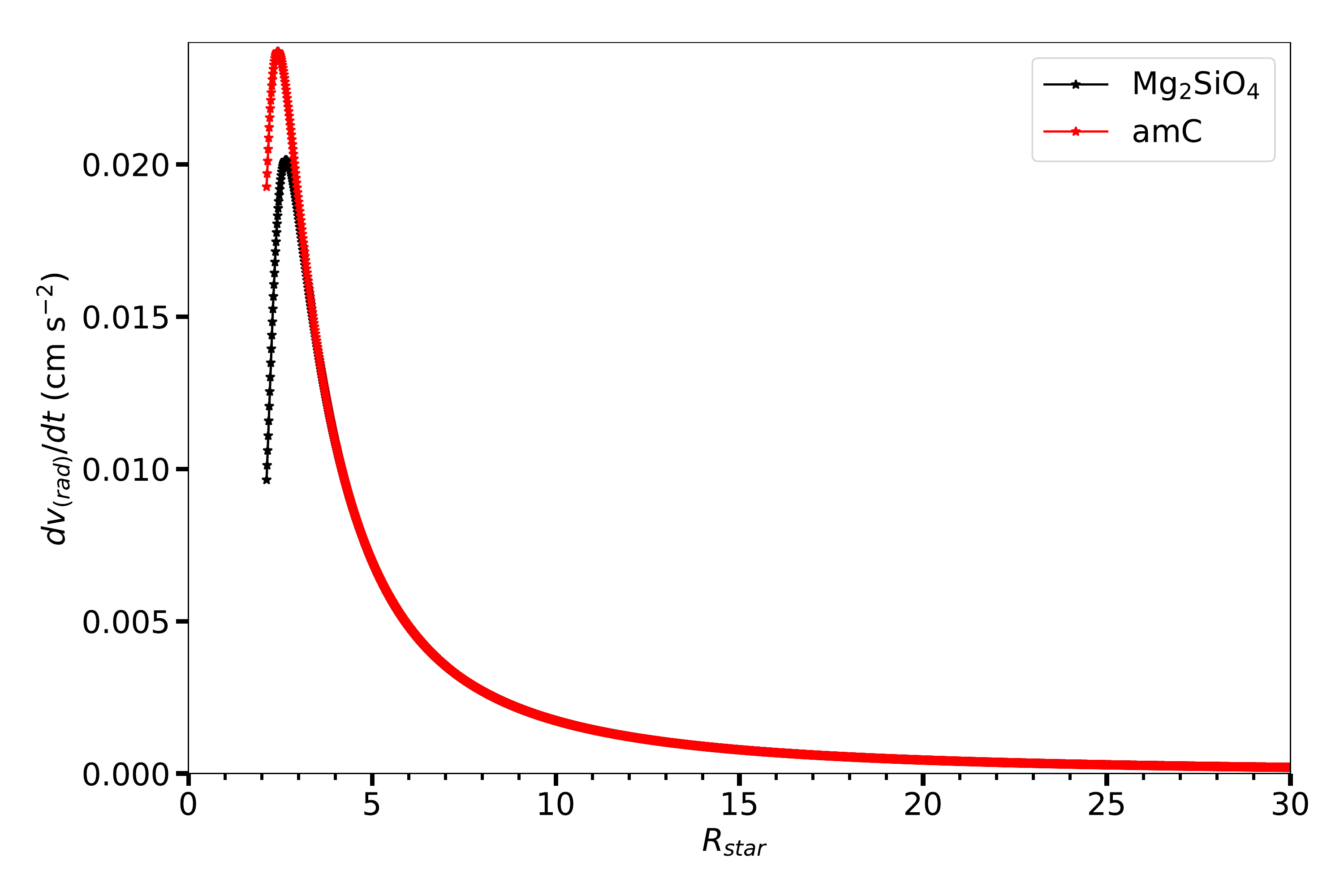}
\caption{\textit{Top:} 2D radial velocity cross-section slices through the orbital plane x-y of a single star after $\approx$ 76 pulsation cycles, with silicate based dust (\textit{left}), and with amorphous carbon dust (\textit{right}; see the right panel of Figure~\ref{Fig:single_star_resolution} for comparison with a single star without dust). \textit{Bottom left:} radial velocity profiles  along the $x$ direction for $-2 <y\,\mathrm{(AU)}< +2$. We remove the particles which constitute the atmosphere of the star to focus on the $V_r$ profile and shock waves. \textit{Bottom right:} the radial acceleration due to radiation pressure on dust, against radius, for silicate-based (Mg$_2$SiO$_4$) and amorphous carbon-based (amC) dust, as derived using the formalism and assumptions made in the code.}
\label{Fig:single_star_dust}
\end{center}
\end{figure*}

\section{Results}
\label{Results}
Our results consist of a set of 40 models. The variables in these models are the $P_{\mathrm{orb}}$, $d_{\mathrm{comp}}$, the mass of the companion ($M_{\mathrm{comp}}$), $U_{\mathrm{amp}}$, and the resonance mode $M$:$N$ (for every $N$ pulsation cycles, the companion completes $\sim M$ orbits). Note that for these resonances $M$ is always $<N$ and $N$ is equal in number to the number of spiral structures ($N_{\mathrm{s}}$). The resonance mode represents the $P_{\mathrm{orb}}$-$P_{\mathrm{puls}}$ resonance. Table~\ref{table:models} lists all the models with their different parameters. Models 1 to 32 led to the formation of multiple structures ($N_{\mathrm{s}} \geq 2$). However, models 33 to 40 led to the formation of one dominant spiral arm ($N_{\mathrm{s}} = 1$). In all the models $P_{\mathrm{puls}}$/$P_{\mathrm{orb}} \approx$ M/N $< 1.0$. Systems where this ratio is larger than 1.0 are beyond the scope of this study and are excluded from our analysis. In such systems, $P_{\mathrm{orb}}$ is smaller than $P_{\mathrm{pul}}$ (394 days) implying that the companion is below the photosphere and hence it is already completely engulfed. 
For computational efficiency, models 1 to 40 do not include acceleration due to radiation pressure on dust. The effect of this process is studied with a small subset of models and we present the results in Section~\ref{dust_section}.

\begin{figure*}
\begin{center}
\includegraphics[width=0.8\textwidth]{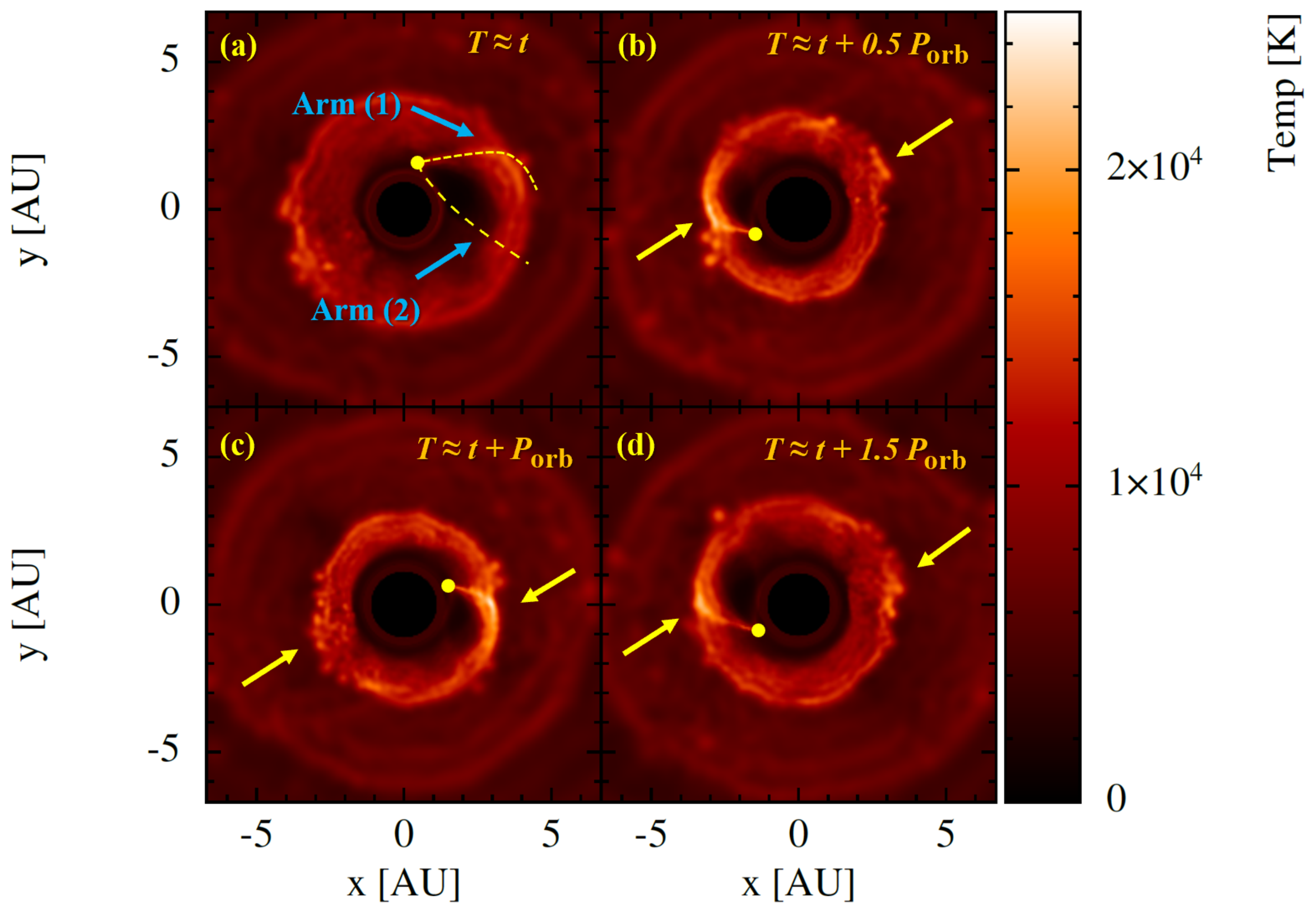}
\caption{2D temperature cross-section slices through the orbital plane from model 9. Figure (a) shows the companion traveling in the dense circumstellar environment creating a wake behind it consisting of two arms, a leading arm denoted (1) and a trailing arm denoted (2). The companion position is identified by a yellow point and the arms are highlighted with dashed yellow lines for guidance. Figure (b) shows two shocks (indicated by the blue arrows) forming due to the interaction of arm (1) with the shock wave on the left-hand side, and the tail of arm (2) with the same shock wave on the right-hand side of the orbital plane. Figure (c) is same as (b) but after $\sim$ 1 pulsation cycle (half an orbital period). Figure (d) is same as (b) but after $\sim$ 2 pulsation cycles (one orbital period). The shocks on both sides cluster to form two spiral arms centered on the AGB star.}
\label{Fig:shocks-comp}
\end{center}
\end{figure*}

\subsection{Formation of periodic shock waves}
\label{sw}
The stellar pulsation creates a cycle of shock waves that travel outwards carrying material away from the star. These shocks form when the outgoing material (due to the pulsation) meets with the in-falling material (due to the gravitational potential of the star). At the shock formation region, the temperature of the gas increases up to $\sim$1.5$\times10^4$\,K and the gas velocity attains $\approx$ 6\,--\,10\,km\,s$^{-1}$.
As the shocks propagate outward, they decelerate and cool down. 

In the case of a single star, the shocks produced by the radial, sinusoidal motion of the piston are spherically symmetric. In Figure~\ref{Fig:shock-waves} we show  cross-sections through the orbital plane of different gas parameters for a single star (model 0). The assumption of spherically symmetry for the piston and shock excitation is idealized (see e.g. \citealt{Freytag_Hofner_2008,Freytag_etal_2017}). The detailed models of \citet{Freytag_etal_2017} show that convection cells and fundamental-mode pulsation will result in chaotic shock wave structures around the AGB star, shifting from spherical symmetry. However, they also show that if the radial velocity is averaged over a sphere around the star, radial periodic shock waves are clearly visible (see figure 5 in \citealt{Freytag_etal_2017}). We will elaborate more in Section~\ref{Disc} on the implications of this assumption for  our results. 

The interpolation at the heart of SPH means that sharp discontinuities require large numbers of particles (high  resolution) to minimize the smoothing region. Although limited due to computational expense, we explore the effect of changing the resolution on the formation of the radial shock waves. We simulate a single star with $2.1 \times 10^5$ and $1.1 \times 10^6$ particles. In Figure~\ref{Fig:single_star_resolution} we show radial velocity cross-sections from these two simulations. The shock waves are better resolved for higher-resolution simulations as expected. Obviously, the increase in the resolution allows for better estimation of the gas properties near discontinuities. In the simulation with the lowest resolution ($2.1 \times 10^5$ particles), the radial periodic shock waves are barely resolved. 

We also test the effect of including the acceleration due to radiation pressure on dust grains on the outflow structure and morphology. In these simulations we use only $1.1 \times 10^6$ particles to reduce the computational time. In Figure~\ref{Fig:single_star_dust} we show radial velocity cross-sections and radial velocity profiles for a single star with acceleration due to radiation pressure on silicate and amorphous carbon grains (see Figure~\ref{Fig:single_star_resolution}-\textit{left} to compare with the model that does not include dust formation). With the addition of dust acceleration, the outflow velocity of the gas increases, as expected. The gas velocity within the shock region increased by a factor of around 2, when adding dust compared to the model without dust. Both models, with Si-based dust grains or amC based dust grains result in similar gas velocity excess. In Figure~\ref{Fig:single_star_dust} we also present the radial acceleration due to dust as a function of radius. Our values are comparable to those presented in figure~2 of \citet{Wang_Willson_2012}.

\begin{figure*}
\centering
\begin{subfigure}{0.49\textwidth}
\centering
\includegraphics[width = \textwidth]{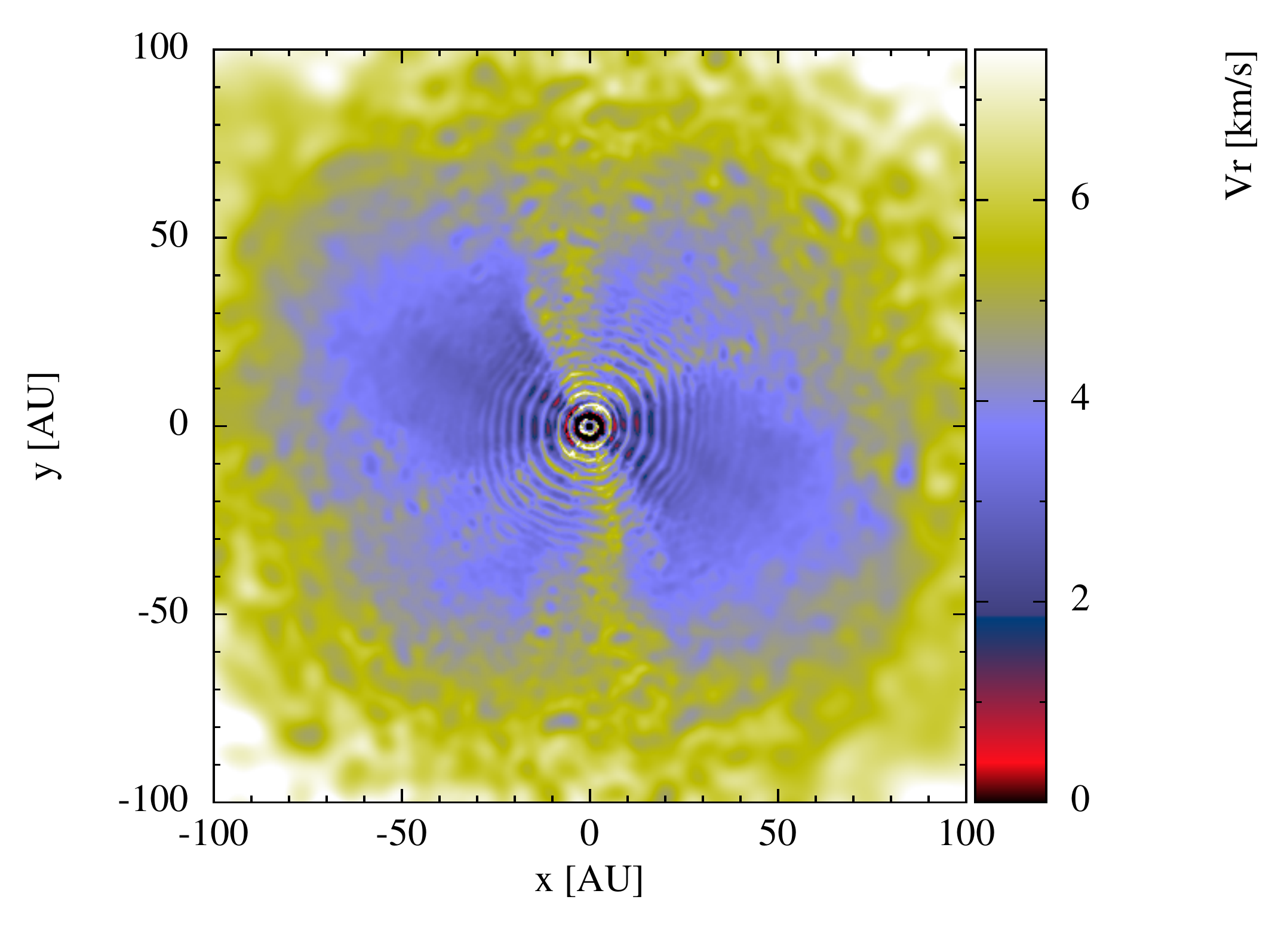}
\label{fig:left}
\end{subfigure}
\begin{subfigure}{0.49\textwidth}
\centering
\includegraphics[width = \textwidth]{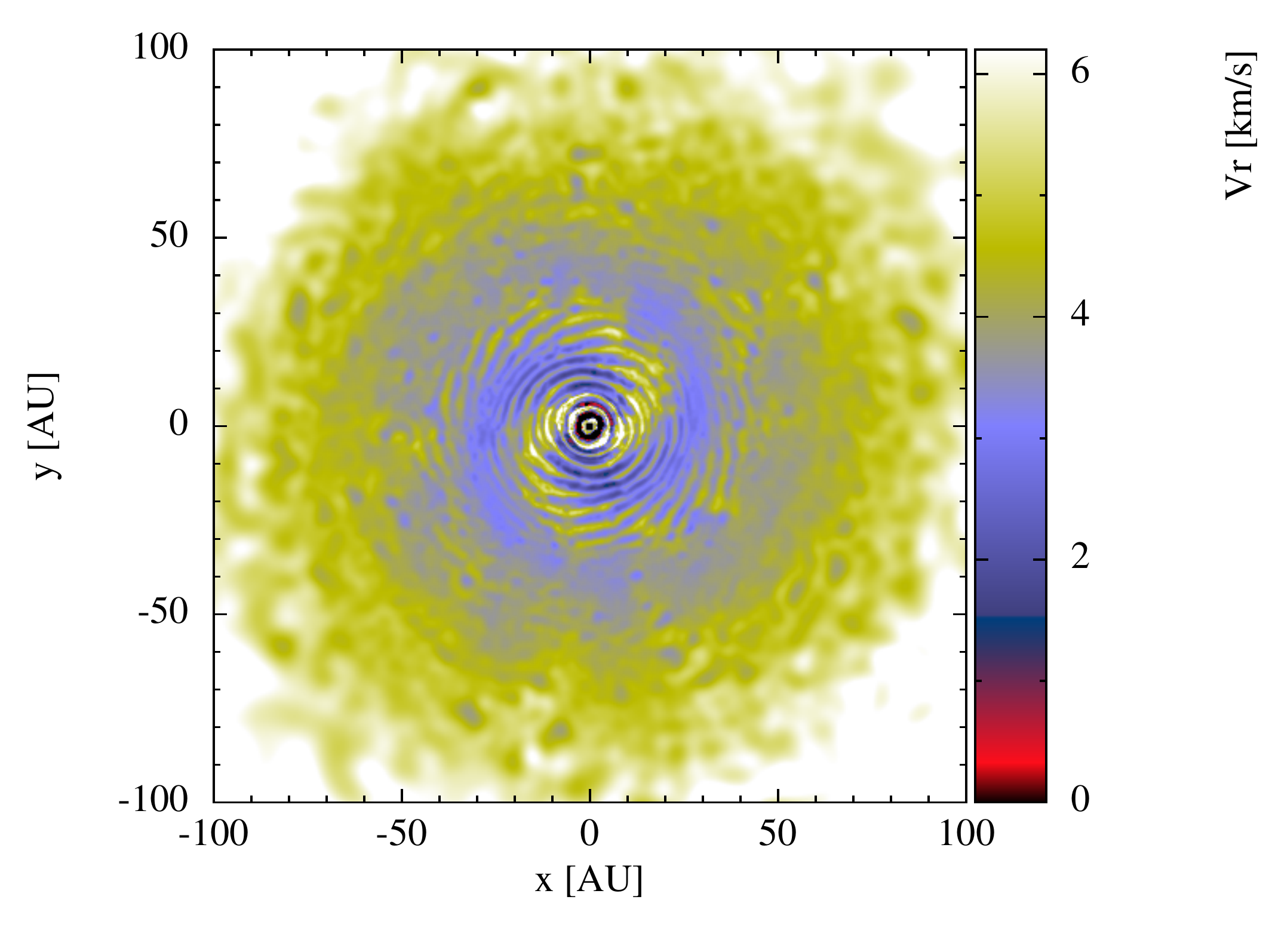}
\label{fig:right}
\end{subfigure}
\begin{subfigure}{0.49\textwidth}
\centering
\includegraphics[width = \textwidth]{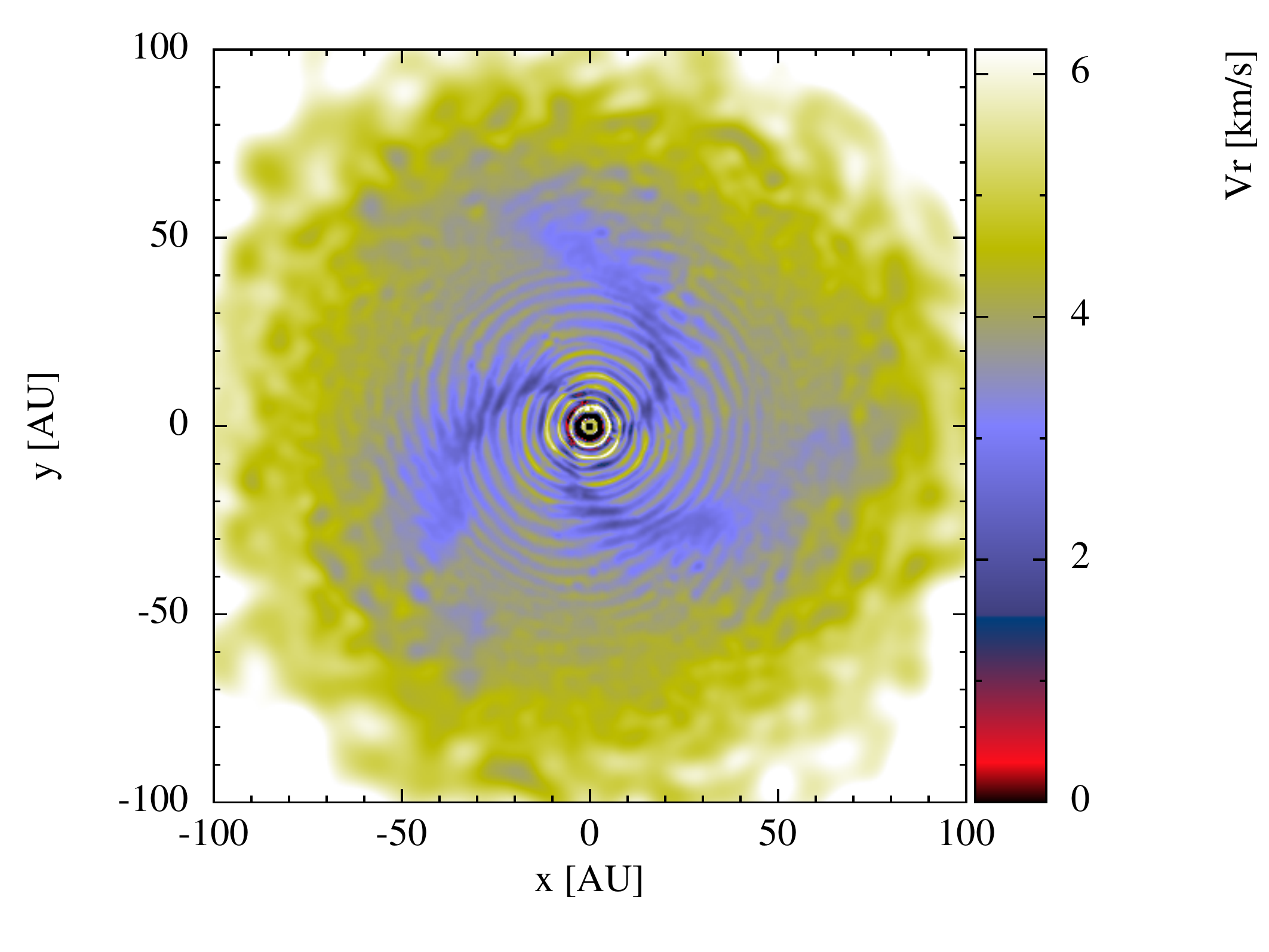}
\label{fig:left}
\end{subfigure}
\begin{subfigure}{0.49\textwidth}
\centering
\includegraphics[width = \textwidth]{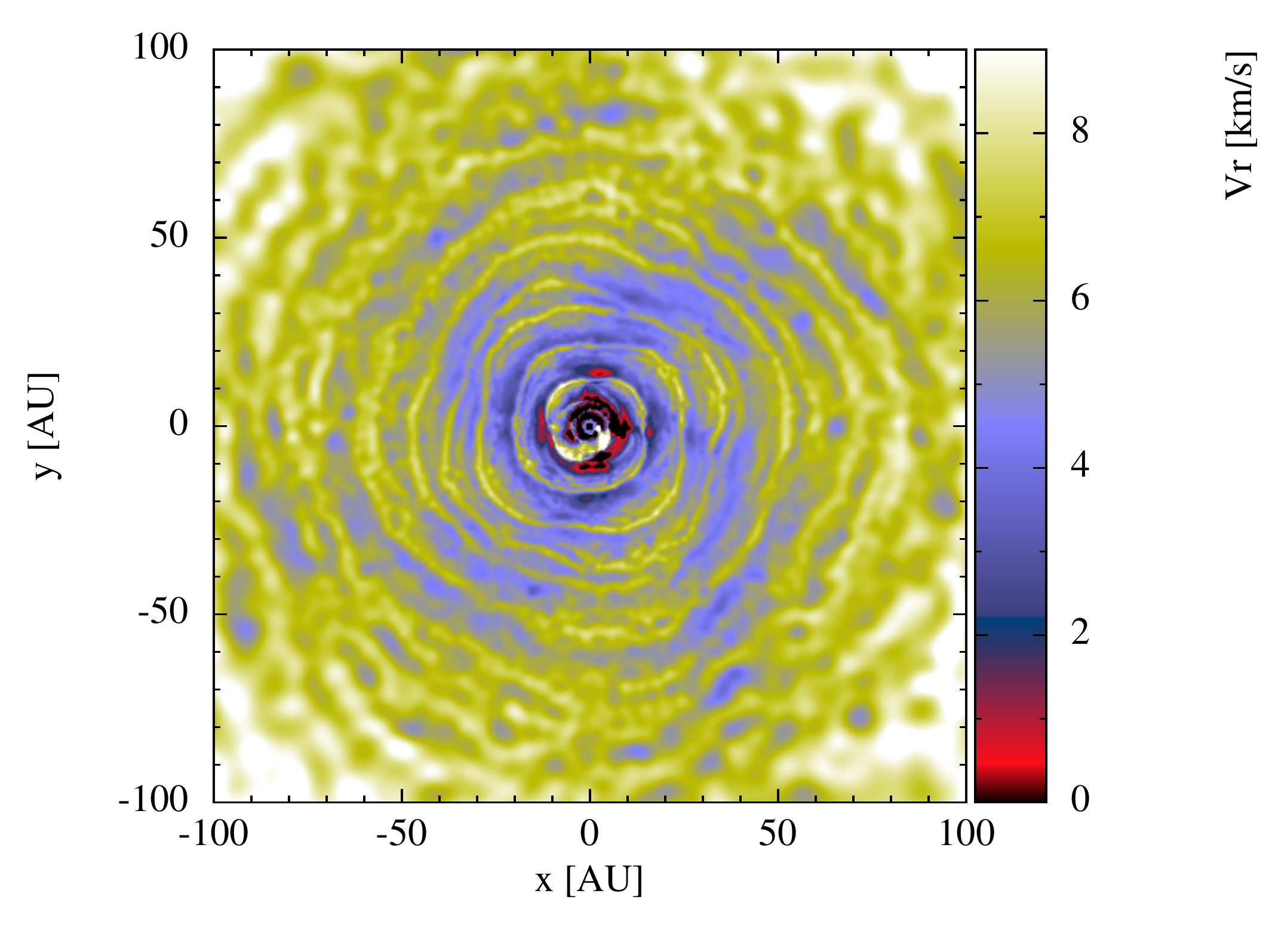}
\label{fig:right}
\end{subfigure}
\vspace{-0.5cm}
\caption{Velocity cross-sections through the orbital plane after $\sim$ 76 pulsation cycles for models 4, 9, 21, and 26, showing the different morphology and numbers of spiral structures depending on the resonance mode between the pulsation and orbital periods. \textit{Top left}: model 4 showing two static bar-like structures due to a resonance exactly equal to 1:2. \textit{Top right}: model 9 showing two spiral structures due to a resonance $\approx$ 1:2. \textit{Bottom left}: model 21 showing 3 spiral structures due to a resonance $\approx$ 1:3. \textit{Bottom right}: model 26 showing 4 spirals due to a resonance $\approx$ 1:4. \textbf{}}
\label{Fig:multiple_spirals}
\end{figure*}

\subsection{Formation of multiple spiral structures}
\label{msa}
If the companion is at a $d_{\mathrm{comp}}$\,$<$\,3$R_*$ and if there is a resonance between $P_{\mathrm{orb}}$ and $P_{\mathrm{puls}}$, multiple spiral structures form in the circumstellar environment centered at the AGB star and extending for more than 70 AU. To elaborate on how these structures form, we choose model 9, where  $P_{\mathrm{orb}}$ is in a resonance mode of around 1:2 with $P_{\mathrm{puls}}$, as an example, and note that the same interaction process occurs in the formation of multiple spiral structures in models 1 to 32  (Table~\ref{table:models}).

While the companion is traveling in the dense circumstellar medium around the star, its gravitational potential creates a wake that trails behind it. A 2D temperature cross-section slice through the orbital plane (panel [a] in Figure~\ref{Fig:shocks-comp}) shows the wake of the companion, consisting of a ``leading arm" denoted arm (1) and a ``trailing arm" (inner arm) denoted arm (2). Due to the close-in circular orbit of the companion, and the resulting high orbital velocity, the two arms have different degrees of curvature, with arm (1) being more curved than arm (2). 

Due to its proximity to the star and the shock formation region, the companion exerts a gravitational pull on the upper atmosphere of the star. This increases the velocity and temperature of the pulled gas below the companion, enhancing the mass loss locally (see, e.g., \citealt{Wang_Willson_2012}). When the upcoming shock waves due to the stellar pulsation meet the wake of the companion, they lead to a strong shock, propagating outward. Particularly, these shocks occur between the shock waves and arm (1) of the wake (blue arrow in panel (a) of Figure~\ref{Fig:shocks-comp}).
At the location of these shocks, the gas temperature rises up to $\approx$ 3 $\times10^4$\,K and the gas velocity increases up to $\approx$ 18\,km\,s$^{-1}$. Compared to its surrounding medium, the shock velocity and temperature are a factor of 3 greater.

After half an orbit, arm (1) of the wake meets a subsequent pulsation cycle, which creates another shock on the other side of the orbital plane (yellow arrow in panel $c$ of Figure~\ref{Fig:shocks-comp}). This repeats with every half orbit and each pulsation cycle. Similar shocks occur between the tail of arm (2) and the stellar pulsations (e.g., yellow arrows in panels $b$, $c$, and $d$ in Figure~\ref{Fig:shocks-comp}).
These shocks cluster on each side of the orbital plane to form two spiral structures, centered on the AGB star, with a long-lasting momentum. The high velocity of the shocks lead to a velocity contrast with the surrounding medium.

Since, the ratio between $P_{\mathrm{puls}}$ and $P_{\mathrm{orb}}$ in model 9 is $\simeq$ 0.48 (slightly offset from $M/N$ = 0.5), this means that the companion meets the shock waves at a slightly earlier location every new pulsation cycle. This is the reason behind the formation of dynamic spiral structures instead of two static bars. We will elaborate more on the formation of static bars in Section~\ref{arms_properties}. 

In Figure~\ref{Fig:multiple_spirals} we show a sample of multiple spiral structures/bars from models 4, 9, 21, and 26. The plots are radial velocity cross-sections through the orbital plane. In each of these models, a certain number of spiral structures (or bars) form around the system depending on the resonance between the orbital and pulsation periods. A time-evolution visualization of the models can be found in the supplementary material.

Our models show spiral structures in the velocity cross-sections, however, the spirals are not clearly resolved in the density cross-section plots (see Figure~\ref{Fig:density}). This is because the density at the shock location is not substantially different compared to the surrounding medium. An increase in the resolution of the simulation might allow us to better resolve the gas properties at the shock location and therefore resolve the spiral structures in the density cross-section. In addition, implementing more realistic cooling regimes will increase the density contrast at the shocks. The 1-D models of \citep{Wang_Willson_2012} show multiple spiral arms in the density cross-sections. This again emphasizes that implementing realistic cooling and increasing the resolution of the simulation are necessary to resolve the spirals in the density plots.

\begin{figure*}
\begin{center}
\includegraphics[width=0.45\textwidth]{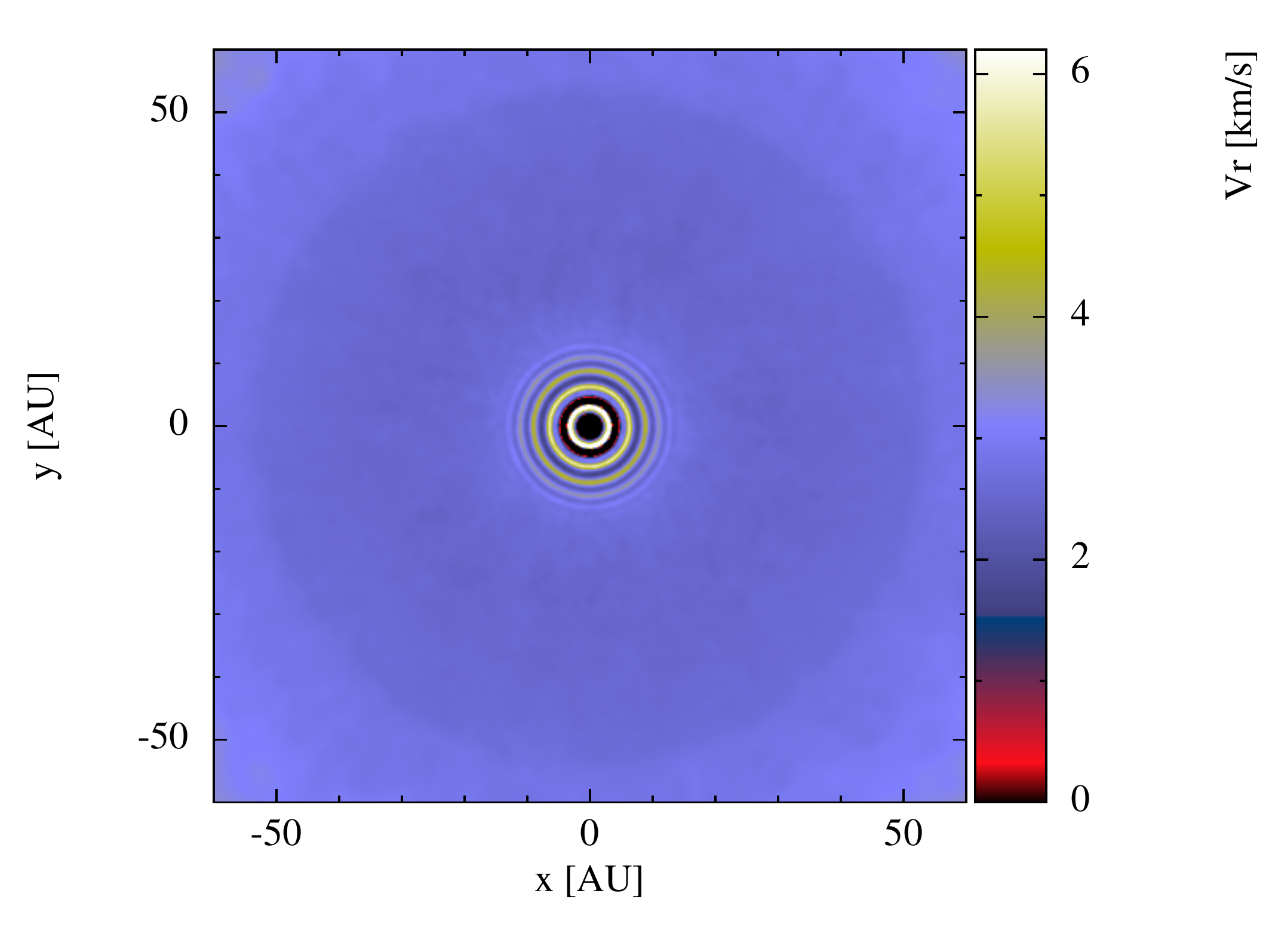}
\includegraphics[width=0.45\textwidth]{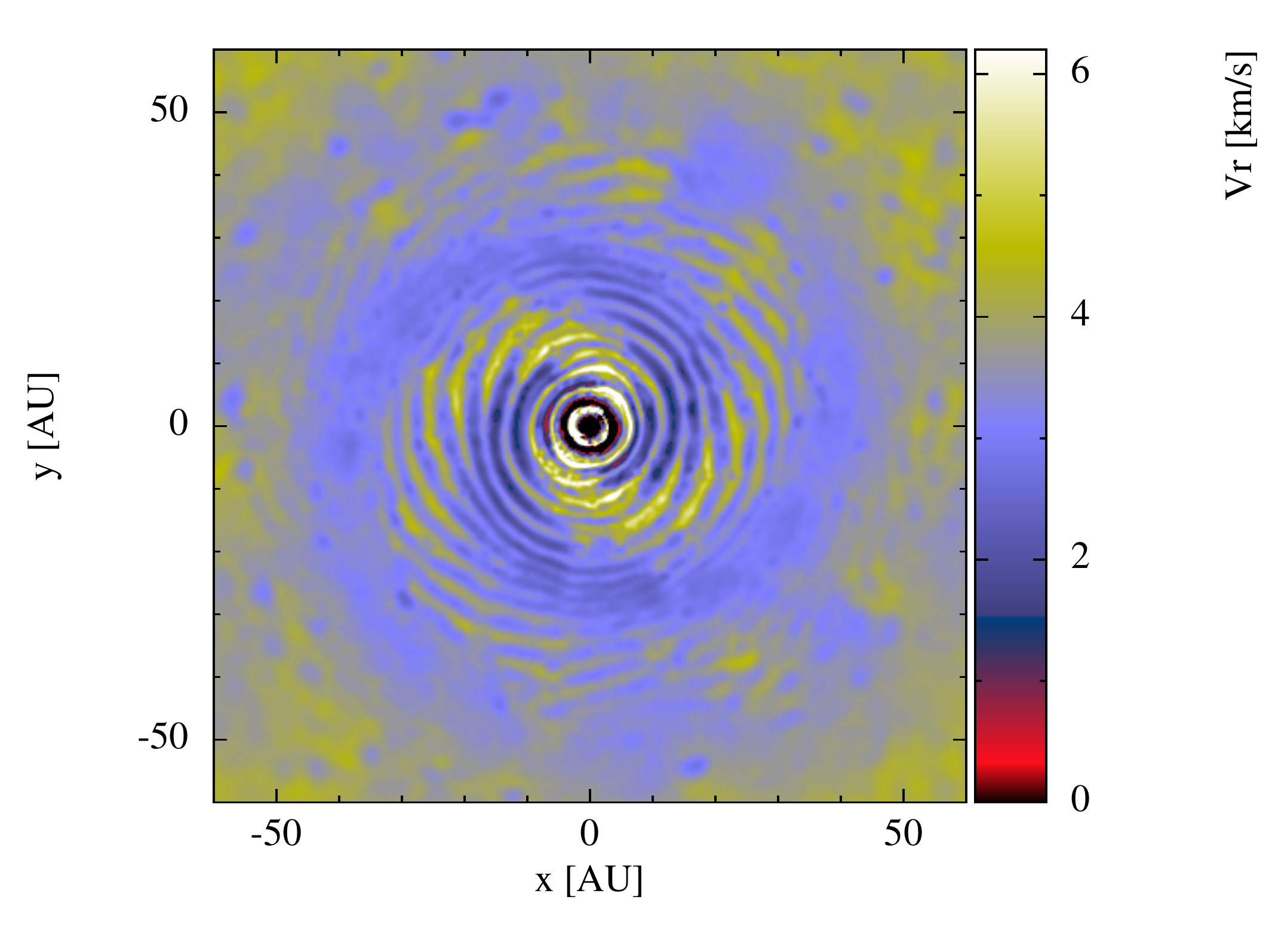}
\includegraphics[width=0.44\textwidth]{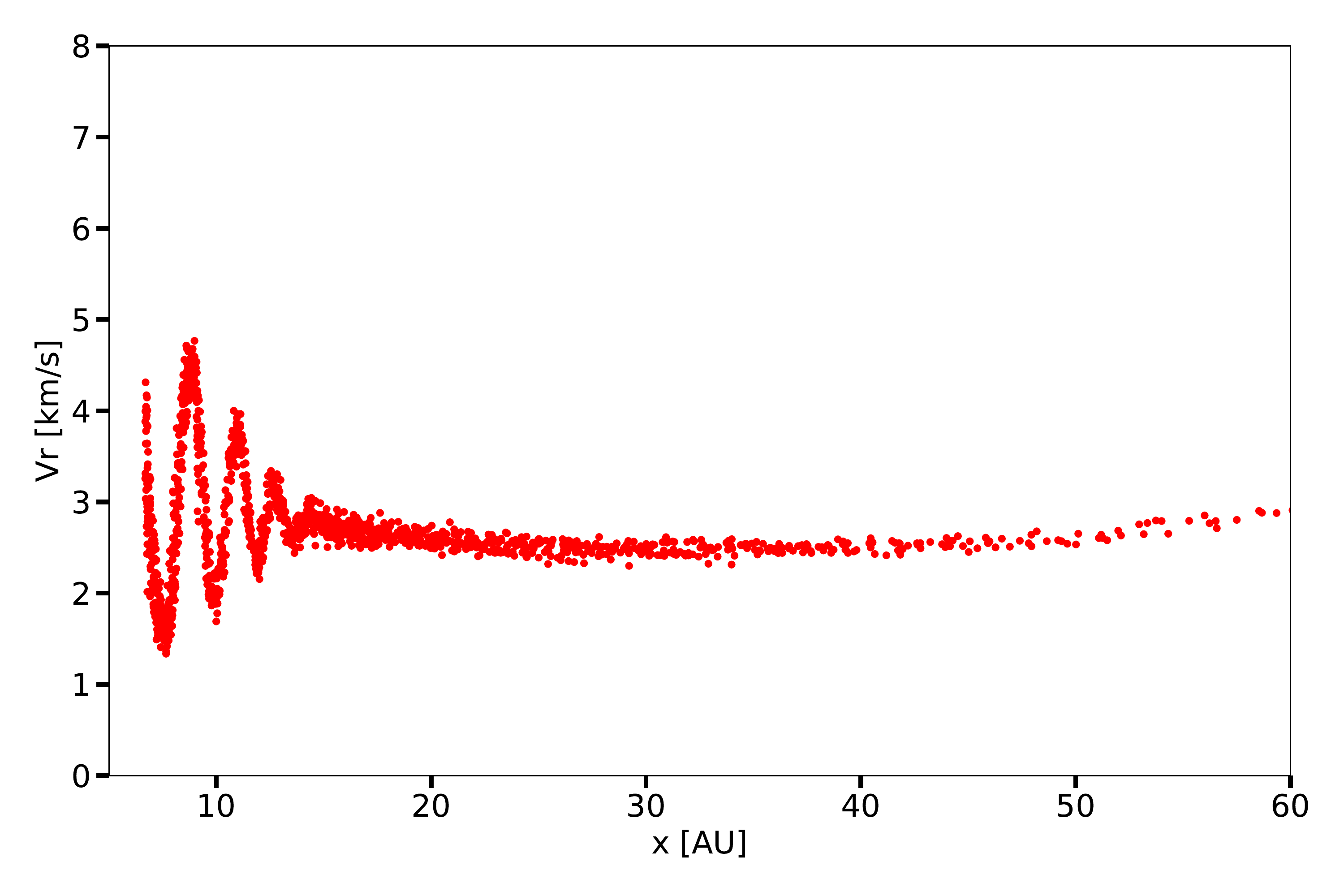}
\includegraphics[width=0.44\textwidth]{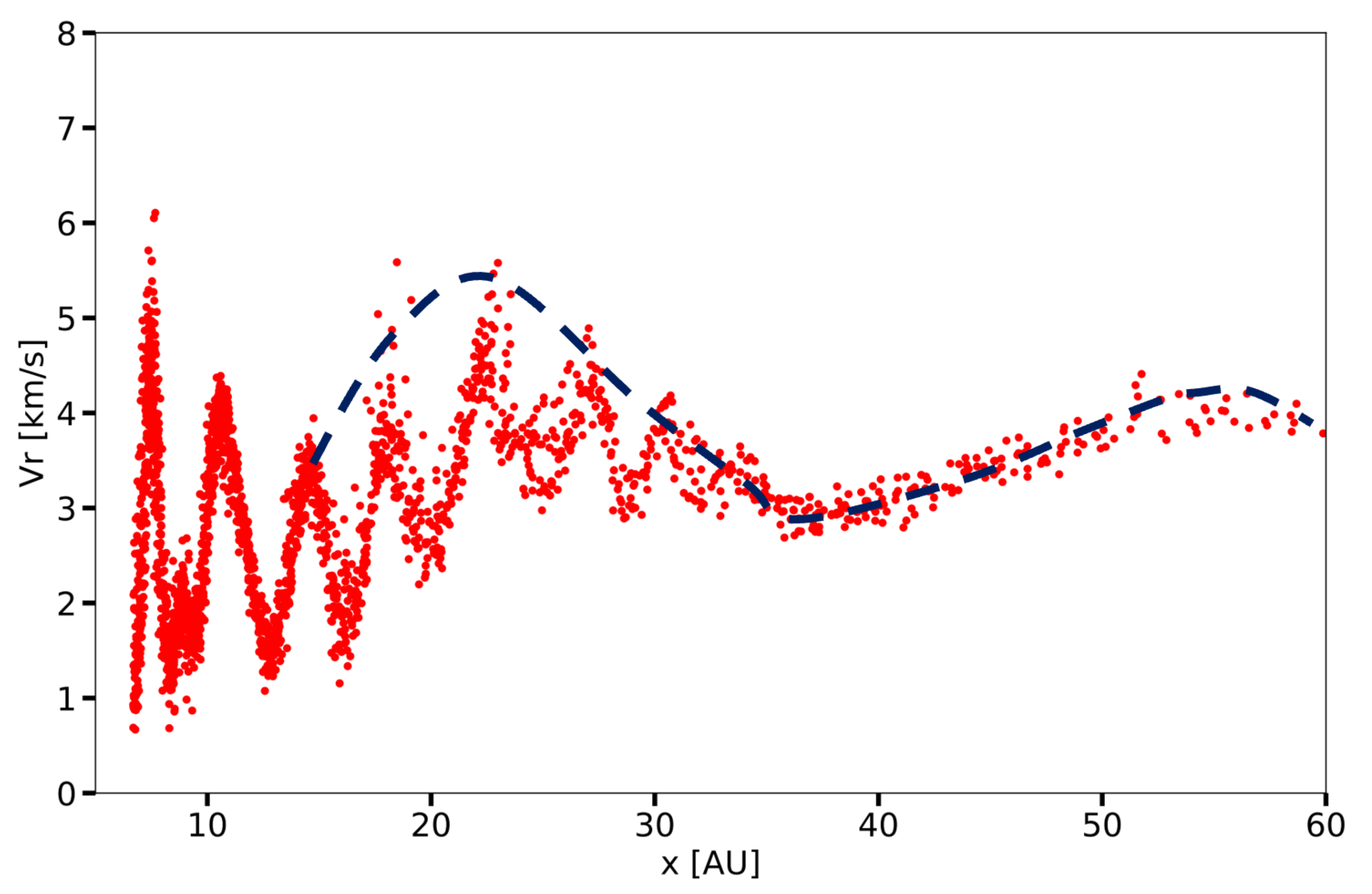}
\includegraphics[width=0.45\textwidth]{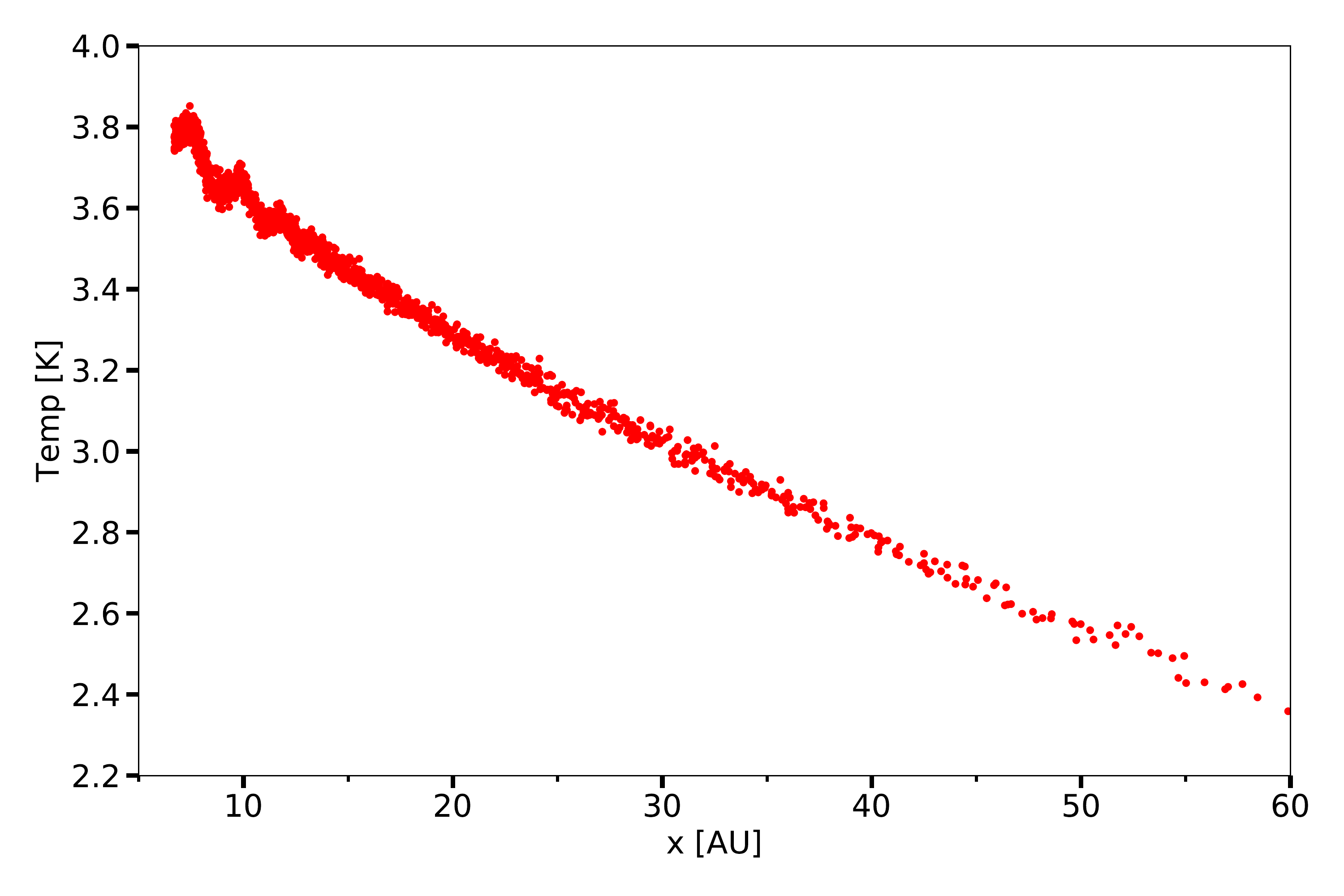}
\includegraphics[width=0.45\textwidth]{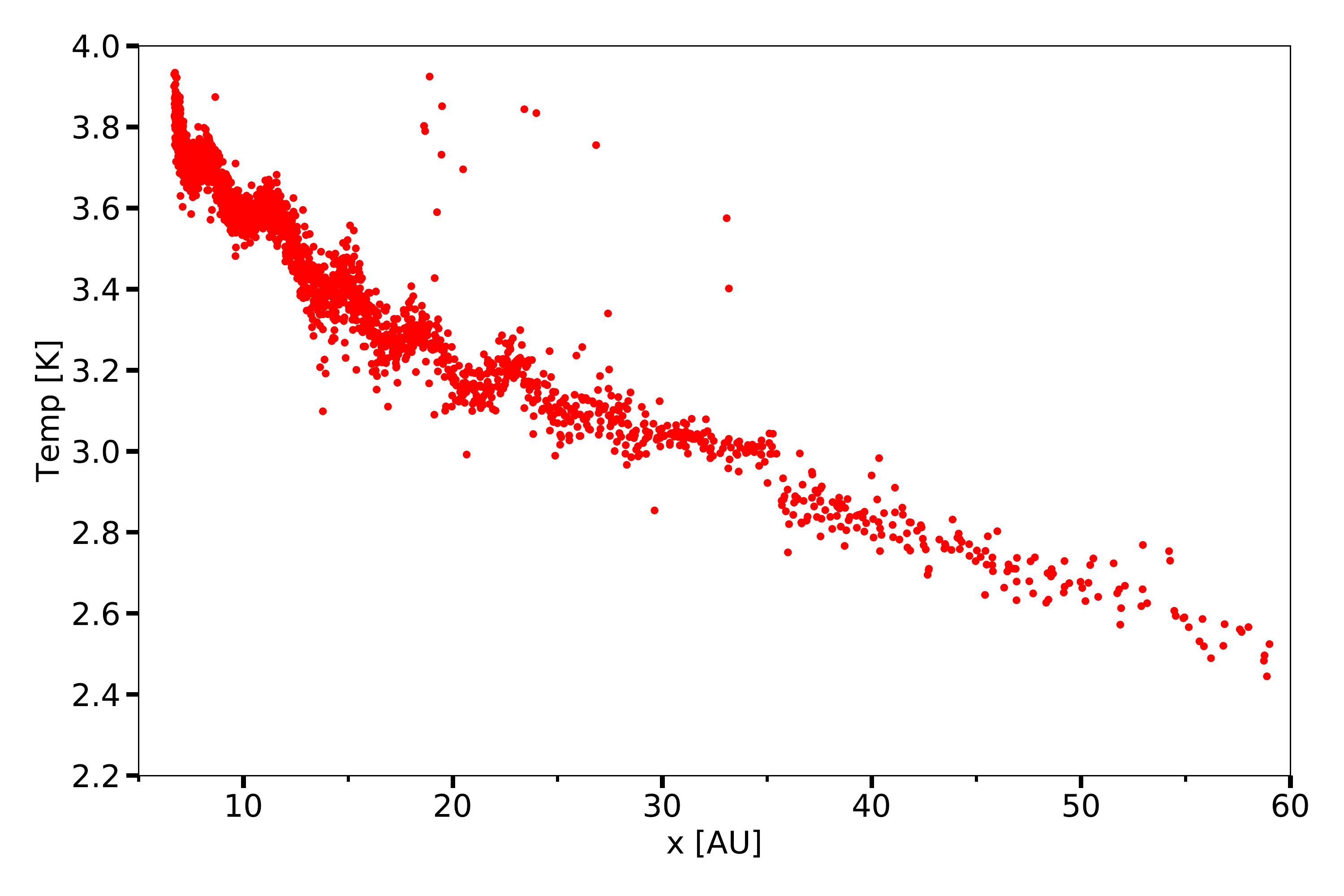}
\includegraphics[width=0.47\textwidth]{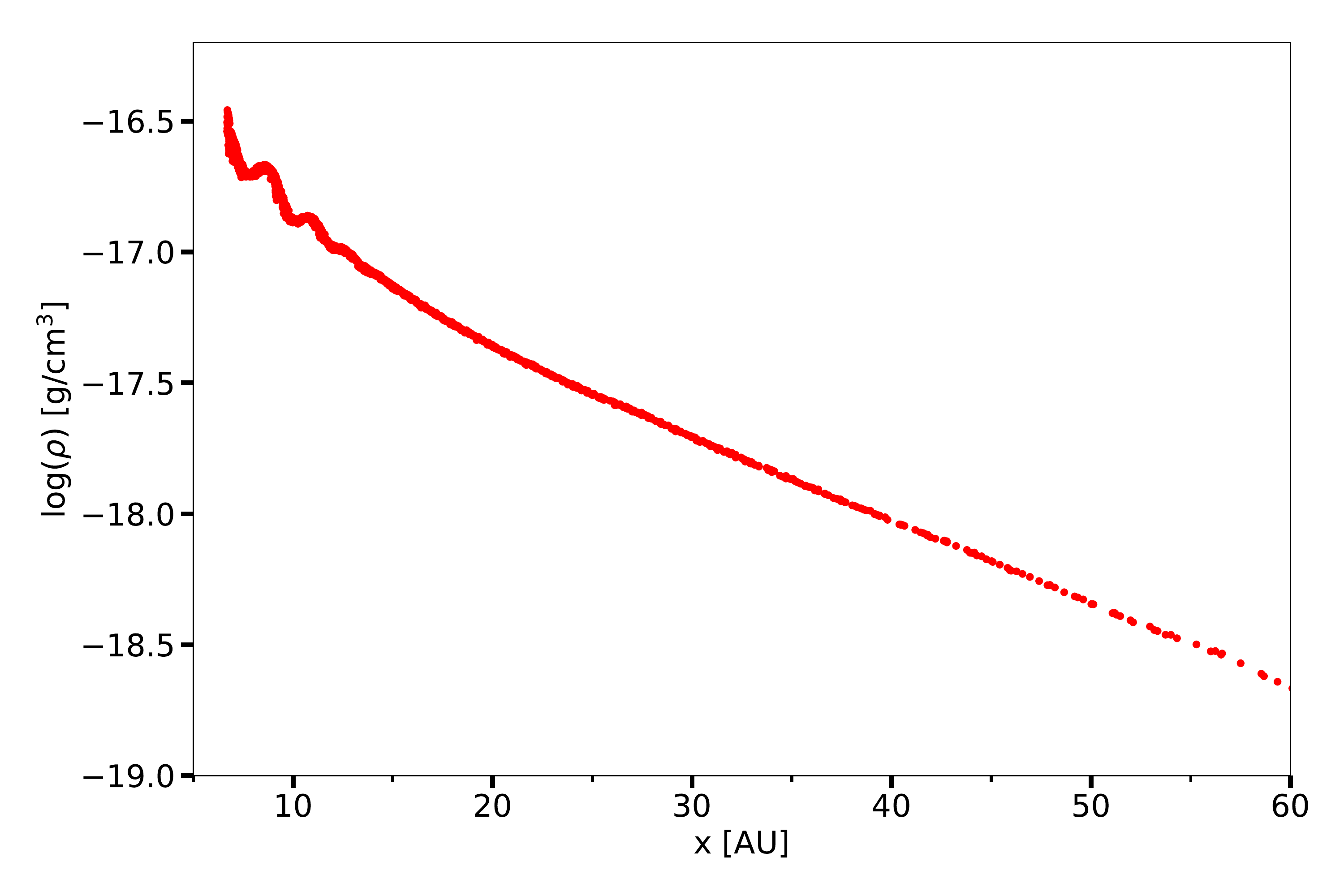}
\hspace{-0.5cm}
\includegraphics[width=0.47\textwidth]{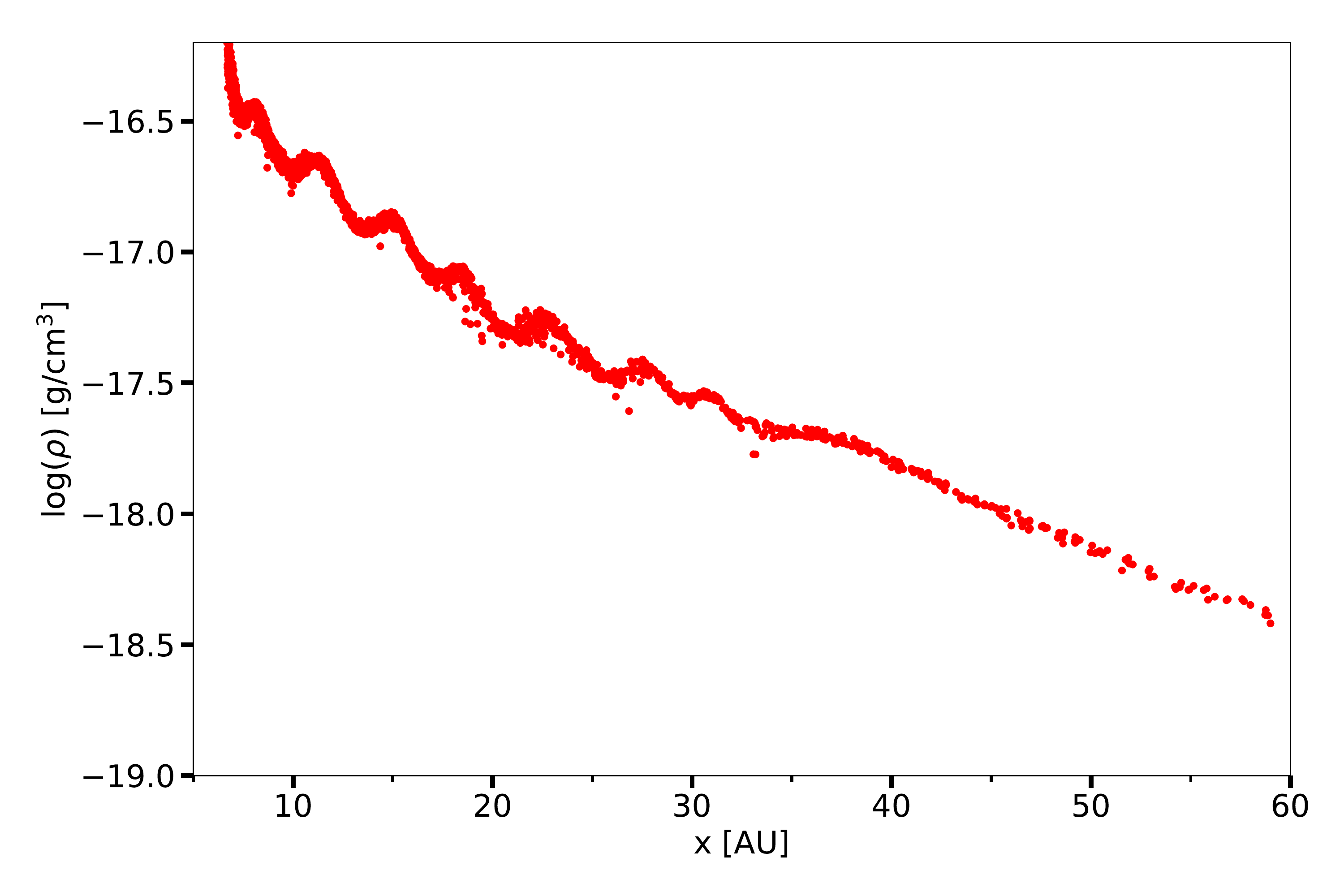}
\caption{From top to bottom: radial velocity cross-section and radial velocity, temperature, and density profiles from models 0 (\textit{left}) and 9 \textit{(right)} after $\approx$ 90 pulsation cycles, along the $x$ direction for $-2 <y\,\mathrm{(AU)}< +2$. We remove the particles which constitute the atmosphere of the star to focus on the $V_r$ profile and spiral structures. We add a blue dashed line for guidance to help the reader track the spiral structure location.}
\label{Fig:model_9_0_profiles}
\end{center}
\end{figure*}

Figure~\ref{Fig:model_9_0_profiles} presents a comparison of radial velocity, temperature, and density profiles along the $x$-axis from models 0 (no companion) and 9, after $\simeq$ 90 pulsation cycles. We choose to show the profiles after 90 pulsations because at this stage the second winding of the spiral arm is parallel to the y-axis, which allows a better representation of the spiral arms in the profiles. The radial velocity profiles are drastically different between the two models. In model 0, the velocity profile shows the stellar pulsations only. However, in model 9, the velocity profile shows (1) the stellar pulsation, which appear as rapid oscillations over short distances ($\sim$ 2-3 AU) and (2) the spiral structures which appear as global oscillations over the extent of the circumstellar medium. Overall, the spiral structures show a velocity contrast of a factor of $\approx$ 2.0 compared to the medium between the spirals. This distinction between models 0 and 9 is not very obvious in the temperature and density profiles, however, there are still major differences between the two models and the shocks between the stellar pulsations and the companions are clearly visible in the profile as temperature and density gradients, particularly between 15 and 35 AU, which is consistent with the location of the spiral arm. 
This indicates the presence of spiral structures in the density profiles but the contrast with the surrounding medium is low, which is the reason these spirals are not resolved in the density cross-sections. 

In Figure~\ref{Fig:entropy}, we show entropy cross-sections for models 9. The two spiral structures are resolved in the entropy cross-section, which emphasizes that the multiple spiral structures are indeed present and are the outcome of clustered shocks between the stellar pulsations and the wake of the companion.

\begin{figure}
\begin{center}
\includegraphics[width=\columnwidth]{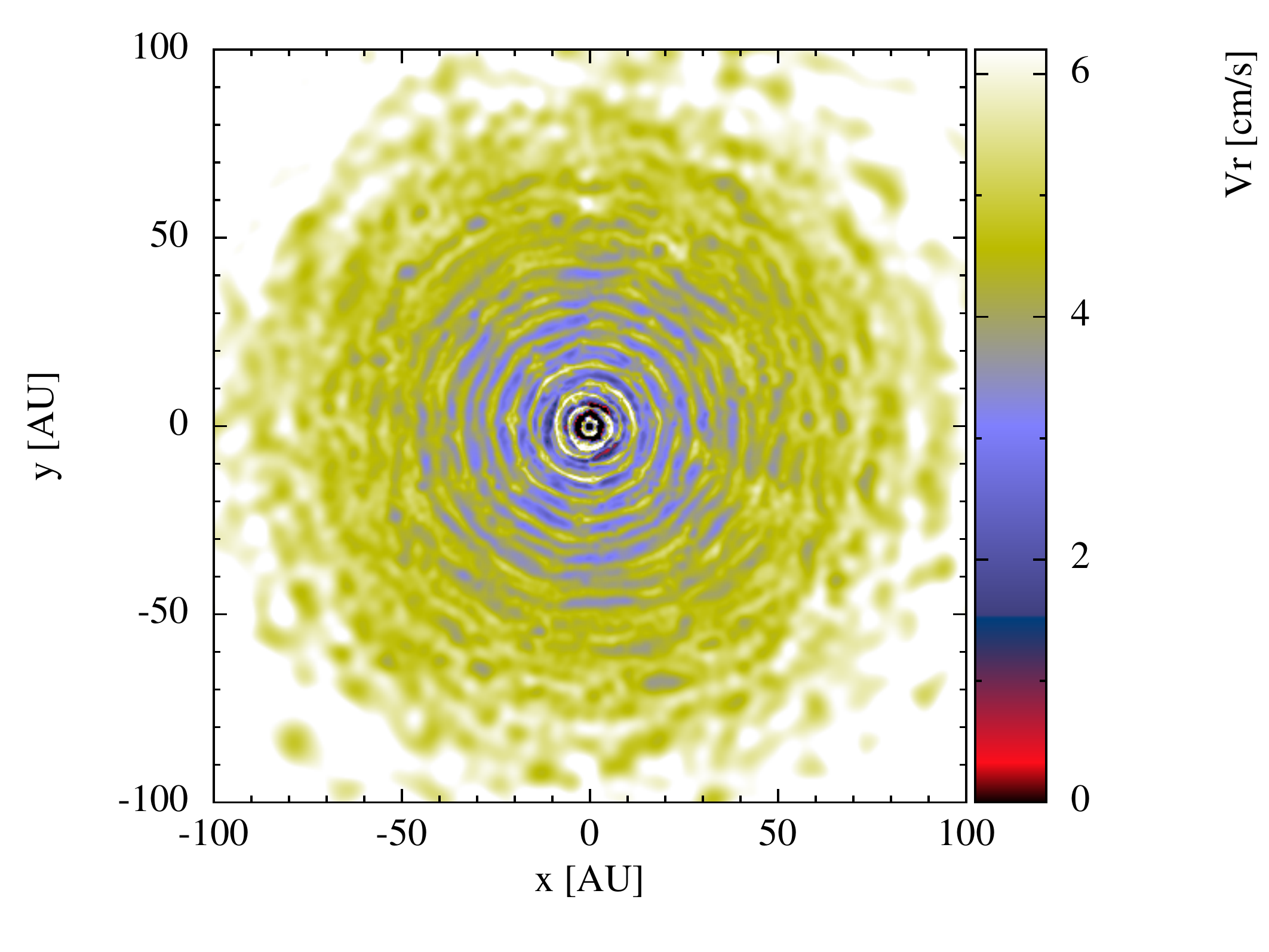}
\caption{Radial velocity cross-section through the orbital plane, after $\approx$ 76 pulsation cycles, from a model that exhibits no $P_{\mathrm{orb}}$-$P_{\mathrm{puls}}$ resonance ($P_{\mathrm{orb}}$ = 1044 days). The shocks do not cluster and hence no spiral structures form.}
\label{Fig:no_spirals}
\end{center}
\end{figure}

If there is no $P_{\mathrm{orb}}$-$P_{\mathrm{puls}}$ resonance, the shocks between the wake of the companion and the pulsations occur at different locations in the orbital plane leading to a random distribution of shocks around the AGB star, hence neither evident spirals nor bars form (see Figure~\ref{Fig:no_spirals}).

\subsection{Properties of the spiral arms}
\label{arms_properties}
The multiple spiral arms forming around the AGB star in the orbital plane are Archimedean spirals propagating in the radial direction. The spirals have the following properties:

\begin{itemize}
\item\textbf{Number and period of the spiral arms:} the number of spiral arms, $N_{\mathrm{s}}$, is directly correlated with the resonance mode and hence with $P_{\mathrm{orb}}$ and $P_{\mathrm{puls}}$. For a resonance mode $M$:$N$, the number of spiral arms is always equal to $N$. We consider the spiral arm period ($P_{\mathrm{sa}}$) as the period of the spiral pattern. It is the time for which the spiral pattern repeats itself. In the case of two spiral arms, the period is one half of a full apparent rotation of a spiral arm. In the case of three spiral arms, the period is one third of a full apparent rotation.
\citet{Wang_Willson_2012} derived a relation between the $P_{\mathrm{sa}}$, $P_{\mathrm{orb}}$, $P_{\mathrm{puls}}$, and $N_{\mathrm{s}}$ based on their models and some analytic considerations. We find the same relation based on our models. For $N_{\mathrm{s}} > 1$, $P_{\mathrm{sa}}$ is derived using:
\begin{equation}
P_{\mathrm{sa}} = \frac{P_{\mathrm{orb}}P_{\mathrm{puls}}}{\left|MP_{\mathrm{orb}} - NP_{\mathrm{puls}}\right|},
\end{equation}
where $M$ and $N$ represent the resonance mode $M$:$N$. In Table~\ref{table:models} we list $P_{\mathrm{sa}}$ for all our models.

\begin{figure*}
\begin{center}
\includegraphics[width=0.8\textwidth]{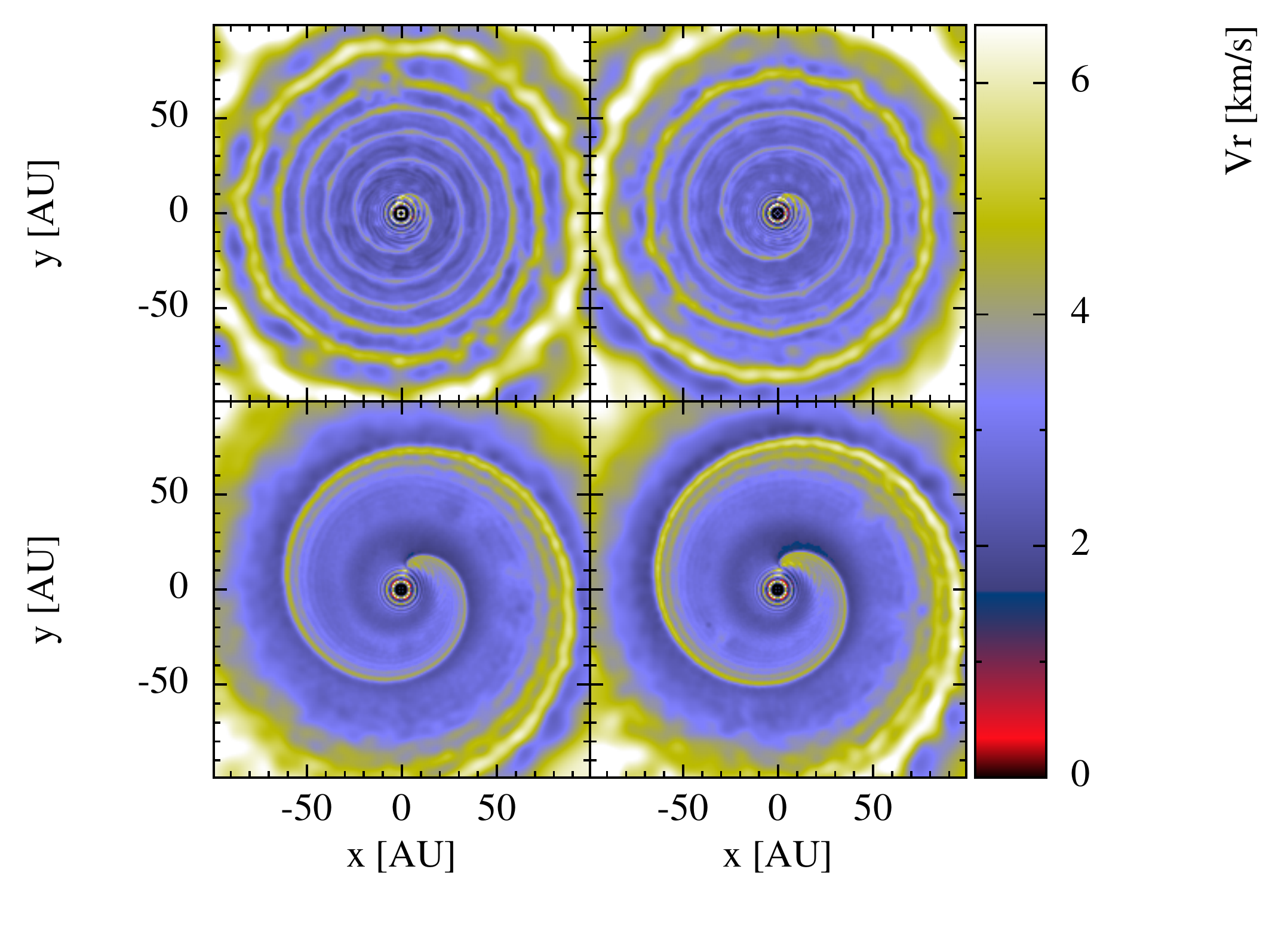}
\vspace{-1.cm}
\caption{\textit{Top}: 2D radial velocity cross-section through the orbital plane from models 33 (\textit{left}) and 36 (\textit{right}), after $\approx$ 76 pulsation cycles. \textit{Bottom}: same for models 39 (\textit{left}) and 40 (\textit{right}), after $\approx$ 88 pulsation cycles. All models show a single spiral in the orbital plane.}
\label{Fig:1spiral}
\end{center}
\end{figure*}

\vspace{0.1cm}
\item \textbf{The winding of the spiral arms:} depending on the orbital period of the companion, the spiral structures show faster or slower winding. We discuss this thoroughly in Section~\ref{Porb}.
Figure~\ref{Fig:arm_spacing_9} represents radial velocity cross-section plots of different time-steps from model 9, which shows the winding of the spiral arms with time. The arms appear to be merging and fading at larger distances, however, given that the spirals are moving radially, we do not expect them to completely merge. The apparent merging and fading of the arms is a numerical effect due to the limited number of particles in our simulations. Injecting more particles in the simulation should maintain the spiral structures over a longer time-span. In addition, our simulations show fading shocks/spirals at distances larger than 70 AU. This is expected given that the shocks are spreading over larger distances in a region of low resolution, near the boundaries of our simulations. Similarly, Gawryszczak et al. (2002) carried out SPH simulations of AGB stars in close binary systems. Their models result in the formation of a single spiral structure due to the orbital motion of the secondary star. In these models they estimate the diffusion (dissipation) radius ($r_d$) and timescale ($t_d$) of the spirals due to density gradient at the boundary of their simulations to be $r_d = t_d \times v_w = P_{\mathrm{orb}} \times v_w^2/c$, where $v_w$ is the wind velocity and $c$ is the local sound speed. However, they also note that due to the decrease in the resolution of the simulation and increase in the smoothing length at its boundary, the dissipation radius of the spirals becomes much smaller then the predicted one and is dominated by the numerical dissipation.

\vspace{0.1cm}
\item \textbf{Equation of motion of the spirals:} although they appear to be rotating in a specific direction, the shocks forming the spiral arms propagate in the radial direction. The apparent rotation of the spiral pattern and the winding of the spiral arms are due to the offset between ratio of $P_{\mathrm{puls}}$/$P_{\mathrm{orb}}$ and $M/N$. The shocks cluster into spiral arms that propagate in the radial direction with a speed ($V_{\mathrm{s}}$) depending on the $M_{\mathrm{comp}}$, $U_{\mathrm{amp}}$, and the local sound speed. The local adiabatic sound speed is density dependent and it is expressed as $c_s^2= \gamma p / \rho = \gamma (\gamma - 1)e$ with $\gamma$ being the adiabatic index, $e$ the gas specific internal energy, $\rho$ the gas density, and $p$ the gas pressure. The gas Mach number ($M_{\mathrm{gas}}$) is then derived by dividing the gas radial velocity by the local sound speed: $M_{\mathrm{gas}} = V_r / c_s$. The Mach number of the spiral structure ($M_{\mathrm{s}}$) is equal to $M_{\mathrm{gas}}$ at the location of the spirals (where the shocks sweep through the gas). In Figure~\ref{Fig:Mach_number_2} we plot the Mach number of the gas for models 1 to 12, after $\simeq$ 76 pulsation cycles. In most of the models, the spiral arms show supersonic speeds. From our models and analytical considerations, we derive the following equation that describes the equation of motion of the spiral arms:
$$
r(t) \sim  2\,\theta (t) \,V_{\mathrm{s}}(r,t) \,d_{\mathrm{comp}} + d_{\mathrm{comp}},
$$
where $r(t)$ defines the radial position of a section of the spiral arm at time $t$ and  $\theta (t)$ is equal to   $(t/P_{\mathrm{sa}}) \pi$.

\vspace{0.1cm}
\item \textbf{Formation of bars:} if the ratio of $P_{\mathrm{puls}}/P_{\mathrm{orb}}$ is $\simeq$ $M/N$ (e.g., model 3, 4, 5), the shocks cluster into bar-like structures instead of dynamic spiral arms (see top-left panel of Figure~\ref{Fig:multiple_spirals}). For such models, the shocks between the pulsation and the companion's wake occur at almost the same location during every pulsation cycle. If the ratio of $P_{\mathrm{puls}}/P_{\mathrm{orb}}$ is exactly equal to $M/N$, the bars are static (e.g., model 4), however, if it is slightly greater or smaller, the bar-like structures are then dynamic, but their period is very long and they eventually turn into spiral arms on a timescale of the order of $\simeq$ $P_{\mathrm{sa}}$ (e.g., model 5; see Section~\ref{Porb}).

\end{itemize}

\subsection{Observing the system at different viewing angles}
One of the advantages of 3D modeling is the ability to observe the system in different directions. Cross-section slices through the polar planes, (z-x) and (z-y), do not show spiral structures, instead they show complex structures, such as arcs and bars around the system (Figure~\ref{Fig:yz} and~\ref{Fig:xz}). The formation, morphology, and symmetry of these structures depends on the resonance between $P_{\mathrm{puls}}$ and $P_{\mathrm{orb}}$.

To describe the formation of these arcs and bars, we will adopt model 9 again as an example. The wake of the companion resembles a cone-like-structure in 3D (Figure~\ref{Fig:shocks_xz}). 
If the companion is traveling through the x-z or y-z plane, a 2D cross-section through the wake of the companion in these planes is a 2D-circle-shape (see Figures~\ref{Fig:shocks_xz} and~\ref{Fig:wake_xz} ), whose  radius depends on the location of the slice relative to the wake.  
As the stellar pulsations meet the wake of the companion, strong shocks form in the x-z or y-z planes, which cluster to form arcs and bars. The multiple bars (top panel of Figure~\ref{Fig:xz}) are mainly formed by the strong shocks between the upper and lower edges of the wake and the stellar pulsation; due to the proximity of the companion, the amplitude of the stellar pulsations is the highest when meeting the upper and lower edges of the wake. Figures~\ref{Fig:shocks_xz} represents a schematic illustration to illustrate the formation of the arcs/bars in the polar planes. 

The bar-like structures can be seen in the y-z or x-z planes, only if the stellar pulsations meet the companion around z $\approx 0$ or y $\approx 0$, respectively. Otherwise, the structures will form at planes with z and y different from zero and slices through z = 0 and y = 0 do not show evident structures (e.g., model 21 in Figures~\ref{Fig:yz} and~\ref{Fig:xz}). Therefore, the symmetry and the shape of these structures depend on the location of the cross-section slice and the orbital resonance.

\subsection{Formation of a single spiral}
\label{1spiral}

If the companion is at a distance $d_{\mathrm{comp}}$\,$\gtrsim$\,4$R_*$, the sub-stellar companion creates a single spiral when traveling through the circumstellar medium of the AGB star. In Figure~\ref{Fig:1spiral} we show cross-section plots through the orbital plane for models 33, 36, 39, and 40, all showing a single spiral. Single spirals have been observed around several AGB stars (e.g., \citealt{Mauron_etal_2006,Maercker_etal_2012,Kim_etal_2012,Kim_etal_2012b,Randall_etal_2020,Decin_etal_2020,Homan_etal_2020}). In these cases the system is composed of a binary system with a large orbital separation (several dozens of astronomical units). The formation of a single spiral in such binary systems is due to the reflex motion of the binary and the gravitational focusing of the mass lost from the giant star (e.g., \citealt{Mastrodemos_Morris_1999, Mohamed_Podsiadlowski_2007,Raga_Canto_2008,Kim_etal_2012,Guerrero_etal_2020,Chen_etal_2020,Castellanos_etal_2021}).

However, in the case of a sub-stellar companion orbiting the AGB star at a moderate distance $d_{\mathrm{comp}}$\,$\sim$\,4\,--\,10$R_*$ and when $P_{\mathrm{orb}}$ $>>$ $P_{\mathrm{puls}}$, the single spiral arm is formed due to the interaction of the wake of the companion with the circumstellar environment and the stellar pulsations. Due to the gravitational potential of the companion and its wider orbit, the companion drags a large number of particles behind it, increasing the density and velocity in the wake substantially. As the companion orbits the star and the gas moves outwards, the trailing wake of the companion forms a spiral structure centered at the companion. In addition, since $P_{\mathrm{orb}}$ is much larger than $P{\mathrm{puls}}$, the companion meets a new cycle of shock waves after traveling a small fraction of its orbit. Each time the wake of the companion meets a new shock wave cycle, a shock is created. These shocks propagate in the radial direction and cluster into the spiral arm, enhancing its momentum. These shocks are imprinted on the spiral arm as enhanced velocity regions. In Figure~\ref{Fig:1spiral_zoom} we show a zoom-in plot on the shock formation regions for these 4 models, to highlight the interaction between the pulsations and the wake of the companion. The shocks can be seen on both the inner and leading arms of the spiral and they are more obvious on the inner one near the stellar pulsations, compared to the smoother leading arm. This is particularly the case for models 33 and 36 where the companion is closer to the star. In the case of simulations that assume continuous outflows with constant velocity winds (e.g., \citealt{Kim_etal_2012}), these shocks are not seen. 
For models 39 and 40, where the companion is at a distance of 10$R_*$, the spiral arms are smoother, however, the leading arm shows multiple components, forming discrete layers, due to the stellar pulsations. These discrete layers are formed by the gradient in velocity caused by the shock waves from the pulsations and are not typically seen in simulations with a constant velocity outflow.

For models with a single spiral arm (models 33 and 40), the spiral structures are well resolved in the density cross-sections (Figure~\ref{Fig:density}). Since in this case the spirals are created by the companion traveling through the circumstellar medium, the density in the wake of the companion is substantially larger compared to the surrounding medium. The single spiral structure (from model 40) is also clearly resolved in the entropy cross-section (Figure~\ref{Fig:entropy}). 

The period of a single spiral arm, $P_{\mathrm{sa}}$, is equal to $P_{\mathrm{orb}}$. In this case, the spiral pattern repeats itself when the companion completes one full orbit. A time-evolution visualization of the models can be found in the supplementary material.

In Figures~\ref{Fig:1spiral_yz} and~\ref{Fig:1spiral_xz} we show cross-section plots of the same models in the polar, y-z and x-z planes, respectively. These plots slice vertically through the orbital plane and therefore instead of a single spiral arm we see multiple arcs on both sides of the star. The arcs in this case are slices through the spiral arm, showing its inner and leading arms. The number of arcs increases with increased winding of the spiral arm and the opening angle of the arcs is smaller for closer companions ($d_{\mathrm{comp}}\sim$4--5$R_*$, e.g, models 33 and 36) compared to more distance companions ($d_{\mathrm{comp}}\sim$10$R_*$, e.g., models 39 and 40). In all the models, the arcs are composed of multiple layers, which again are caused by the interaction of the wake with the stellar pulsations. The pulsations are clearly visible inside the first arc as spherical shock waves (see Figure~\ref{Fig:1spiral_zoom}).

\begin{figure*}
\begin{center}
\includegraphics[width=0.49\textwidth]{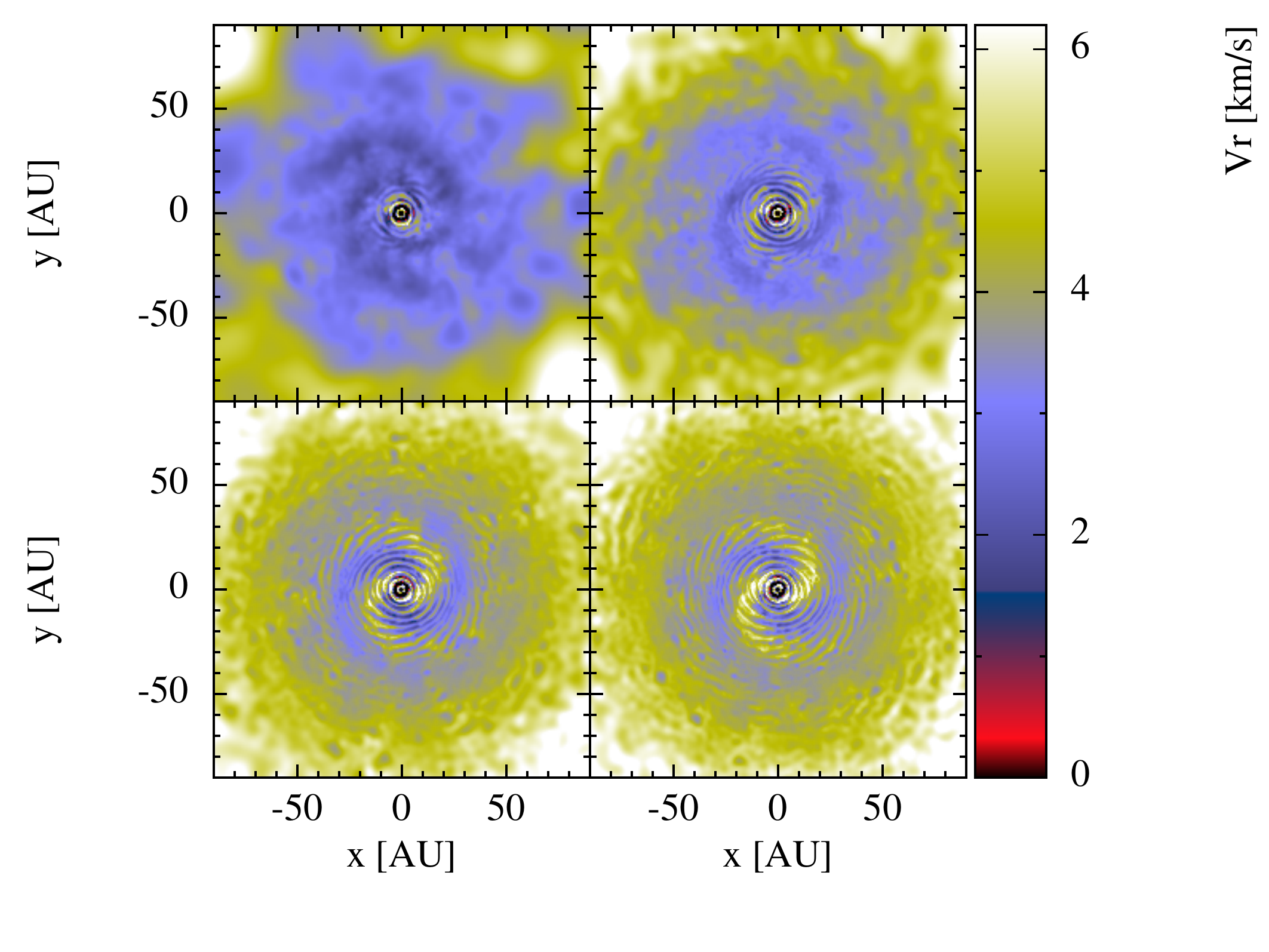}
\includegraphics[width=0.49\textwidth, height= 6.5cm]{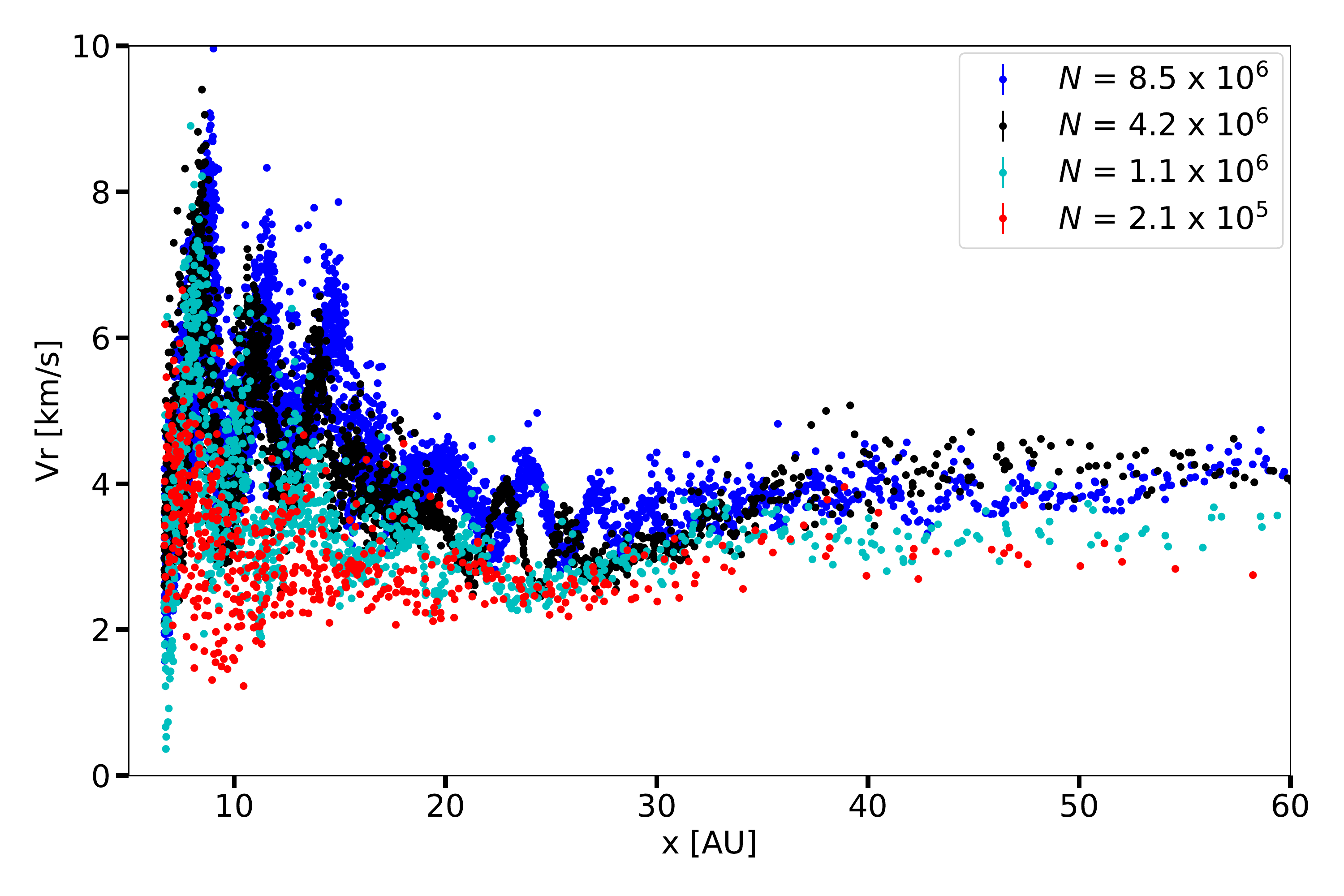}
\caption{\textit{Left:} 2D radial velocity cross-section through the orbital plane for model 9 after 76 pulsation cycles, but using different resolutions . The models from left to right, top to bottom have the following number of particles: $2.1\times10^5$, $1.1\times10^6$, $4.2\times10^6$, and $8.5\times10^6$, respectively. The spirals are clearly better resolved when we increase the resolution of the simulation. \textit{Right:} radial velocity profiles of the same simulations, color coded based on the resolution of the simulation as indicated in the legend.}
\label{Fig:resolution}
\end{center}
\end{figure*}

\begin{figure*}
\begin{center}
\includegraphics[width=0.75\textwidth]{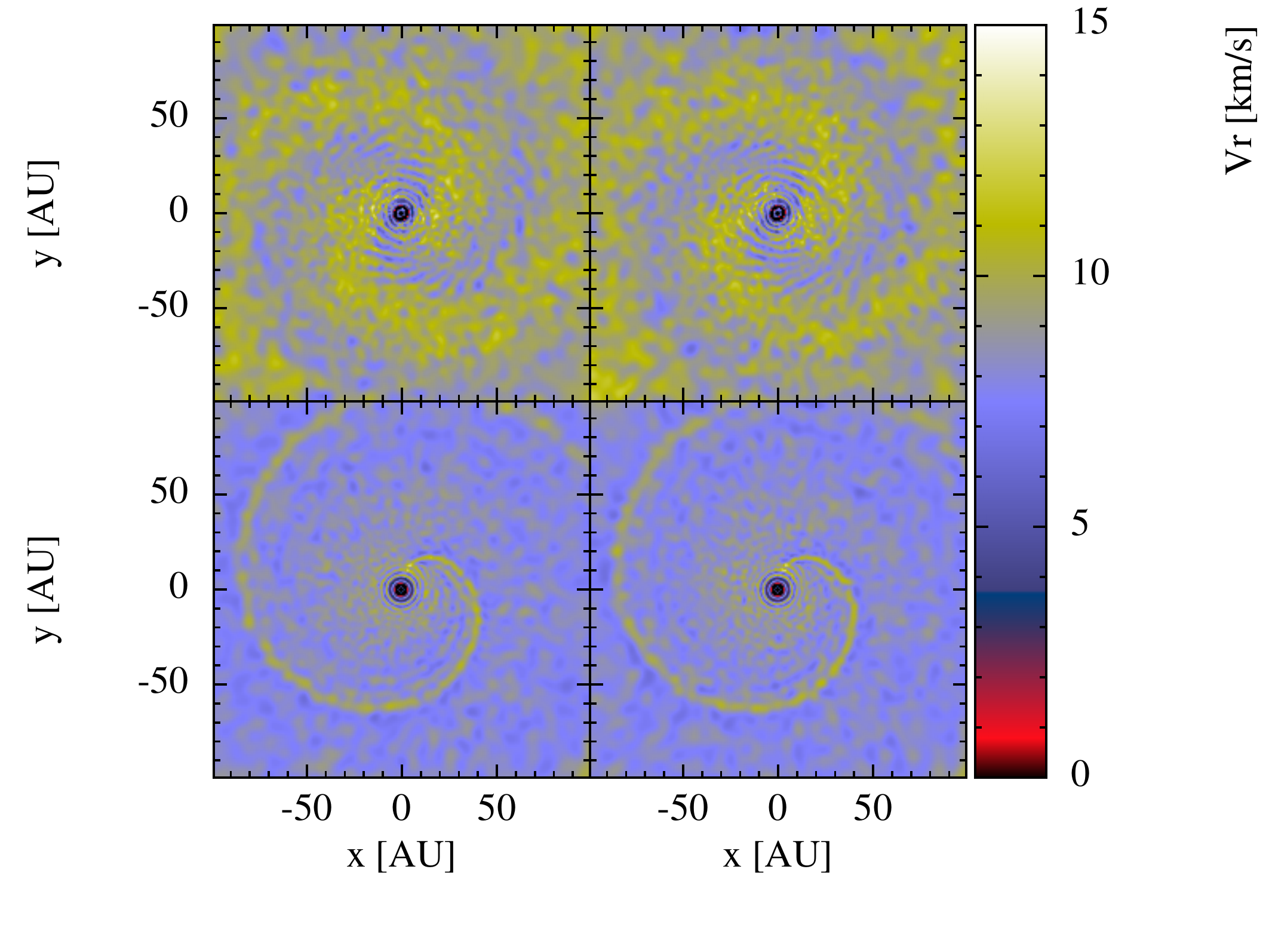}
\caption{2D radial velocity cross-section through the orbital plane for models 9 (\textit{top}) and 39 (\textit{bottom}), but this time including acceleration from silicate-based (\textit{left}) and amorphous carbon (right) dust.}
\label{Fig:dust_9_39}
\end{center}
\end{figure*}

\subsection{Testing different resolutions}
\label{sec_resolution}
We explore the effect of resolution on the outcome of our models. In Figure~\ref{Fig:resolution} we show 2-D radial velocity cross-section plots for 4 simulations with parameters similar to model 9 but for different resolutions:  $2.5 \times 10^5$, $1.1 \times 10^6$, $4.2 \times 10^6$ (original model), and $8.5 \times 10^6$ particles.
It is clear that the spirals are better resolved for higher number of particles. Increasing the number of particles in the simulation reduces the scale over which the particles' properties are smoothed, and therefore allows for better estimates of the properties of the particles at density and velocity discontinuities, such as the location of the shocks between the stellar pulsations and the wake of the companion. We also present radial velocity profiles for the four models in Figure~\ref{Fig:resolution}, showing a higher amplitude in the velocity of the gas at the shocks, which again implies that a higher resolution allows us to better determine the gas properties at shock discontinuities and to better resolve the spiral arms. Figure~\ref{Fig:resolution_density} shows the density cross-section for the same models. The shocks are slightly better resolved for the simulations with higher resolution, yet the spiral structures are not evident, as discussed previously.

\subsection{Including acceleration due to dust}
\label{dust_section}
We include acceleration due to dust in a few models to test its effect on the formation of spiral arms. For these models we use $1.1 \times 10^6$ particles for computational efficiency. We experiment with both silicate and carbonaceous dust, as detailed in Section~\ref{sec_dust_calc}. In Figure~\ref{Fig:dust_9_39} we present velocity cross-sections from 4 simulations. These simulations have the same parameters as models 9 and 39, but with acceleration from silicate-based and carbonaceous-based dust grains. In the simulations that have the same parameters as model 9, two spiral arms form. However, the spiral structures are noisier compared to the models that do not include dust. This is expected given the added dust acceleration. We also observe a single spiral arm in the models that have the same parameters as model 39. Again, the spiral structures are noisier due to the added dust acceleration. If we use higher resolution, e.g., $2.4 \times 10^6$ particles, similar to the rest of the models, we expect that the spiral structures will be better resolved (see Section~\ref{sec_resolution}). We discuss the effect of dust acceleration on the possibility of observing multiple spiral structures in Section~\ref{sec_observation}.

\begin{figure*}
\begin{center}
\includegraphics[width=0.49\textwidth]{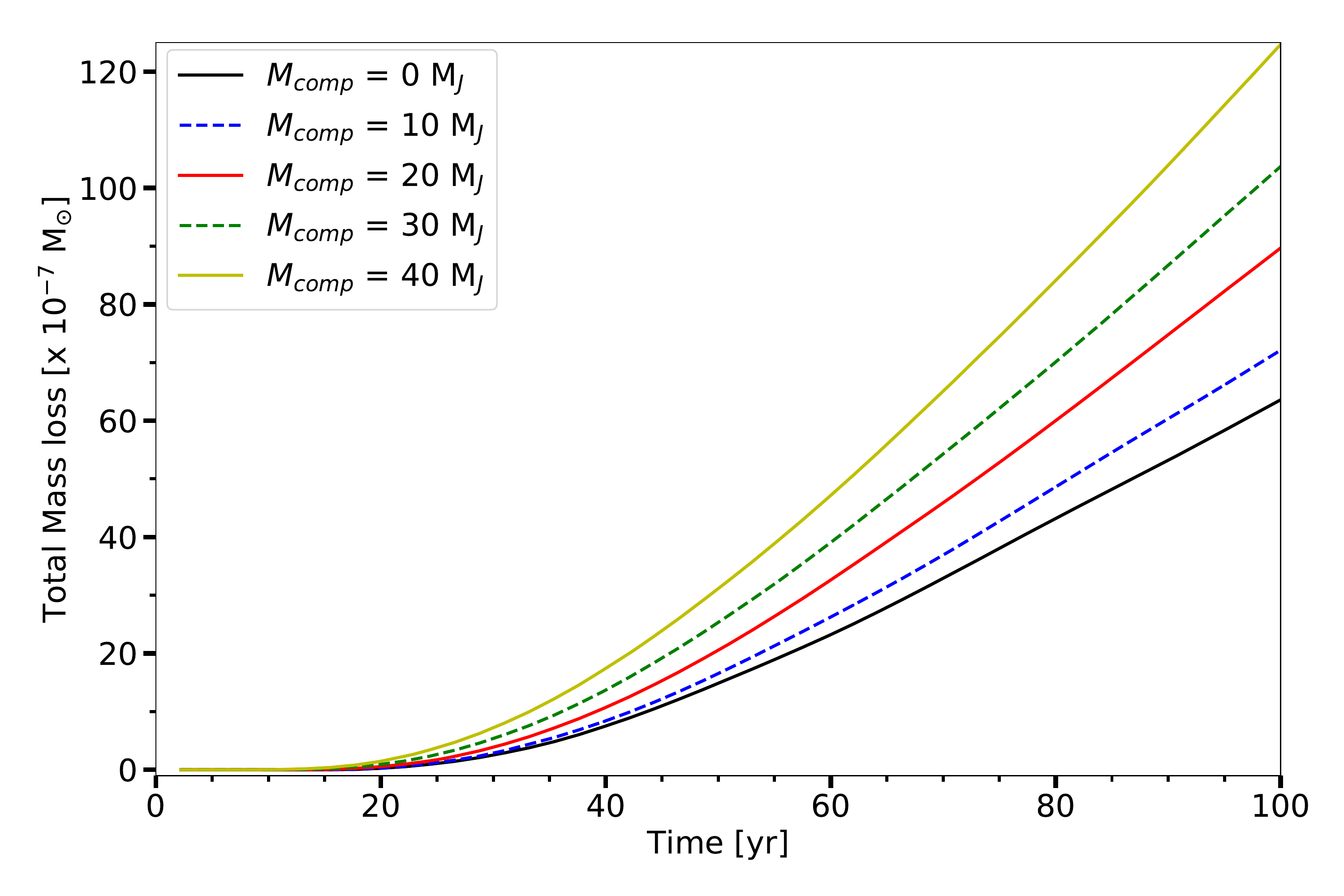}
\includegraphics[width=0.49\textwidth]{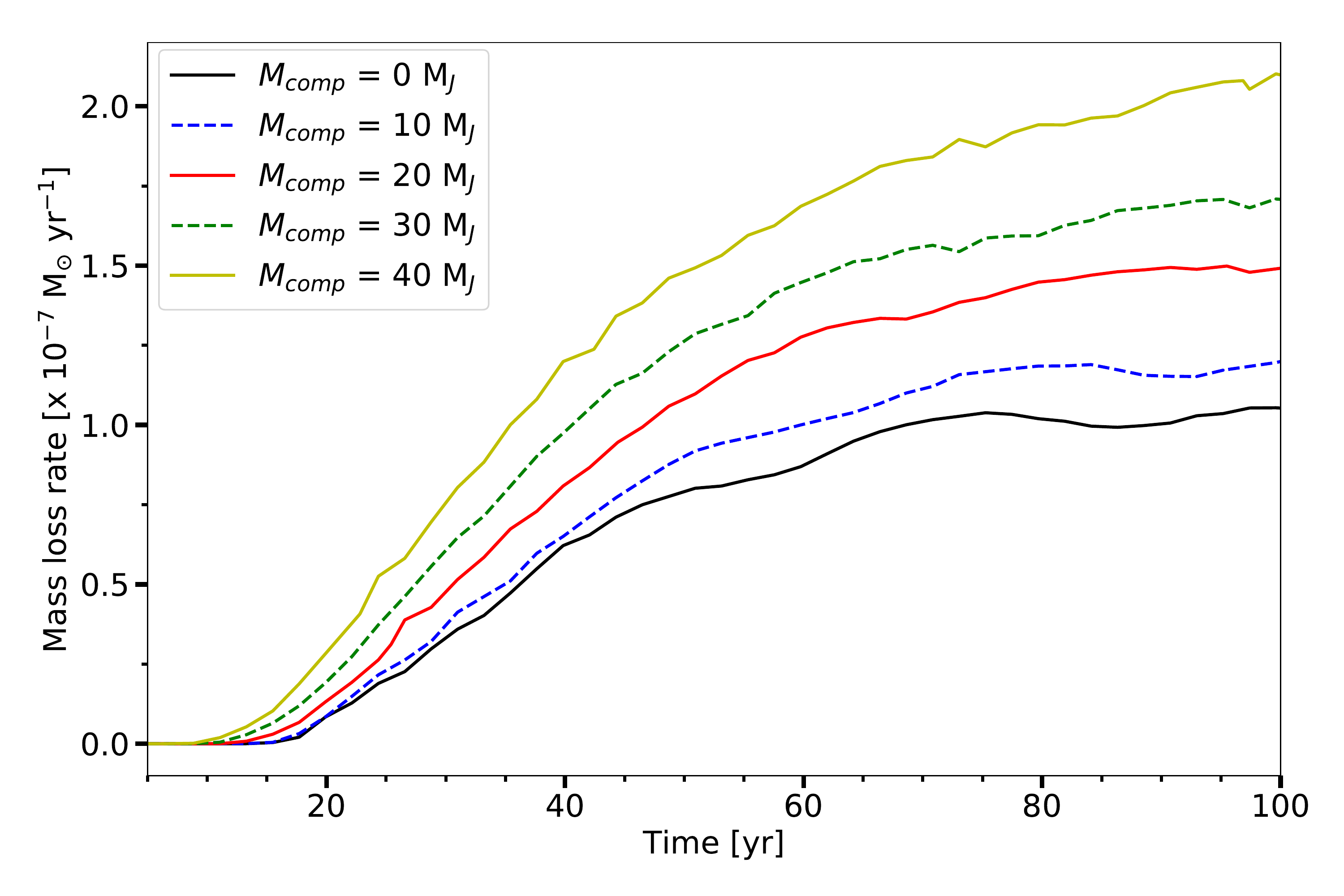}
\caption{The evolution of the total mass loss (\textit{left}) and the mass-loss rate (\textit{right}) per year per unit solar mass, for different $M_{\mathrm{comp}}$ at the same distance from the star. The models presented are model 0 (black solid line), model 13 (blue dashed line), model 14 (red solid line), model 9 (green dashed line), and model 15 (yellow solid line). The only parameter changed in these models is the companion mass. Both the total mass loss and mass-loss rate increase with increasing $M_{\mathrm{comp}}$.}
\label{Fig:mass_loss_M}
\end{center}
\end{figure*}

\begin{figure*}
\begin{center}
\includegraphics[width=0.49\textwidth]{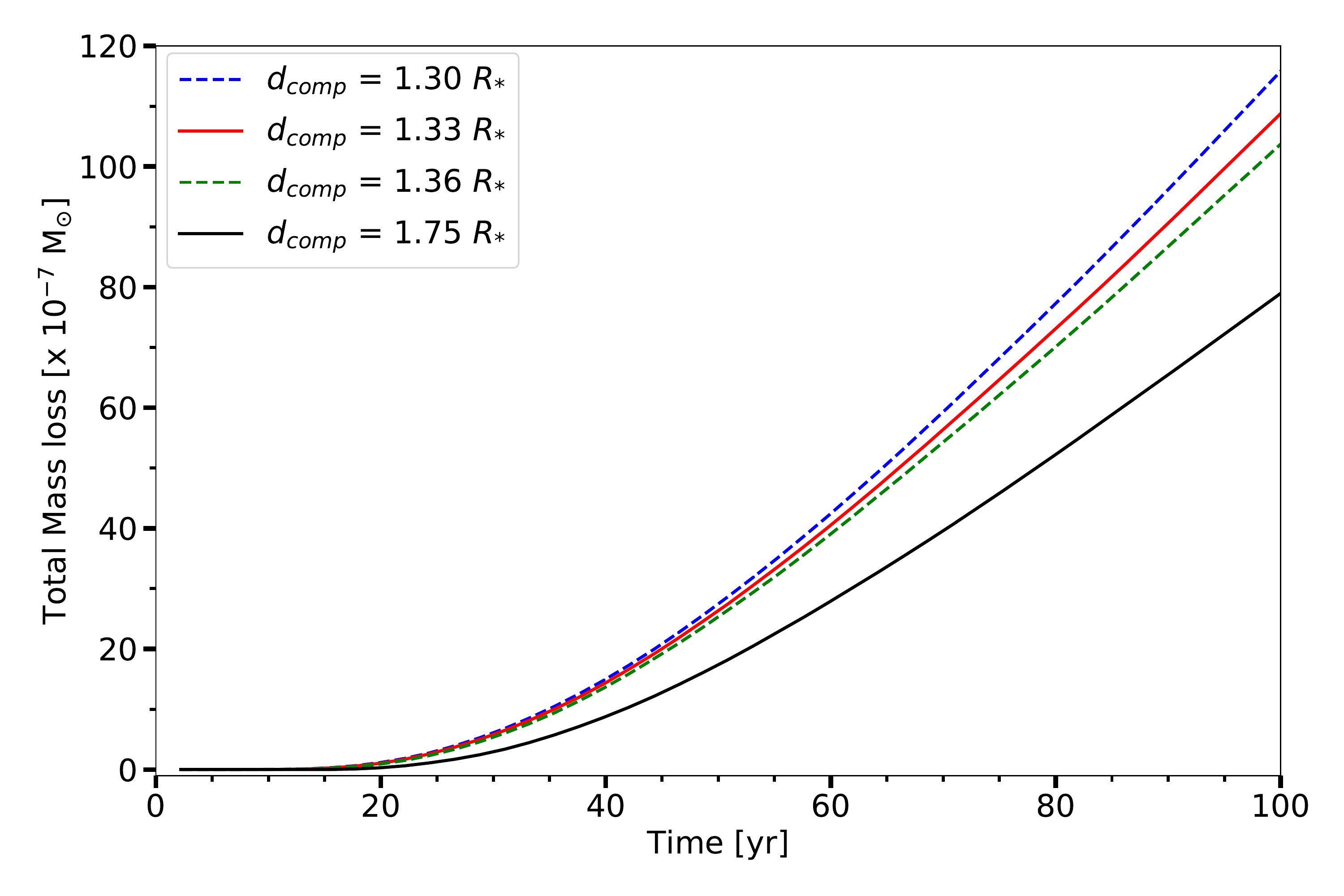}
\includegraphics[width=0.49\textwidth]{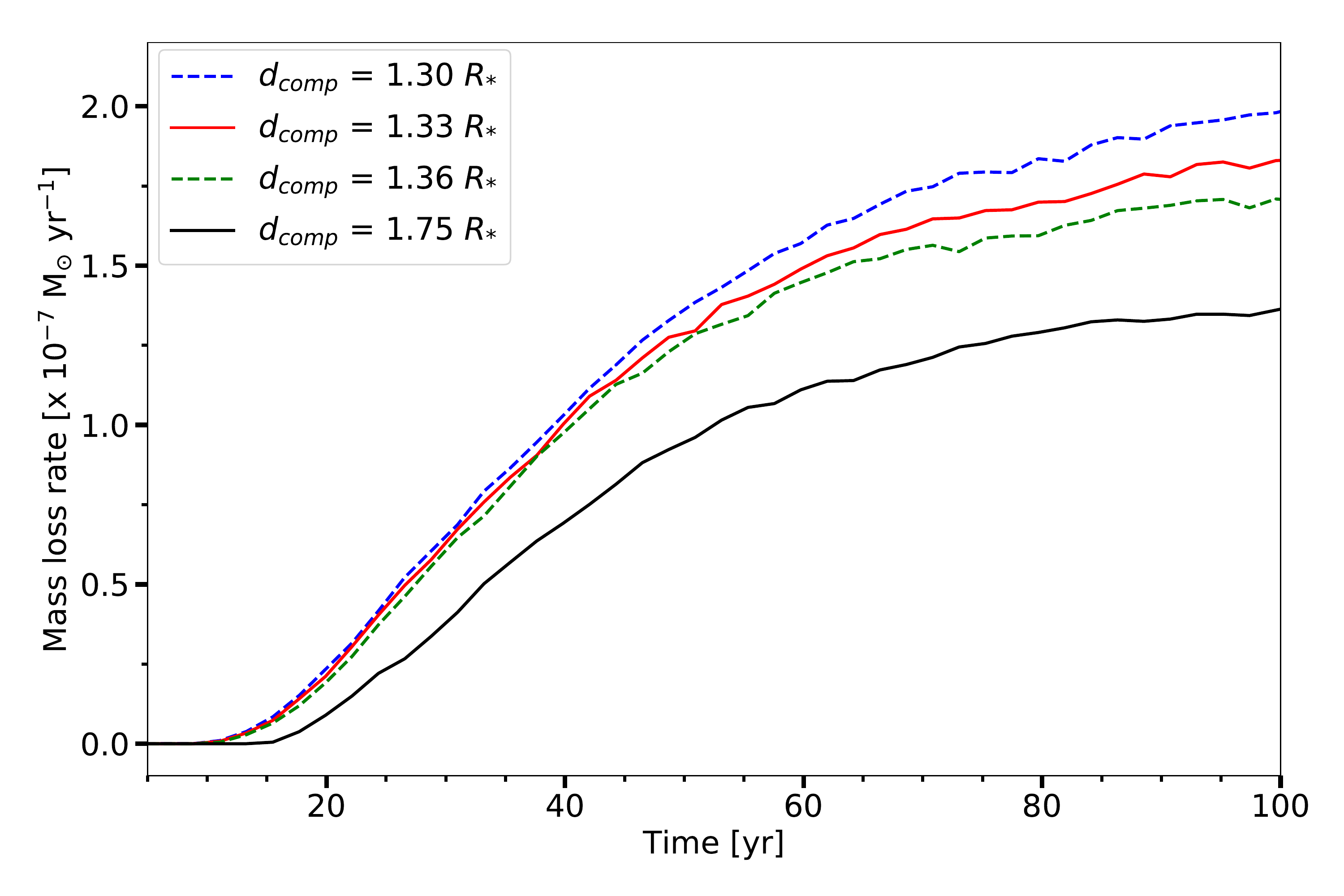}
\caption{The evolution of the total mass loss (\textit{left}) and the mass-loss rate (\textit{right}), per year per unit solar mass, for different $d_{\mathrm{comp}}$ with the same companion mass ($M_{\mathrm{comp}}$ = 30\,M$_{\mathrm{J}}$). The models presented are model 1 (blue dashed line), model 4 (red solid line), model 9 (green dashed line), and model 20 (black solid line). The only parameter changed in these models is $d_{\mathrm{comp}}$. Both the total mass loss and mass-loss rate decrease with increasing $d_{\mathrm{comp}}$.}
\label{Fig:mass_loss_D}
\end{center}
\end{figure*}

\section{Discussion}
\label{Disc}

\subsection{The effect of the companion on the mass loss}
Our models show that the presence of a close-in companion around the pulsating star enhances the mass loss. This is due to: (1) the gravitational pull exerted by the companion on the gas in the upper layers of the atmosphere and (2) the high-velocity and high-temperature shocks between the wake of the companion and the stellar pulsation. These shocks lift material away from the star and enhance the outflow/mass loss. 

Since the gravitational potential of the companion is the main factor at play, there is a trend of higher mass-loss rate with increasing $M_{\mathrm{comp}}$ and decreasing $d_{\mathrm{comp}}$. These trends are illustrated in Figures~\ref{Fig:mass_loss_M} and~\ref{Fig:mass_loss_D}. The higher the $M_{\mathrm{comp}}$, the greater the pull on the gas in the atmosphere and the stronger the shocks. The same effects manifest when the companion is closer to the star. While we show here the effect of one factor (presence of a companion) on the mass loss rate, there are multiple factors that affect mass loss in AGB stars, such as the pulsation mode, radiation pressure on dust, cooling, etc (e.g., \citealt{Hofner_2008,Wood_2015, McDonald_Zijlstra_2016, Groenewegen_Sloan_2018,Hofner_Hans_2018}). 
In addition, changes in the atmosphere of the star caused by the nearby companion might lead to changes within the star's inner layers and eventually affect the mass loss processes (e.g., \citealt{Privitera_etal_2016,Privitera_etal_2016_II,Staff_etal_2016,Rao_etal_2018,Stephan_etal_2020,Rapoport_etal_2021}). Therefore, it is worth noting the limitations of our models in revealing any such effect on the stellar interior given that we only simulate the atmosphere and circumstellar environment of the star.

\subsection{Effect of the companion mass and distance on the spiral arms}
\label{Mcomp}
In models 9, 12, 13, 14, and 15 we study the effect of $M_{\mathrm{comp}}$ on the structure of the spiral arms and their contrast with the surrounding medium. For these models we adopt different companion masses (5, 10, 30, 40, and 50\,M$_{\mathrm{J}}$) for the same orbital period and orbital separation. In this case the orbital period and separation are intentionally fixed just to test the effect of different companion masses.

\begin{figure*}
\begin{center}
\includegraphics[width=\textwidth]{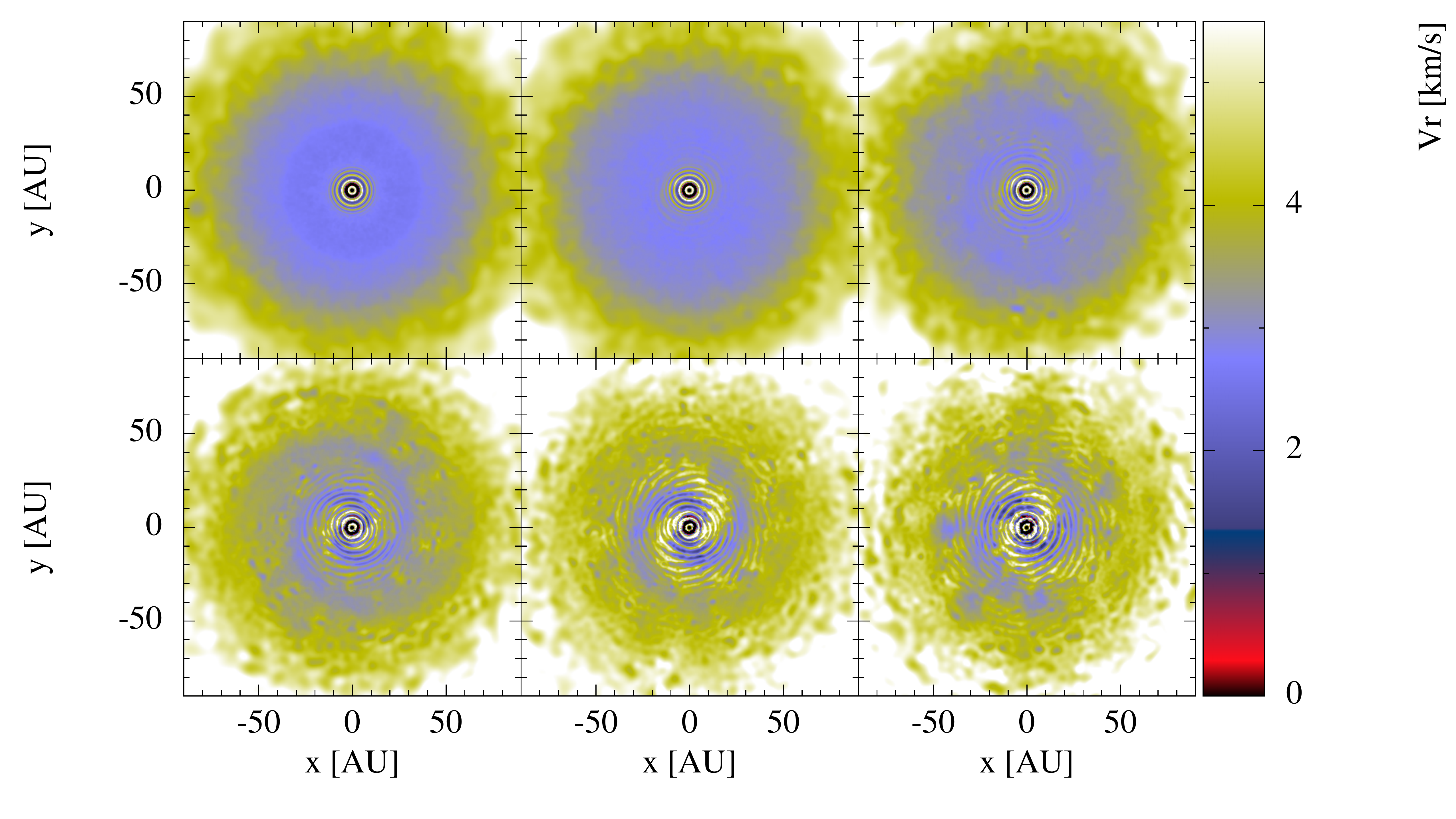}
\vspace{-0.8cm}
\caption{2D radial velocity cross-section through the orbital plane from model 0 ($M_{\mathrm{comp}}$ = 0), 12 ($M_{\mathrm{comp}}$ = 5\,M$_{\mathrm{J}}$), 13 ($M_{\mathrm{comp}}$ = 10\,M$_{\mathrm{J}}$), 14 ($M_{\mathrm{comp}}$ = 20\,M$_{\mathrm{J}}$), 9 ($M_{\mathrm{comp}}$ = 30\,M$_{\mathrm{J}}$), and 40 ($M_{\mathrm{comp}}$ = 50\,M$_{\mathrm{J}}$), going from left to right, top to bottom, after $\sim$ 76 pulsation cycles. There is a trend of increasing velocity contrast between the spiral structures and their surrounding medium with increasing $M_{\mathrm{comp}}$.}
\label{Fig:diff_mass}
\end{center}
\end{figure*}

\begin{figure}
\begin{center}
\includegraphics[width=\columnwidth]{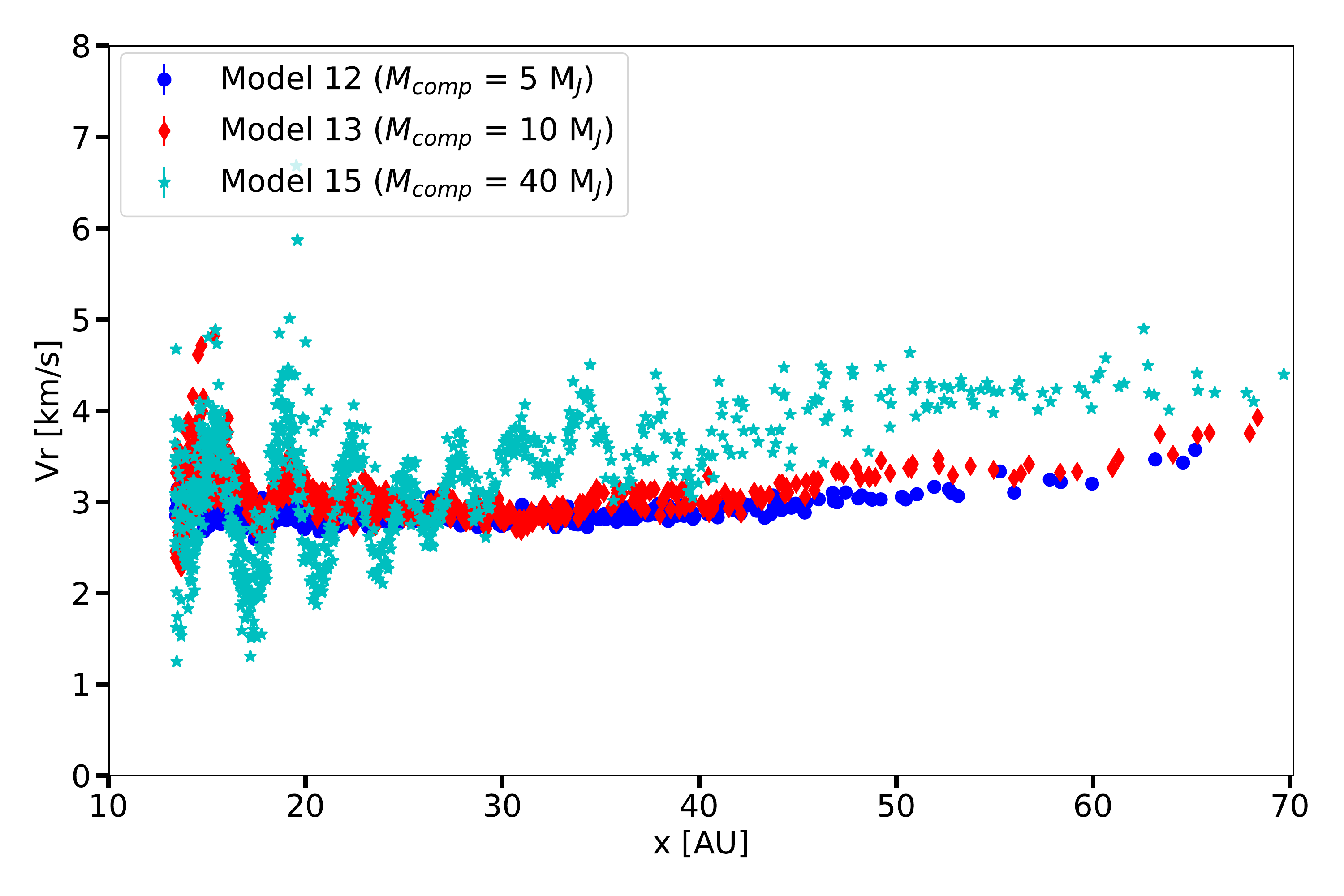}
\caption{Radial velocity profiles along the x-axis from models 12 (blue circles), 13 (red diamonds), and 15 (cyan stars), after $\approx$ 76 pulsation cycles. The profiles show higher velocities for higher mass companion.}
\label{Fig:vr_profiles}
\end{center}
\end{figure}

The models show that the velocity contrast of the spiral arms with the surrounding medium is higher for higher $M_{\mathrm{comp}}$ (see Figure~\ref{Fig:diff_mass}). The greater gravitational pull exerted by the heavier companion on the gas in the circumstellar environment and in the upper atmosphere of the star leads to stronger shocks of higher velocity and temperature. This is highlighted in Figure~\ref{Fig:vr_profiles} where we plot the radial velocity profiles from some of these models. The profiles show that the radial velocity contrast and the amplitude of the shocks is higher for higher $M_{\mathrm{comp}}$. For model 11, the presence of a higher mass companion leads to an increase in the radial velocity of the gas by a factor of $\approx$ 2, compared to models 0 and 12 (Figure~\ref{Fig:vr_profiles}). 

In all the models presented in Figure~\ref{Fig:diff_mass}, two spiral arms form, with the exception of model 12 ($M_{\mathrm{comp}}$ = 5\,M$_{\mathrm{J}}$). In this model the gravitational pull exerted by the companion on the surrounding gas is relatively weak, leading to weak shocks and unresolved spiral arms. An increase in the resolution of the simulation might help resolve the spiral arms in such a case (see Section~\ref{sec_resolution}). Nevertheless, The 1-D models of \citet{Wang_Willson_2012} show multiple spiral structures for a close-in companion with a mass = 5\,M$_{\mathrm{J}}$. Similarly, a more distant companion with a mass = 5\,M$_{\mathrm{J}}$ also gives rise to a single spiral structure in the grid-based models of \citet{Kim_etal_2012}. This again emphasizes that the absence of spiral structures in our models for a companion mass $\lesssim$ 5\,M$_{\mathrm{J}}$ is likely a numerical effect.

\begin{figure}
\begin{center}
\includegraphics[width=\columnwidth]{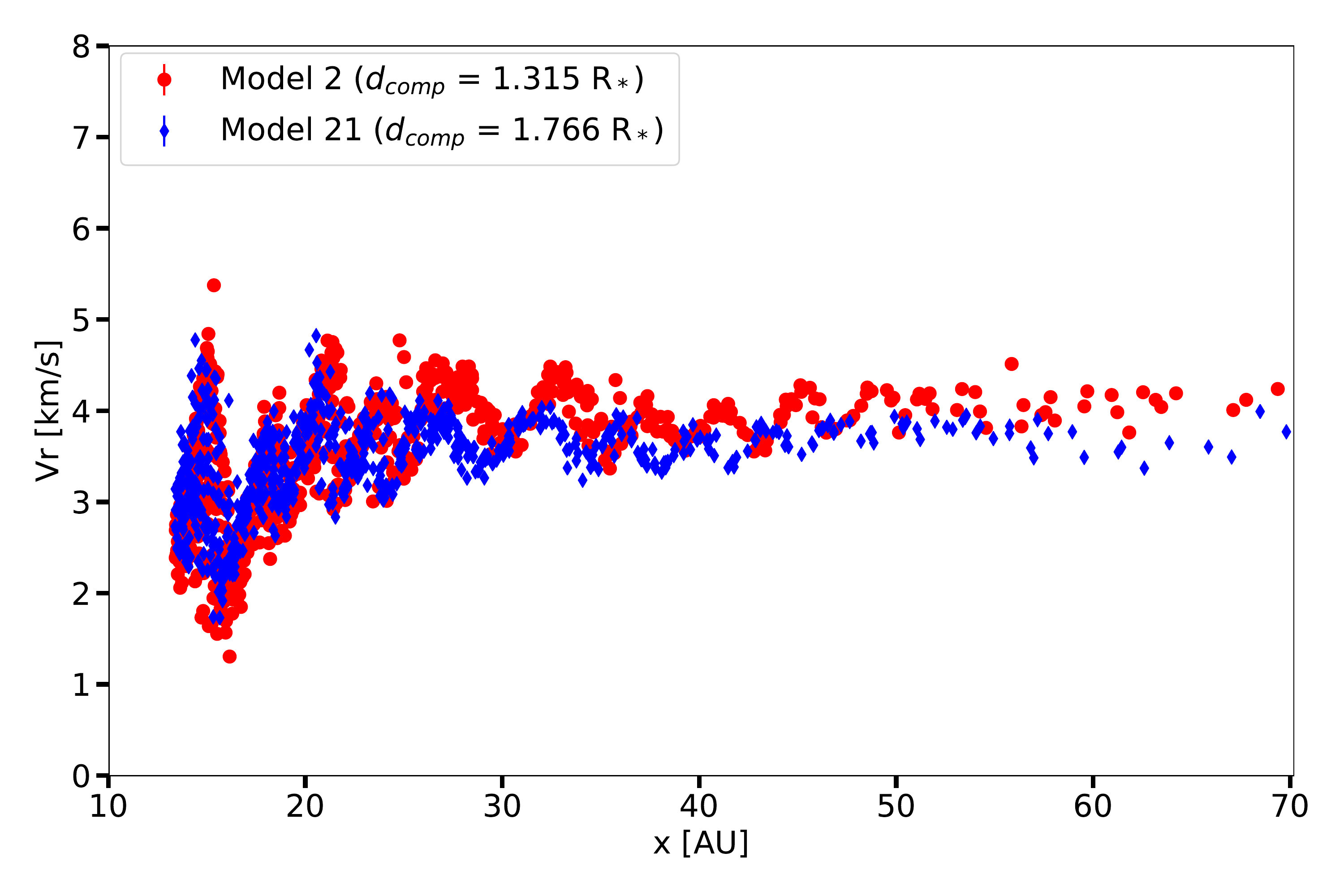}
\caption{Radial velocity profiles along the x-axis from model 2 (red) and model 21 (blue) after $\approx$ 76 pulsation cycles. The shock velocities are slightly higher for model 2 where the companion is closer to the star's atmosphere.}
\label{Fig:diff_dist}
\end{center}
\end{figure}

Our models also show that if the companion is closer to the star,  the shocks between the companion's wake and the stellar pulsations are stronger --- showing higher velocity and higher temperature.
Figure~\ref{Fig:diff_dist} represents the velocity profiles of model 2 and 21. For model 2 where the companion is closer to the star, the velocity contrast due to the shocks is higher compared to model 21. 

\begin{figure*}
\begin{center}
\includegraphics[width=0.9\textwidth]{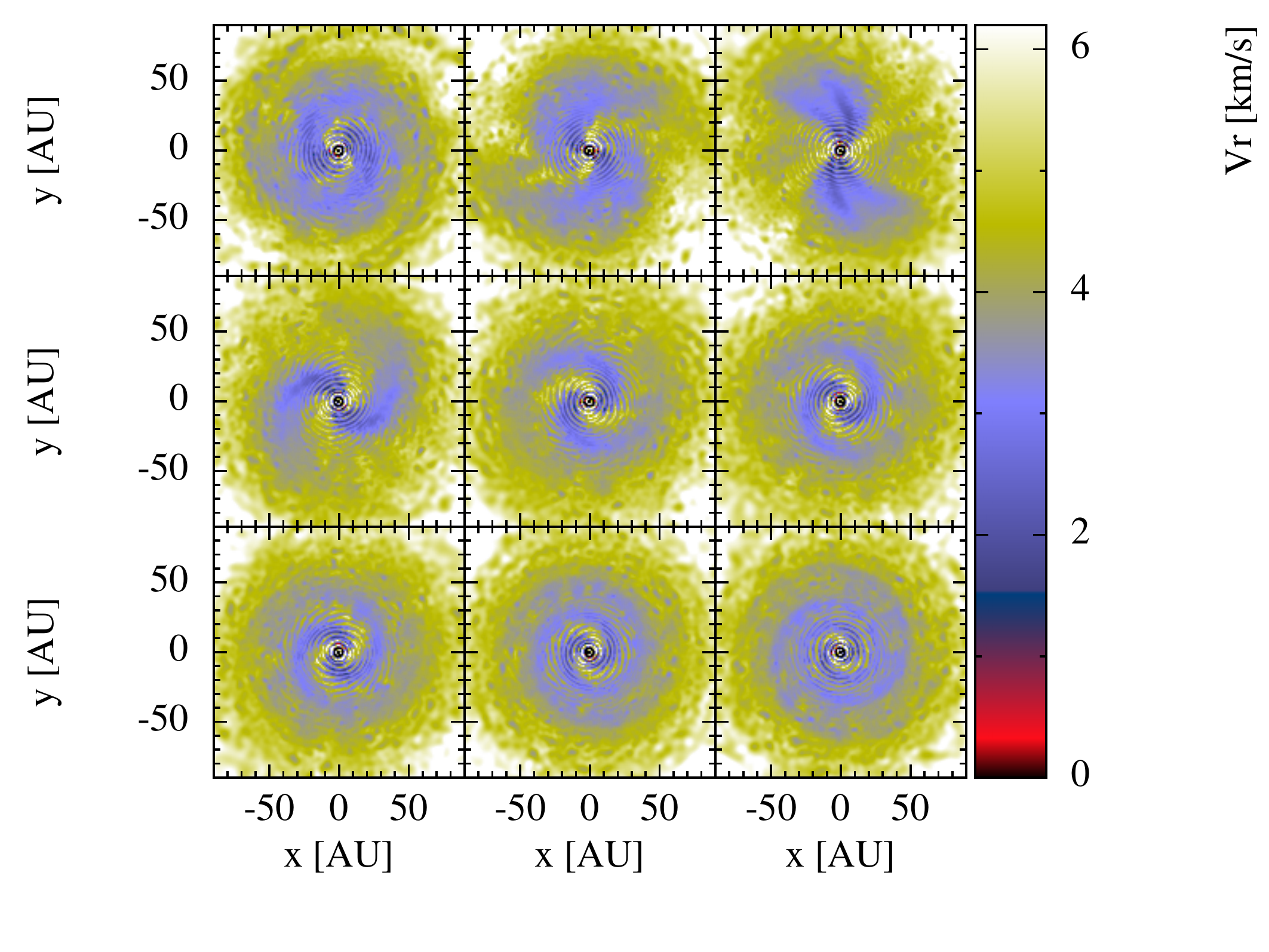}
\vspace{-1.0cm}
\caption{2D radial velocity cross-section through the orbital plane from models 1 ($P_{\mathrm{orb}}$ = 761), 2 ($P_{\mathrm{orb}}$ = 775), 5 ($P_{\mathrm{orb}}$ = 791), 6 ($P_{\mathrm{orb}}$ = 801), 7 ($P_{\mathrm{orb}}$ = 808), 8 ($P_{\mathrm{orb}}$ = 812), 9 ($P_{\mathrm{orb}}$ = 815), 10 ($P_{\mathrm{orb}}$ = 823), and 11 ($P_{\mathrm{orb}}$ = 833), going from left to right, top to bottom, after $\approx$ 76 pulsation cycles. The models show the changing shape of the spiral structures with changing $P_{\mathrm{orb}}$.}
\label{Fig:diff_orb_dist}
\end{center}
\end{figure*}

\subsection{Effect of the orbital period and resonance mode on the spiral arms}
\label{Porb}
The $P_{\mathrm{orb}}$ and its resonance with $P_{\mathrm{puls}}$ plays the major role in determining the properties and morphology of the spiral arms (e.g. number, period, winding, etc). The higher the offset between $P_{\mathrm{orb}}$/$P_{\mathrm{puls}}$ and $M/N$, the shorter the period of the arms and vice versa. In Figure~\ref{Fig:diff_orb_dist} we highlight the effect of $P_{\mathrm{orb}}$ on the period of the spiral arms and their winding. At the same time-step of the simulation, the spirals experience different windings depending on the $P_{\mathrm{orb}}$ and therefore $P_{\mathrm{sa}}$. Model 11 (with $P_{\mathrm{sa}}$ = 7293\,d) shows more than one complete winding after $\approx 76$ pulsation cycles, with a winding angle $\omega >2\pi$\,rad. However, for model 5 ($P_{\mathrm{sa}}$ = 103885\,d) the winding angle $\omega <\pi/4$\,rad after $\approx$ 76 pulsation cycles.

As seen pole-on from the z$>0$ direction, the spiral arms appear to rotate following a specific direction in the orbital plane. The direction of the apparent rotation depends on the ratio of $P_{\mathrm{puls}}$ to $P_{\mathrm{orb}}$. If $P_{\mathrm{puls}}/P_{\mathrm{orb}}$ is less than $M/N$, the apparent rotation of the spiral arms is in the counter-clockwise direction, and if $P_{\mathrm{puls}}/P_{\mathrm{orb}}$ is greater than $M/N$, then it is in the clockwise direction. Figure~\ref{Fig:rot_spirals} illustrates both cases, using models 2 and 6 as examples, showing the spiral arm locations at different time-steps. In model 2 where $P_{\mathrm{puls}}/P_{\mathrm{orb}}$ = 0.508, the spiral arms appear to rotate in the counter-clockwise direction, however in model 6 where $P_{\mathrm{puls}}/P_{\mathrm{orb}}$ = 0.492, the spiral arms appear to rotate in the clockwise direction. Nevertheless, the shocks that form the spiral arms are moving in the radial direction in both cases.

\subsection{Can the spiral arms be observed?}
\label{sec_observation}

The early (more than two decades ago) theoretical studies of single spirals forming in the outflows of evolved stars laid a solid foundation for further simulations and for our understanding and interpretation of the observations of such structures with current facilities. There are two main considerations in determining whether the multiple spirals simulated here will be observable: first, we need to consider the companion survivability and the different factors that can affect the formation of evident spirals; second we need to consider the capabilities of current facilities/instruments to resolve such structures. 

Several studies have been carried out to investigate the survival of planets around giant stars and their orbital evolution (e.g., \citealt{ Villaver_Livio_2007,Villaver_Livio_2009,Nordhaus_etal_2010,Kunitomo_etla_2011,Villaver_etal_2014}). \citet{Nordhaus_Spiegel_2013} showed that most companions with orbits $<$ 6\,AU will be engulfed by stars with masses between 1 and 3\,M$_{\odot}$. However, various parameters and mechanisms affect the survival of the planet (e.g. tidal drag, mass loss, the system parameters), hence this topic is still an extensive field of research. Studying the survival of the companion and its orbital evolution are of course not the focus of this work. Our simulations assume that the companion is not engulfed by the star and therefore we exclude any decay in the orbit of the companion and we assume that the companion survives and is in a stable orbit over the simulation timescale --- all reasonable assumptions \citep{Duncan_Lissauer_1998,Villaver_Livio_2007,Batygin_etaL_2008,Kervella_etal_2017,Vanderburg_etal_2020}. However, if there is a change in the orbital period/separation (e.g., due to orbital decay), there would be implications for the possibility of observing the spiral structures: if the timescale of the formation of the spirals is shorter than the timescale of a stable orbit, it is likely that the spirals would not be observable. Whilst, if the timescale of a stable orbit is longer than the timescale of the formation of spirals, such spirals could form and might last for a few thousands of years.

The presence of more than one sub-stellar companion around the AGB star should lead to the interaction of both companions with the circumstellar environment and the stellar winds. This will likely cause even more complex structures, with the possibility of multiple, `interfering' spiral arms, increasing the challenge to observe and distinguish the structures. The interaction of the stellar pulsations with multiple companions is not considered here, but will be the subject of our follow-up study. 

Furthermore, our models are idealized - we assume a spherical, radially pulsating AGB star. While this assumption is reasonable, in reality the morphology of the atmosphere and the shock structures are much less spherically symmetric \citep{Freytag_etal_2017}. In a realistic outflow, we expect that the spiral arms might deviate from symmetric structures in the orbital plane and show more complexity, e.g., if the pulsations are more chaotic in a particular direction, it is possible that shocks cluster in a spiral arm only on one side of the orbital plane.  This adds to the challenges of observing the spiral structures and determining the influence of a close-in companion on the circumstellar environment of the AGB star. 

While the presence of acceleration due to dust can lead to less symmetric and noisier spiral structures, both \citet{Wang_Willson_2012}'s and our models show that despite of the added acceleration due to radiation pressure on dust, observable spiral structures can form around the star. However, a more realistic formalism for dust acceleration is needed to better assess the formation of spiral arms in this case and the potential of observing them.

In several of our models, the spiral arms complete their first winding on a time-scale of 20 to 150 years (e.g., models 1, 2, 6, 7, 8, 9, 10, 11; see Figure~\ref{Fig:diff_orb_dist}). Therefore, we expect that in these cases the spiral arms will achieve several winding on a timescale of the order of hundreds of years, which is short compared to an AGB star lifetime. This could make it difficult to distinguish and observe the spiral arms. However, in the case of $P_{\mathrm{puls}}/P_{\mathrm{orb}}$ being very close to $M/N$ (e.g., models 3, 4, and 5), the winding timescale of the spiral arms jumps to a few thousand years, making it easier to observe and distinguish such multiple spiral arms. For Mira variables with pulsation periods of the order of a few thousand days, the formation and winding timescales of the spiral arms become much longer ($\sim$ thousands of years), increasing the chance of spotting multiple spiral arms in their circumstellar environments.

The inclination of the system also plays an important role in determining the observability of the structures, near-face-on spirals might be easier to detect compared to the arcs and bars that result looking edge-on at the orbital plane. 

The multiple spiral structures in our models extend out to $~$70 AU, scales that are not far off for current and imminent high-resolution, high-sensitivity facilities. Facilities like ALMA which have revealed a diversity of structures around Galactic AGB stars \citep{Lagadec_Chesneau_2015,Kervella_etal_2017,Brunner_etal_2019,Lagadec_2019,Homan_etal_2020}, instruments with extreme adaptive optics such as SPHERE, intended to resolve the circumstellar environment of young stars and proto-planetary systems \citep{Sissa_etal_2018,Avenhaus_etal_2018}, and future instruments, such as the mid- and near-infrared cameras and imager on the James Webb Space telescope \citep{Gardner_etal_2006} could offer us a chance to observe the multiple spiral structures. Given the difficulty to resolve the structures in density-space in our simulations, and that the outflows are cold and dusty, observations in velocity-space and at mm/submm and infrared wavelengths, stand the best chance of detecting the circumstellar features. 

\subsection{Implications for our future solar system}

Our Sun will evolve into an AGB star in around 5 billion years from now, pulsating  with a period of the order of a year and with an outer stellar radius that could extend beyond Earth's orbit and as far as that of Mars \citep{Iben_Renzini_1983}. This process will likely be destructive for the inner planets and might affect the orbits of the gas-giants, particularly their semi-major axis and eccentricity \citep{Debes_Sigurdsson_2002,Veras_etal_2012}. The orbit of the gas-giants in our solar system will possibly be stable for several billion years during the post-main sequence phase (e.g., \citealt{Duncan_Lissauer_1998,Batygin_etaL_2008} and references therein). If the pulsation period of the Sun and the orbital period of, e.g., Jupiter achieve a certain resonance, the interaction between Jupiter and the solar pulsations might lead to the formation of multiple spiral structures, similar to the ones we obtained in our models. If Jupiter ends up in a much more extended orbit compared to the solar radius \citep{Debes_Sigurdsson_2002}, a single spiral arm could form due to the interaction of the wake of Jupiter with the circumsolar environment, similar to models 33 to 40.

In our simulations we did not resolve spiral structures for models with a companion mass $\lesssim 5$\,M$_{\mathrm{J}}$. But a significant increase in the resolution of the simulations (e.g., an order of magnitude more particles) should allow us to resolve spiral structures for a companion mass of 1\,--\,5\,M$_{\mathrm{J}}$. However, we did not attempt this given the substantial increase in the computational time that might result from such increase in the resolution of the simulation. Nevertheless, the 1-D models of \citet{Wang_Willson_2012} did result in the formation of multiple spiral arms for a companion mass of 5\,M$_{\mathrm{J}}$.

\section{Conclusions and summary}
\label{Concl}
We have carried out 3D hydrodynamic models of mass-losing, pulsating AGB stars to simulate the effect of a close-in sub-stellar companion on the morphology of the circumstellar environment of the star. Below is a summary of our results:

\begin{itemize}
    \item If there is a resonance between the orbital period of the companion and the stellar pulsation, and if the companion is at a distance $d_{\mathrm{comp}} < 3R_*$, the interaction between the wake of the companion and the stellar pulsations leads to the formation of strong shocks, which cluster in multiple spiral arms centered on the system. The number of the spiral arms is based on the resonance between the orbital and pulsation periods.
    
    \item Similar to the models of \citet{Wang_Willson_2012}, we found that the period of the spiral structure can be expressed as: $P_{\mathrm{sa}} = \frac{P_{\mathrm{orb}}P_{\mathrm{puls}}}{\left|MP_{\mathrm{orb}} - NP_{\mathrm{puls}}\right|}$, where $M$ and $N$ represent the resonance mode $M:N$.

    \item Our models demonstrate that the presence of a close-in companion enhances the mass loss from the giant star.
    
    \item Our 3D models show that complex structures such as bars- and arc-like structures form in the polar planes. 
    
    \item If the companion is at a distance $d_{\mathrm{comp}} > 4R_*$, the interaction between the companion and the stellar outflow leads to the formation of a single spiral arm.
    
    \item We tested the effect of dust formation on the morphology of the outflow and we found that multiple or single spiral arms are still resolved when adding acceleration due to dust, but they are noisier.
    
    \item Future high-resolution, high-sensitivity instruments might allow us to image these multiple spiral structures around AGB stars.
    
    \item We also speculate that the interaction between Jupiter and the solar pulsations when our Sun turns into an AGB star might lead to the formation of spiral structures similar to those obtained in our simulations.
\end{itemize}

In this work, we make our first steps in investigating the effect of a  sub-stellar companion on the circumstellar environment of a pulsating AGB star in 3D. We plan to introduce more physically motivated mechanisms into our models, such as radiative cooling, radiative transfer and different approaches to mimic a dust driven wind. We also plan to study the effect of multiple sub-stellar companions and planetary systems on the circumstellar environment of AGB stars.

\section*{Acknowledgments}

We acknowledge with thanks Patricia Whitelock for valuable comments.
E.A. acknowledges NSF award AST-1751874, NASA awards 11-Fermi 80NSSC18K1746, 13-Fermi 80NSSC20K1535, 16-Swift 80NSSC21K0173, and a Cottrell fellowship of the Research Corporation.\\ S.M. acknowledges funding from the South African National Research Foundation and the University of Cape Town VC2030 Future Leaders Award (PI: S.M.). The simulations were run on the SAAO computing cluster.

All rendered figures are made using the interactive visualisation tool for Smoothed Particle Hydrodynamics simulations \textsc{SPLASH} \cite{Price_etal_2007}. We thank all contributors and developers of \textsc{SPLASH}. Analysis made significant use of \textsc{python} 3.7.4, and the associated packages \textsc{numpy}, \textsc{matplotlib}, \textsc{pandas}, \textsc{seaborn}, and \textsc{scipy}. 

\section*{Data availability}
The data are available upon request, particularly due to the large size of the output of the simulations (the size of each snapshot from the simulation is several 100s MB. The total size of one simulation is several GB).

\bibliography{biblio}

\appendix

\section{Companion properties}
\label{appA}
For a circular orbit with $M_{\mathrm{comp}}$\,$<<$\,$M_*$, the constant relative velocity is given by:
\begin{equation}
v_{\mathrm{comp}} = (GM_*/d_{\mathrm{comp}})^{\frac{1}{2}},
\end{equation}
where  $G$ is the gravitational constant, $M_*$ is the mass of the star, and $d_{\mathrm{comp}}$ is the orbital radius of the companion. The angular momentum of an object is in general given by $\vec{L}$ = $\vec{r} \times \vec{p}$, where $\vec{r}$ is the vector from the center of rotation and $\vec{p}$ is the linear momentum. For a circular orbit, the position vector is always tangential to the orbit, thus, the angular momentum can be written as $L = rp = rmv$. The angular momentum of the companion can then be derived using: $L = m (GM_*d_{\mathrm{comp}})^{\frac{1}{2}}$. We can write then:
\begin{equation}\label{equ:d}
d_{\mathrm{comp}} = \frac{L^2}{M^2_{\mathrm{comp}}GM_*} 
\end{equation}
From Kepler's third law we can write:
\begin{equation}\label{equ:L}
L^3 = \frac{P_{\mathrm{orb}} G^2 M^3_{\mathrm{comp}}M^2_*}{2\pi}  
\end{equation}
Using equations (\ref{equ:d}) and (\ref{equ:L}) we can then derive $d_{\mathrm{comp}}$ for every $P_{\mathrm{orb}}$ and vice versa. The model  parameters along with the companion properties are stored in Table~\ref{table:models}.

\begin{table*}
\centering
\caption{Models parameters.}
\begin{tabular}{rrrrrrrrr}
\hline
Model  & $M_{\mathrm{comp}}$ & $d_{\mathrm{comp}}$ & $P_{\mathrm{orb}}$ & $P_{\mathrm{puls}}/P_{\mathrm{orb}}$ &  $M$:$N$ & $N_{\mathrm{sa}}$ & $P_{\mathrm{sa}}$ & $U_{\mathrm{amp}}$\\ 
 &  (M$_{\mathrm{J}}$) & ($R_*$) & (days)& & &  & (days) & (km\,s$^{-1}$)\\
\hline
1 & 30 & 1.299 & 761 & 0.517 & 1:2 & 2 & 11105 & 4.0\\
2 & 30 & 1.315 & 775 & 0.508 & 1:2 & 2 & 23488 & 4.0\\
3 & 30 & 1.326 & 785 & 0.501 & 1:2 & 2 & 103097 & 4.0\\
4 & 30 & 1.330 & 788 & 0.500 & 1:2 & 2 & - & 4.0\\
5 & 30 & 1.333 & 791 & 0.498 & 1:2 & 2 & 103885 & 4.0\\
6 & 30 & 1.344 & 801 & 0.492 & 1:2 & 2 & 24276 & 4.0\\ 
7 & 30 & 1.352 & 808 & 0.487 & 1:2 & 2 & 15918 & 4.0\\
8 & 30 & 1.356 & 812 & 0.485 & 1:2 & 2 & 13330 & 4.0\\
9 & 30 & 1.360 & 815 & 0.483 & 1:2 & 2 & 11893 & 4.0\\
10 & 30 & 1.369 & 823 & 0.478 & 1:2 & 2 & 9265 & 4.0\\
11 & 30 & 1.380 & 833 & 0.472 & 1:2 & 2 & 7293 & 4.0\\
\hline
12 & 5 & 1.360 & 815 & 0.483 & 1:2 & 2 & 11893 & 4.0\\
13 & 10 & 1.360 & 815 & 0.483 & 1:2 & 2 & 11893 & 4.0\\
14 & 20 & 1.360 & 815 & 0.483 & 1:2 & 2 & 11893 & 4.0\\
15 & 40 & 1.360 & 815 & 0.483 & 1:2 & 2 & 11893 & 4.0\\
\hline
16 & 20 & 1.091 & 586 & 0.672 & 2:3 & 3 & 23088 & 4.0\\
17 & 20 & 1.104 & 596 & 0.661 & 2:3 & 3 & 23482 & 4.0\\
18 & 30 & 1.721 & 1160 & 0.339 & 1:3 & 3 & 20775 & 4.0\\
19 & 30 & 1.731 & 1170 & 0.336 & 1:3 & 3 & 38415 & 4.0\\
20 & 30 & 1.754 & 1194 & 0.329 & 1:3 & 3 & 39203 & 4.0\\
21 & 30 & 1.766 & 1204 & 0.326 & 1:3 & 3 & 21563 & 4.0\\
22 & 40 & 1.766 & 1204 & 0.326 & 1:3 & 3 & 21563 & 4.0\\
\hline
23 & 20 & 1.010 & 522 & 0.754 & 3:4 & 4 & 20567 & 4.0\\
24 & 20 & 1.019 & 529 & 0.744 & 3:4 & 4 & 18948 & 4.0\\
25 & 30 & 1.019 & 529 & 0.744 & 3:4 & 4 & 18948 & 4.0\\
26 & 50 & 2.093 & 1556 & 0.253 & 1:4 & 4 & 30653 & 4.0\\
27 & 40 & 2.100 & 1564 & 0.251 & 1:4 & 4 & 51351 & 4.0\\
28 & 40 & 2.122 & 1588 & 0.248 & 1:4 & 4 & 52139 & 4.0\\
29 & 50 & 2.129 & 1596 & 0.246 & 1:4 & 4 & 31441 & 4.0\\
30 & 50 & 2.138 & 1606 & 0.245 & 1:4 & 4 & 21092 & 4.0\\
\hline
31 & 30 & 1.360 & 815 & 0.483 & 1:2 & 2 & 11893 & 5.0\\
32 & 30 & 1.360 & 815 & 0.483 & 1:2 & 2 & 11893 & 6.0\\
\hline
33 & 40 & 4.0 & 4033 & 0.069 & - & 1 & 4033 & 4.0\\
34 & 50 & 4.0 & 4015 & 0.069 & - & 1 & 4015 & 4.0\\
35 & 40 & 5.0 & 5637 & 0.069 & - & 1 & 5637 & 4.0\\
36 & 50 & 5.0 & 5611 & 0.069 & - & 1 & 5611 & 4.0\\
37 & 40 & 10.0 & 15943 & 0.069 & - & 1 & 15943 & 4.0\\
38 & 50 & 10.0 & 15870 & 0.069 & - & 1 & 15870 & 4.0\\
39 & 80 & 10.0 & 15658 & 0.069 & - & 1 & 15658 & 4.0\\
40 & 100 & 10.0 & 15521 & 0.069 & - & 1 & 15521 & 4.0\\
\hline
\end{tabular}
\label{table:models}
\end{table*}

\clearpage

\section{Complementary plots}

In this appendix we present complementary plots.

\begin{figure*}
\centering
\centering
\includegraphics[width = 0.9\textwidth]{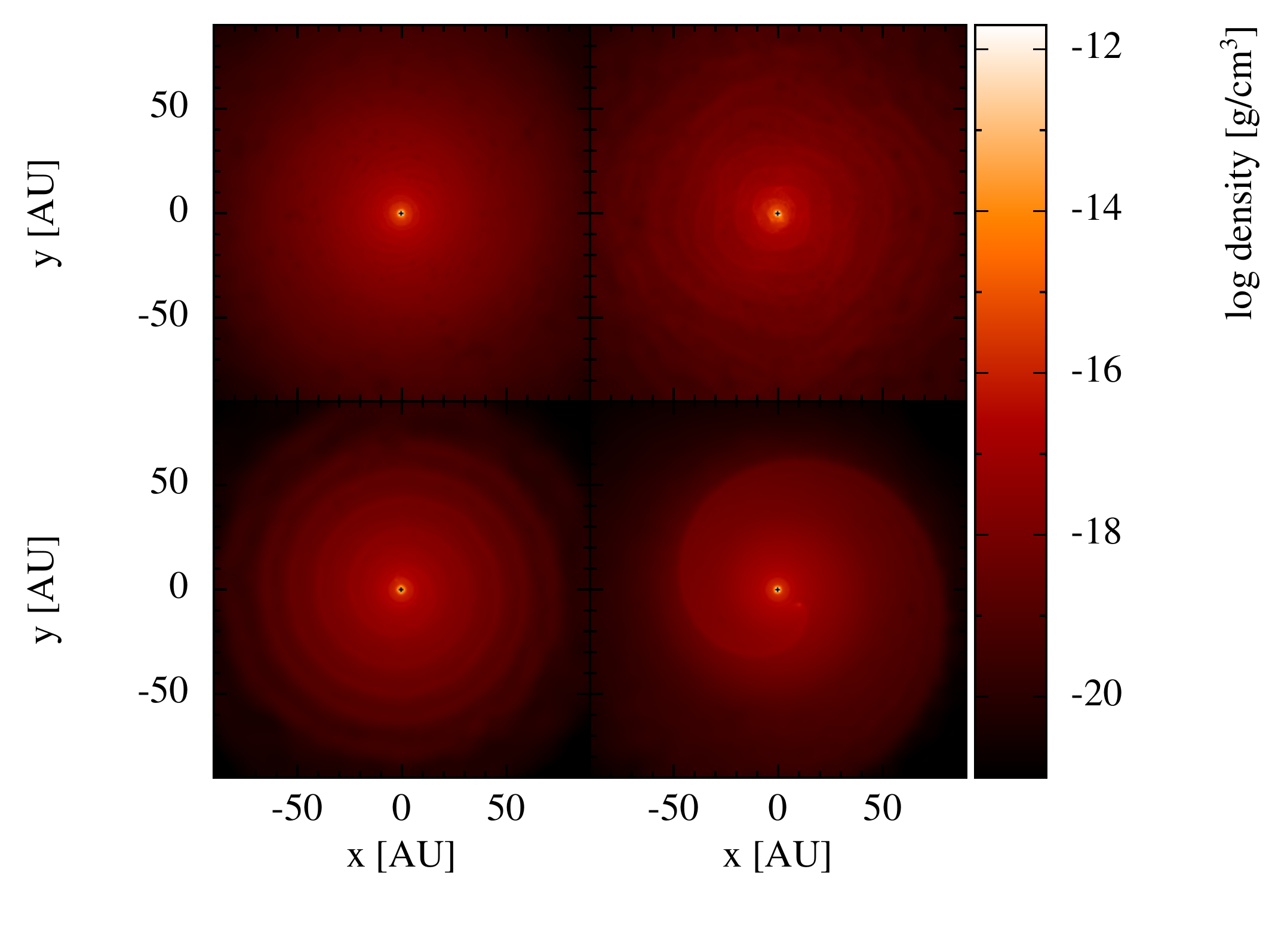}
\vspace{-0.9cm}
\caption{2D density cross-section slices through the orbital plane after $\simeq$ 76 pulsation cycle for model 9 (\textit{Top left}), model 26 (\textit{Top right}), model 33 (\textit{Bottom left}), and model 40 (\textit{Bottom right}). In models where multiple spiral arms form (top panels), the spiral structures are not well resolved in the density plots. However, in models where a single spiral arm forms (bottom panels), the spiral structures are well resolved in the density plots.}
\label{Fig:density}
\end{figure*}

\begin{figure*}
\begin{center}
\includegraphics[width=0.49\textwidth]{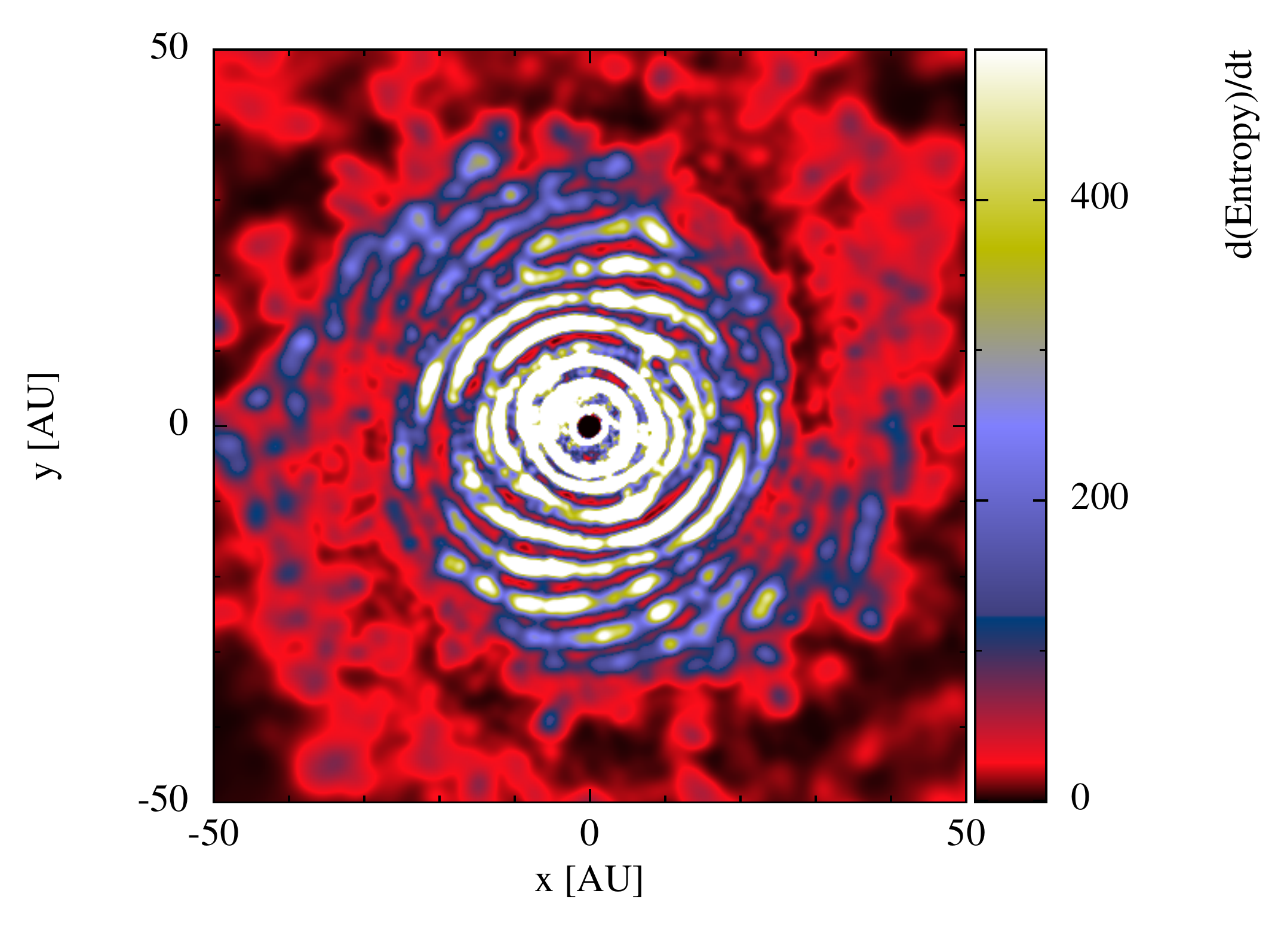}
\includegraphics[width=0.49\textwidth]{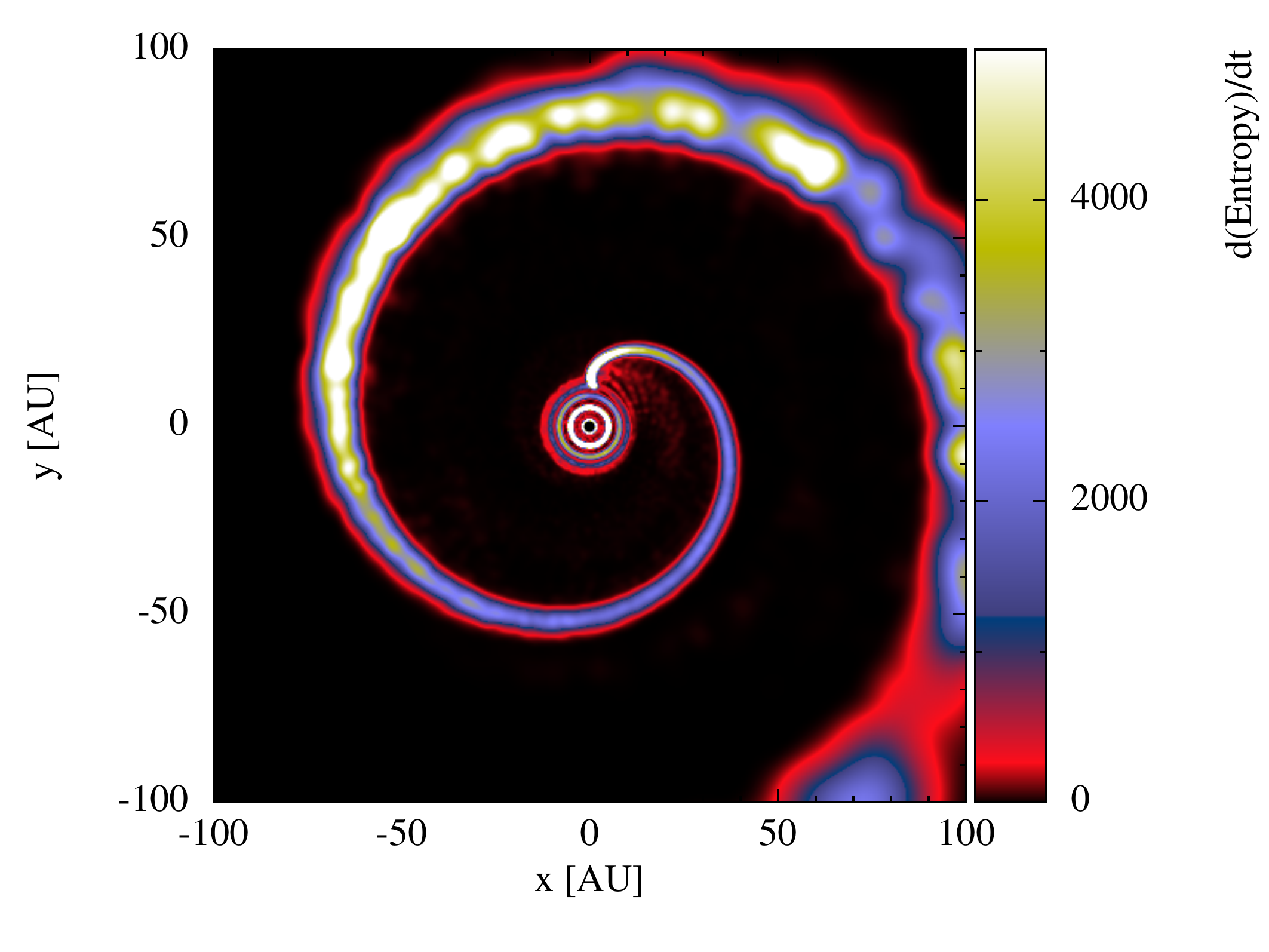}
\caption{2D entropy cross-section plots from models 9 (\textit{left}) and 40 (\textit{right}), after $\approx$ 76 and $\approx$ 89 pulsation cycles, respectively.}
\label{Fig:entropy}
\end{center}
\end{figure*}

\begin{figure*}
\begin{center}
\includegraphics[width=\textwidth]{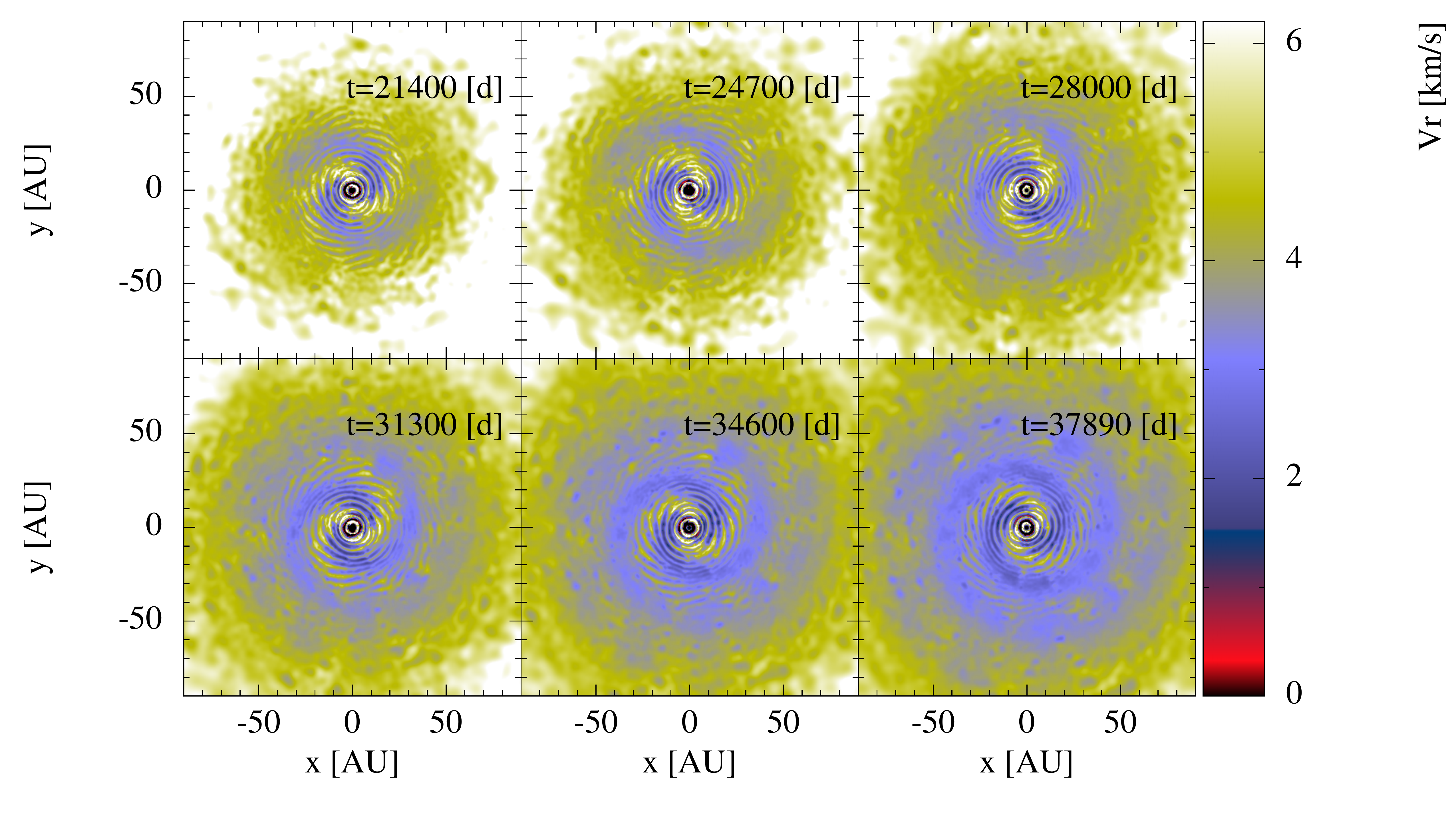}
\vspace{-0.8cm}
\caption{2D radial velocity cross-section slices through the orbital plane from model 9 at different time-steps, in ascending order from left to right, top to bottom, showing the winding of the spiral arms as a function of time. $t$ is the time in days elapsed since the start of the simulation.}
\label{Fig:arm_spacing_9}
\end{center}
\end{figure*}

\begin{figure*}
\begin{center}
\includegraphics[width=\textwidth]{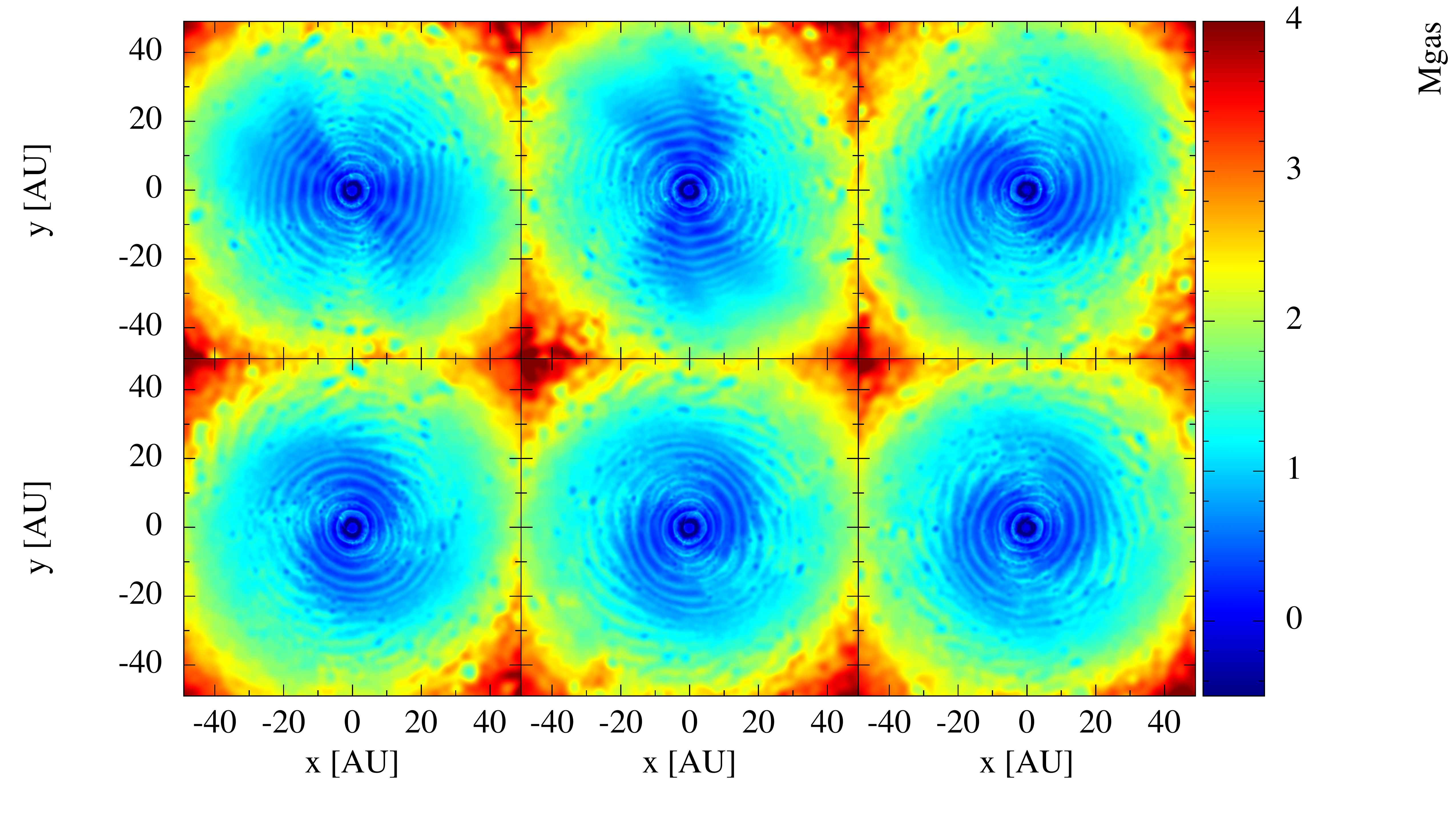}
\vspace{-0.8cm}
\caption{2D $M_{\mathrm{gas}}$ cross-section slices through the orbital plane after $\approx$ 76 pulsation cycles for models 4, 5, 6, 7, 8, and 9, going from left to right, top to bottom. The spiral arms are propagating in the radial direction close to a supersonic speed ($M_{\mathrm{sa}} \gtrsim$ 1) in most of the models.}
\label{Fig:Mach_number_2}
\end{center}
\end{figure*}

\begin{figure*}
\begin{center}
\includegraphics[width=0.8\textwidth]{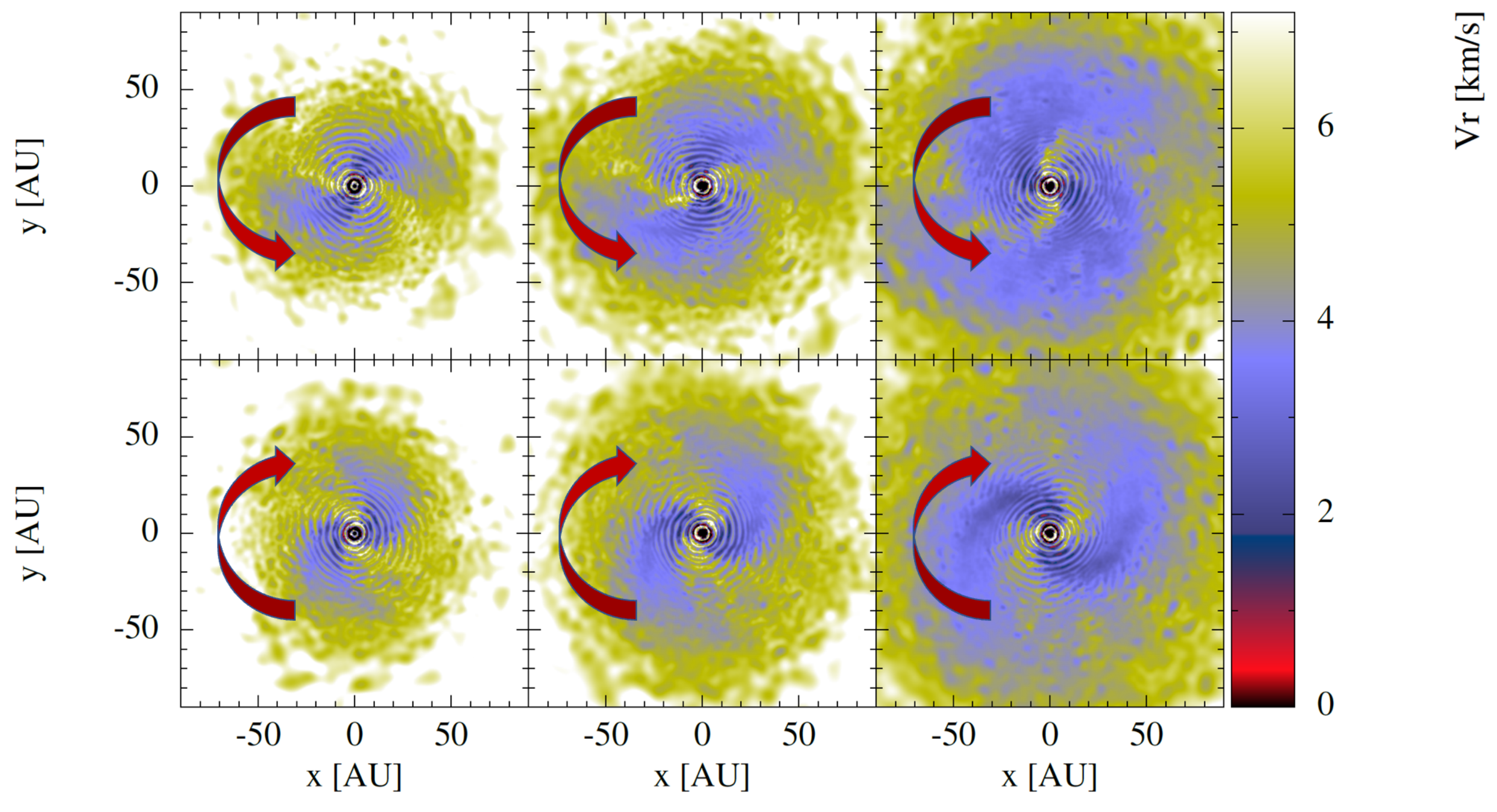}
\caption{2D radial velocity cross-section through the orbital plane from models 2 (\textit{top}) and 6 (\textit{bottom}) after 46, 59, and 80 pulsation cycles, from left to right. The red arrows indicate the apparent direction of rotation of the spiral arms. In model 2 the spiral arms are rotating in the counter-clockwise direction, however in model 6 the spiral arms are rotating in the clockwise direction.}
\label{Fig:rot_spirals}
\end{center}
\end{figure*}

\begin{figure*}
\centering
\begin{subfigure}{0.49\textwidth}
\centering
\includegraphics[width = \textwidth]{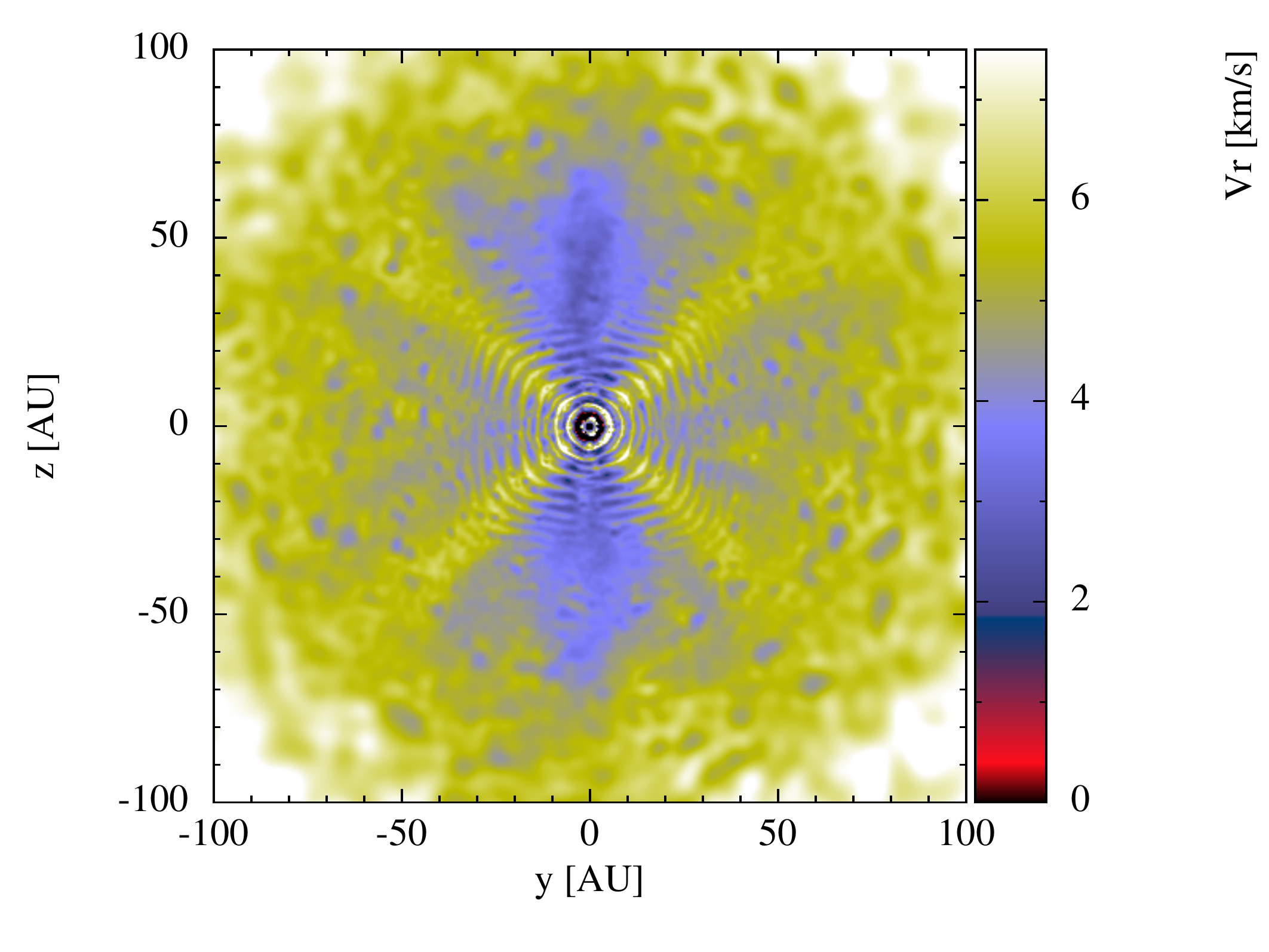}
\label{fig:left}
\end{subfigure}
\begin{subfigure}{0.49\textwidth}
\centering
\includegraphics[width = \textwidth]{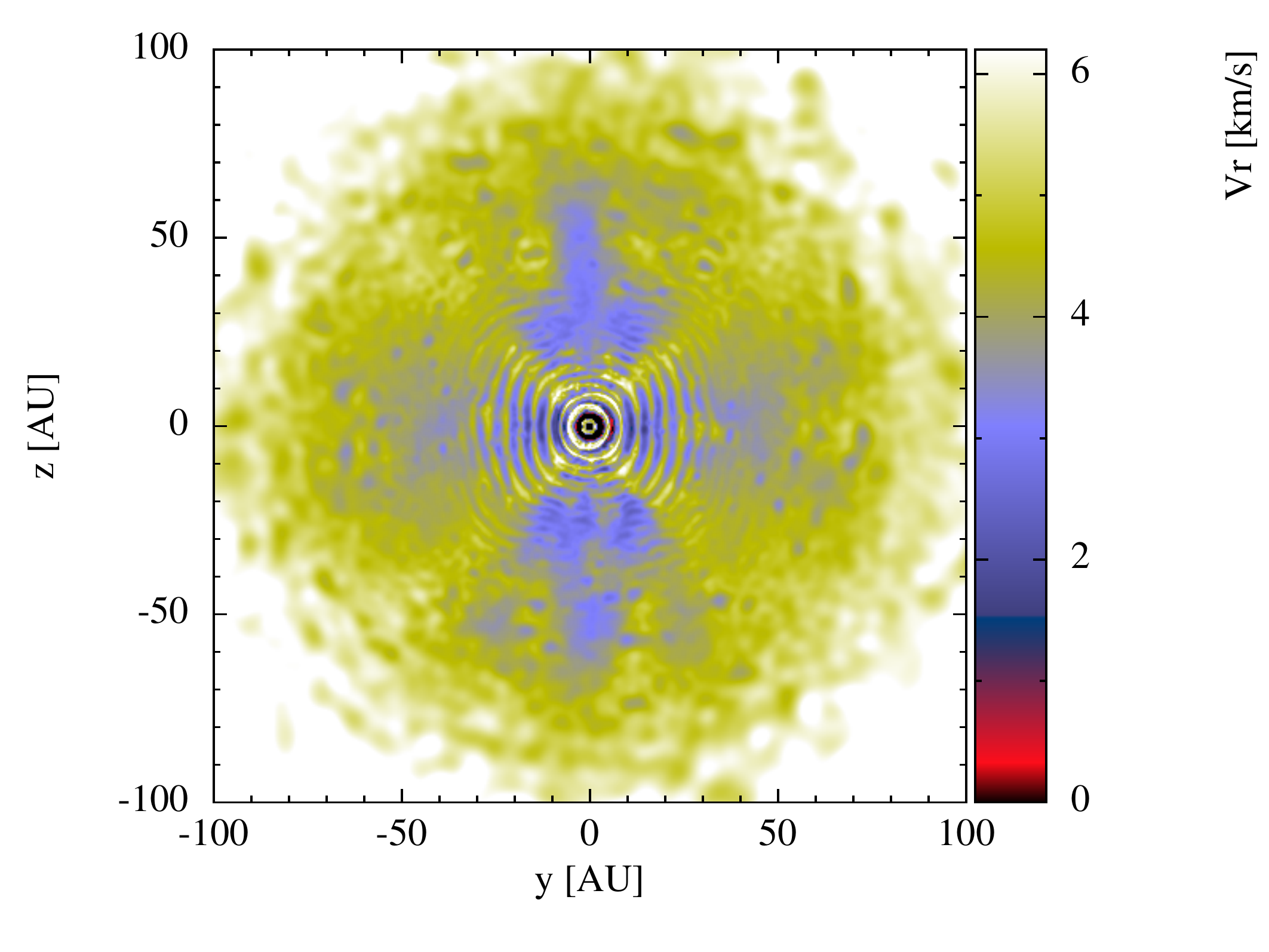}
\label{fig:right}
\end{subfigure}
\begin{subfigure}{0.49\textwidth}
\centering
\includegraphics[width = \textwidth]{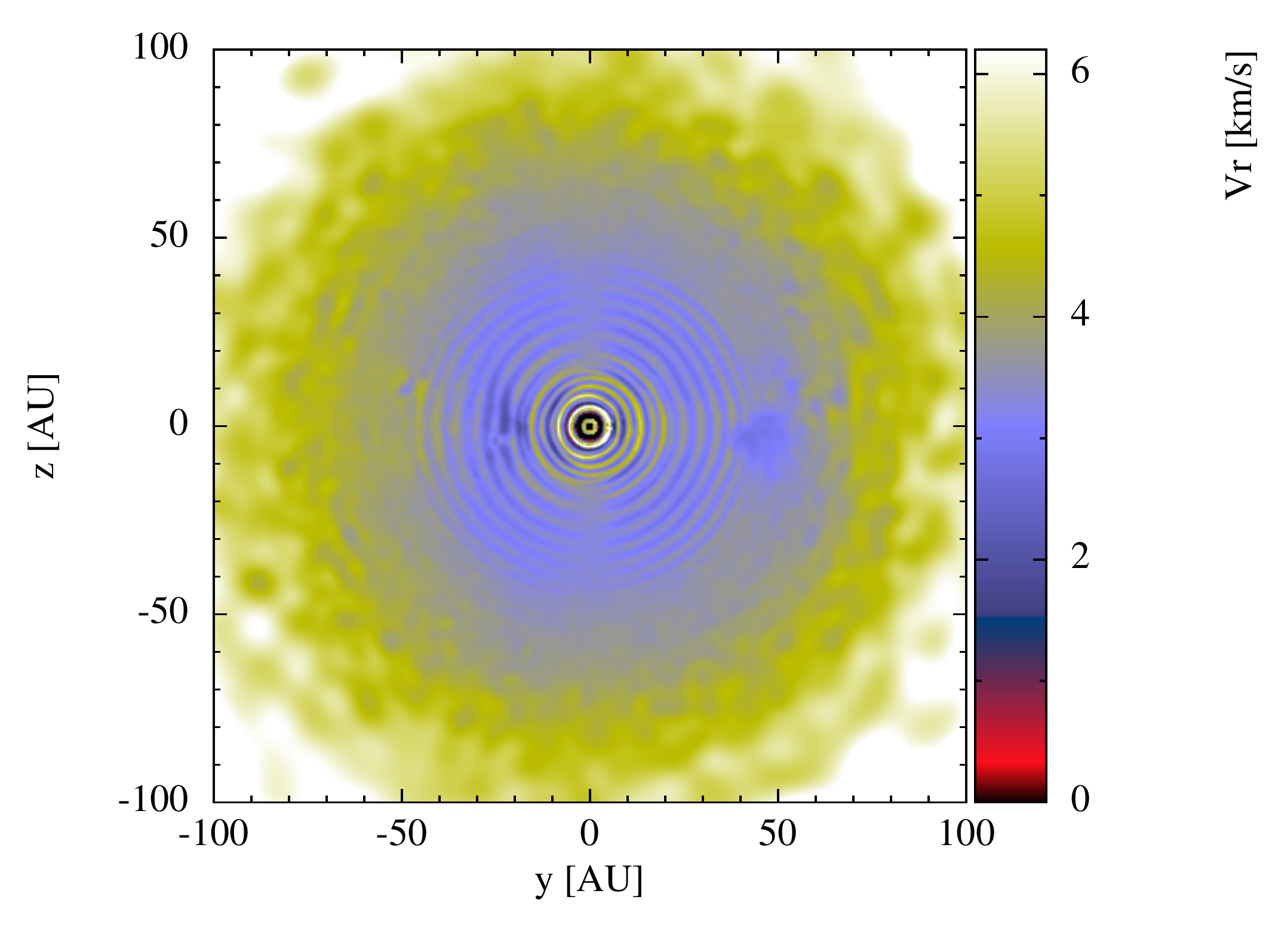}
\label{fig:left}
\end{subfigure}
\begin{subfigure}{0.49\textwidth}
\centering
\includegraphics[width = \textwidth]{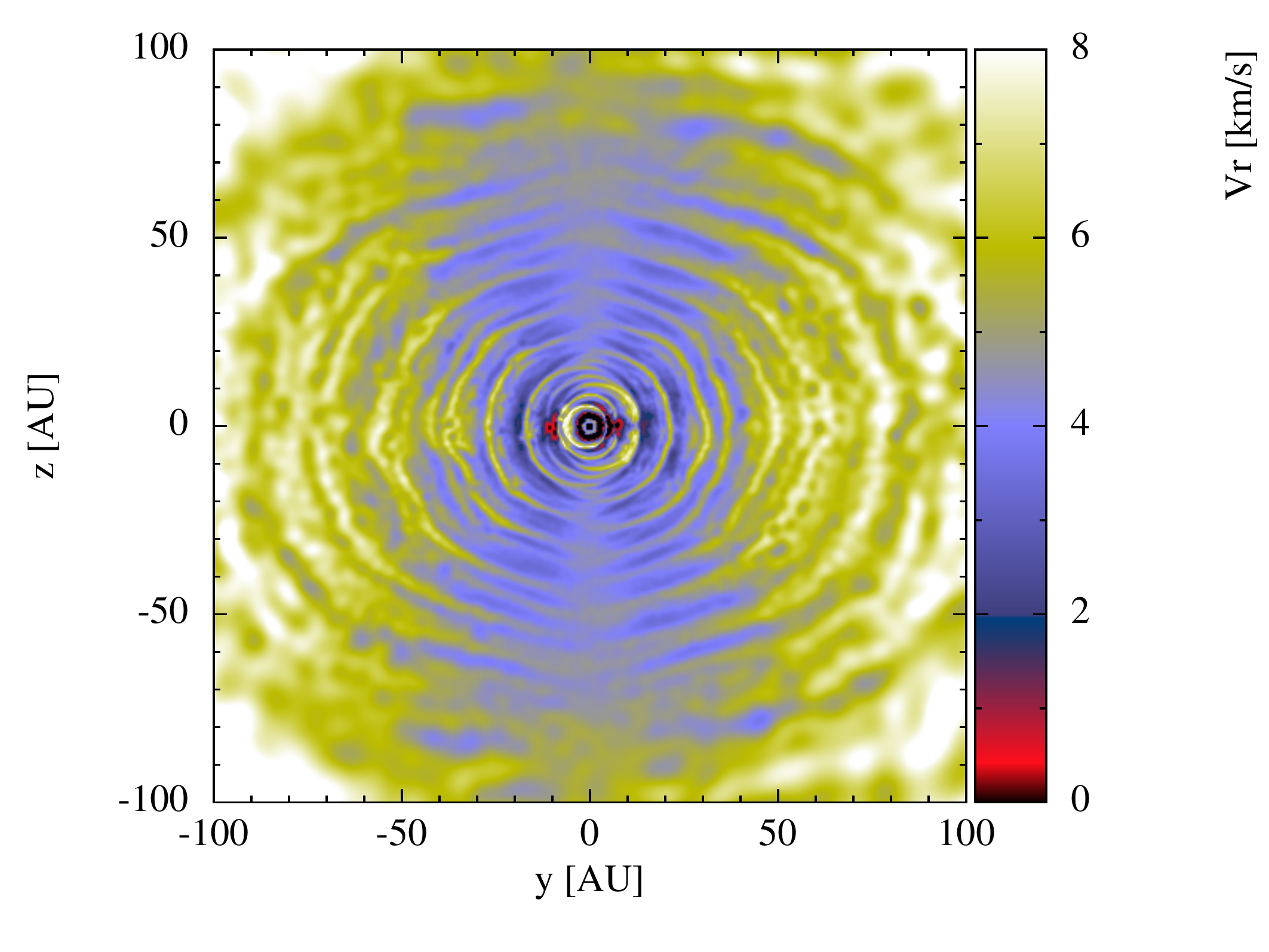}
\label{fig:right}
\end{subfigure}
\vspace{-0.5cm}
\caption{2D velocity cross-section slices through the y-z plane after $\sim$ 76 pulsation cycle for models 4 (\textit{Top left}), 9 (\textit{Top right}), 21 (\textit{Bottom left}), and 26 (\textit{Bottom right}).}
\label{Fig:yz}
\end{figure*}

\begin{figure*}
\centering
\begin{subfigure}{0.49\textwidth}
\centering
\includegraphics[width = \textwidth]{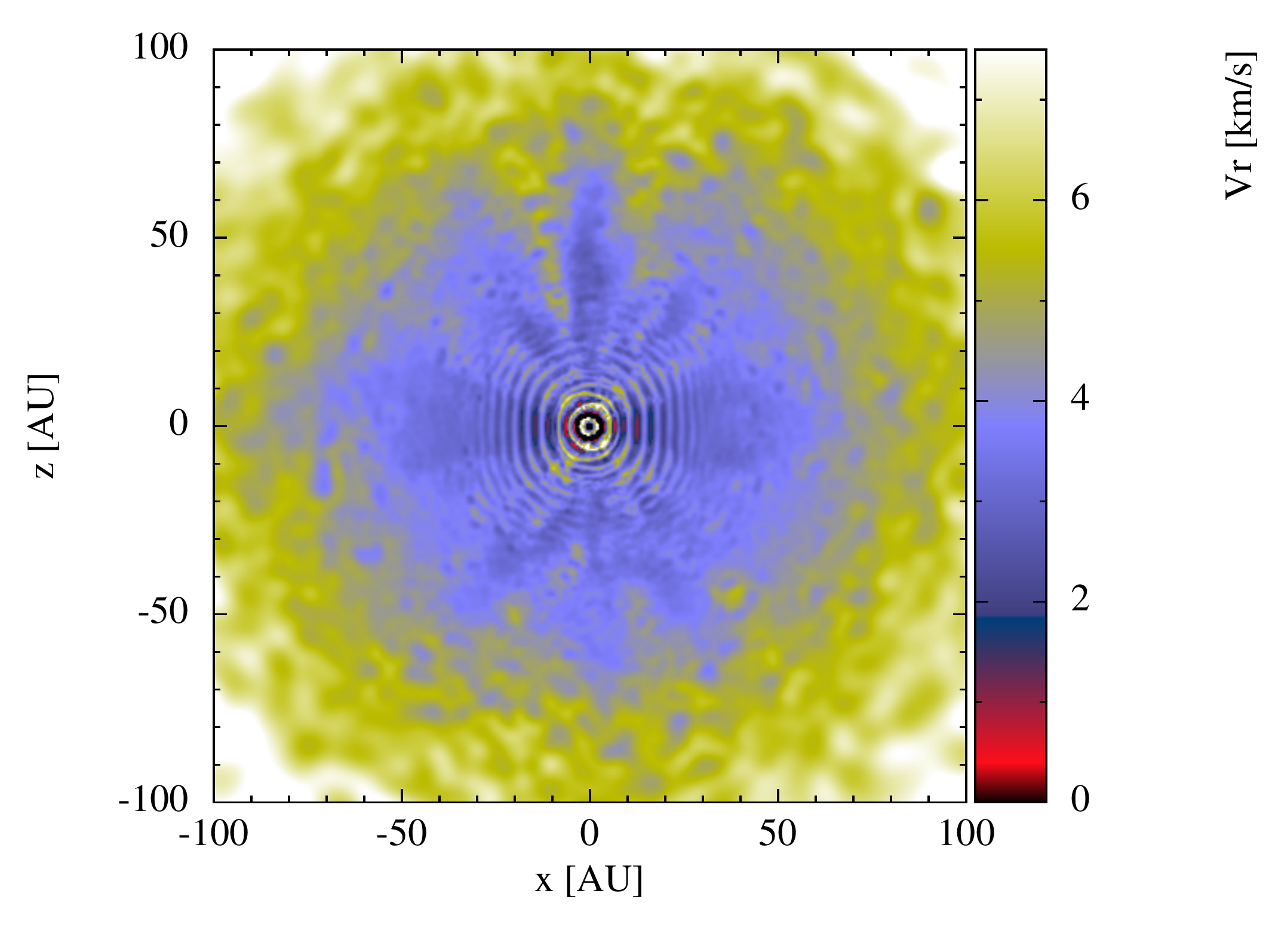}
\label{fig:left}
\end{subfigure}
\begin{subfigure}{0.49\textwidth}
\centering
\includegraphics[width = \textwidth]{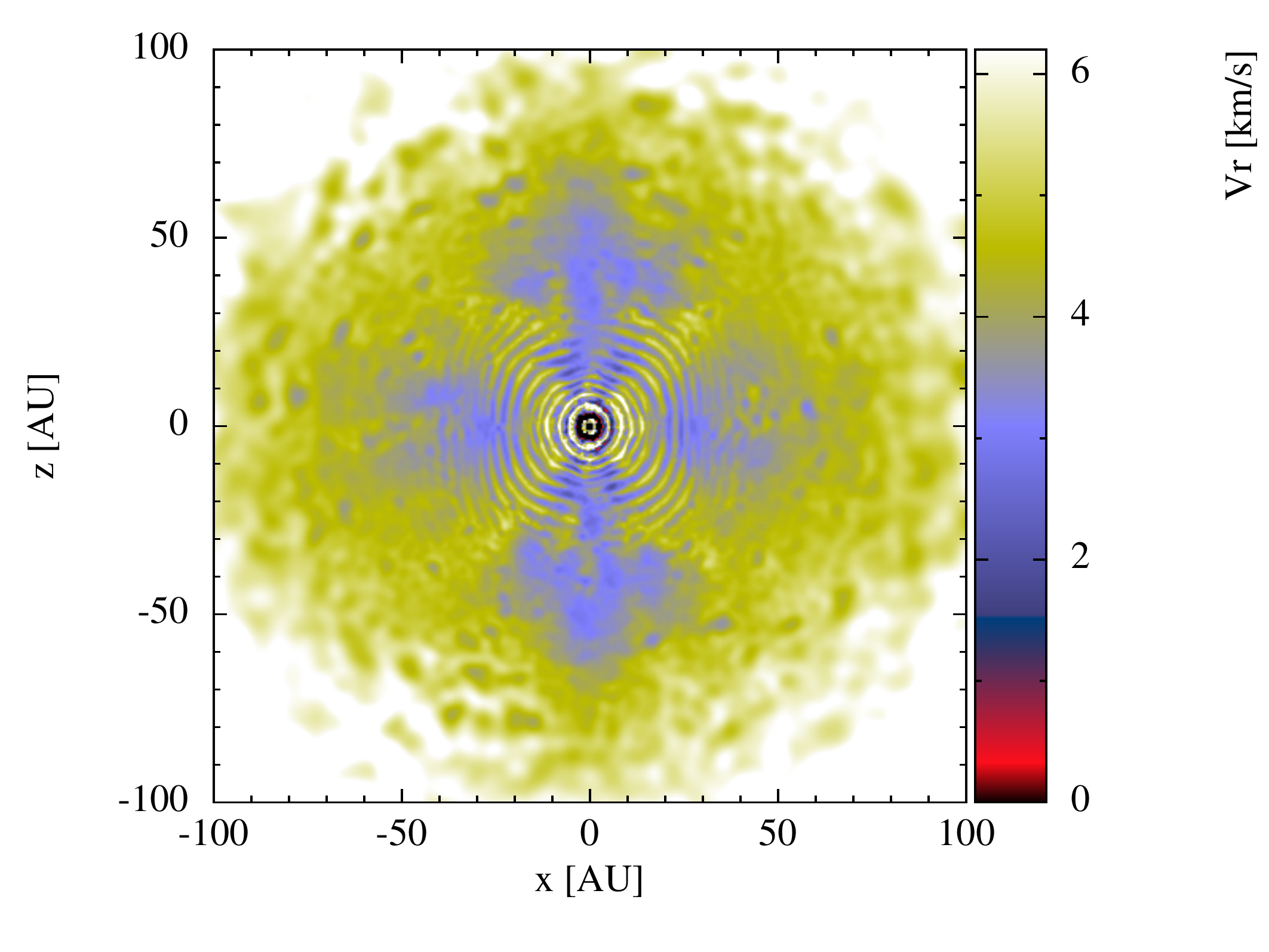}
\label{fig:right}
\end{subfigure}
\begin{subfigure}{0.49\textwidth}
\centering
\includegraphics[width = \textwidth]{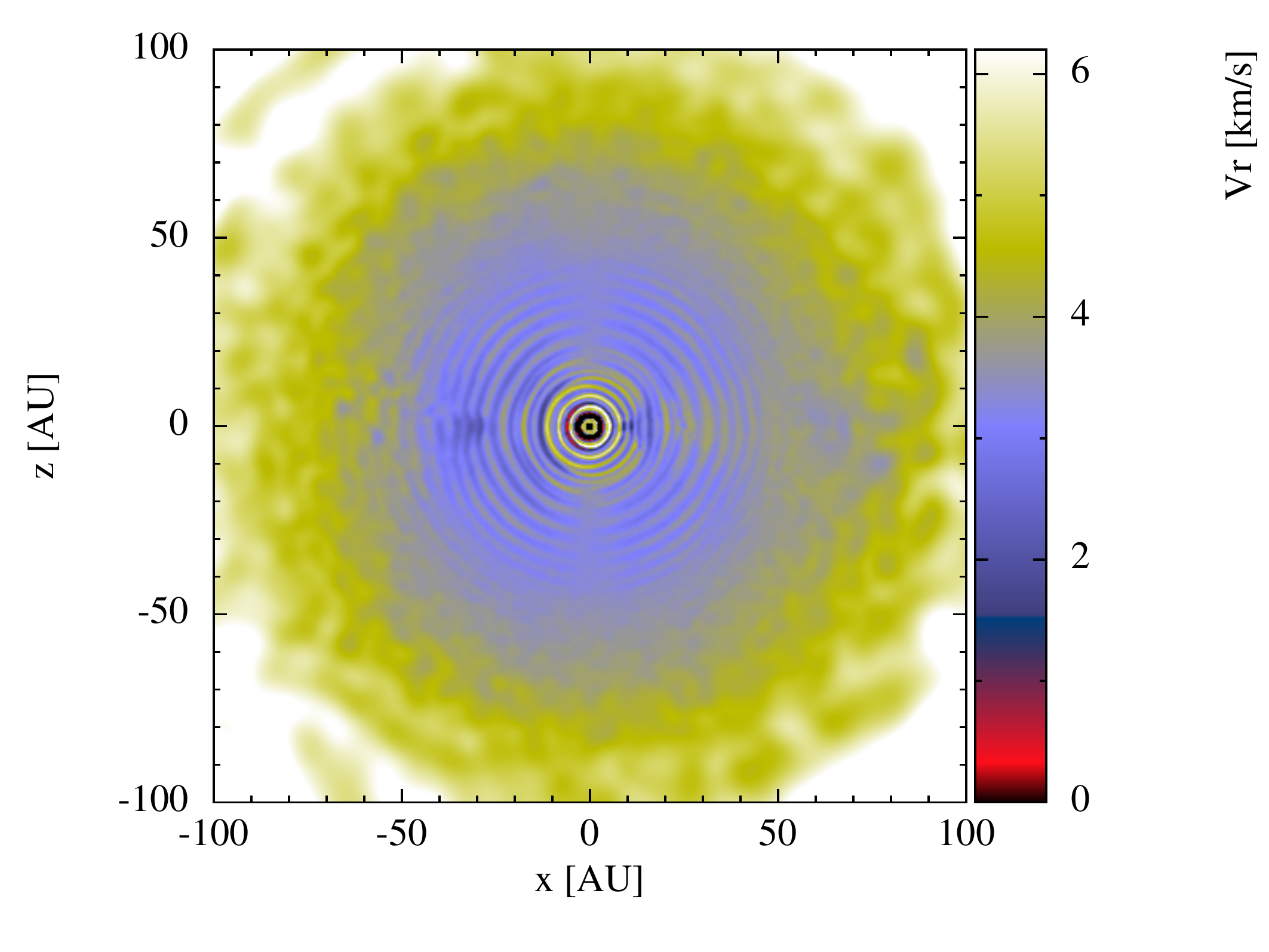}
\label{fig:left}
\end{subfigure}
\begin{subfigure}{0.49\textwidth}
\centering
\includegraphics[width = \textwidth]{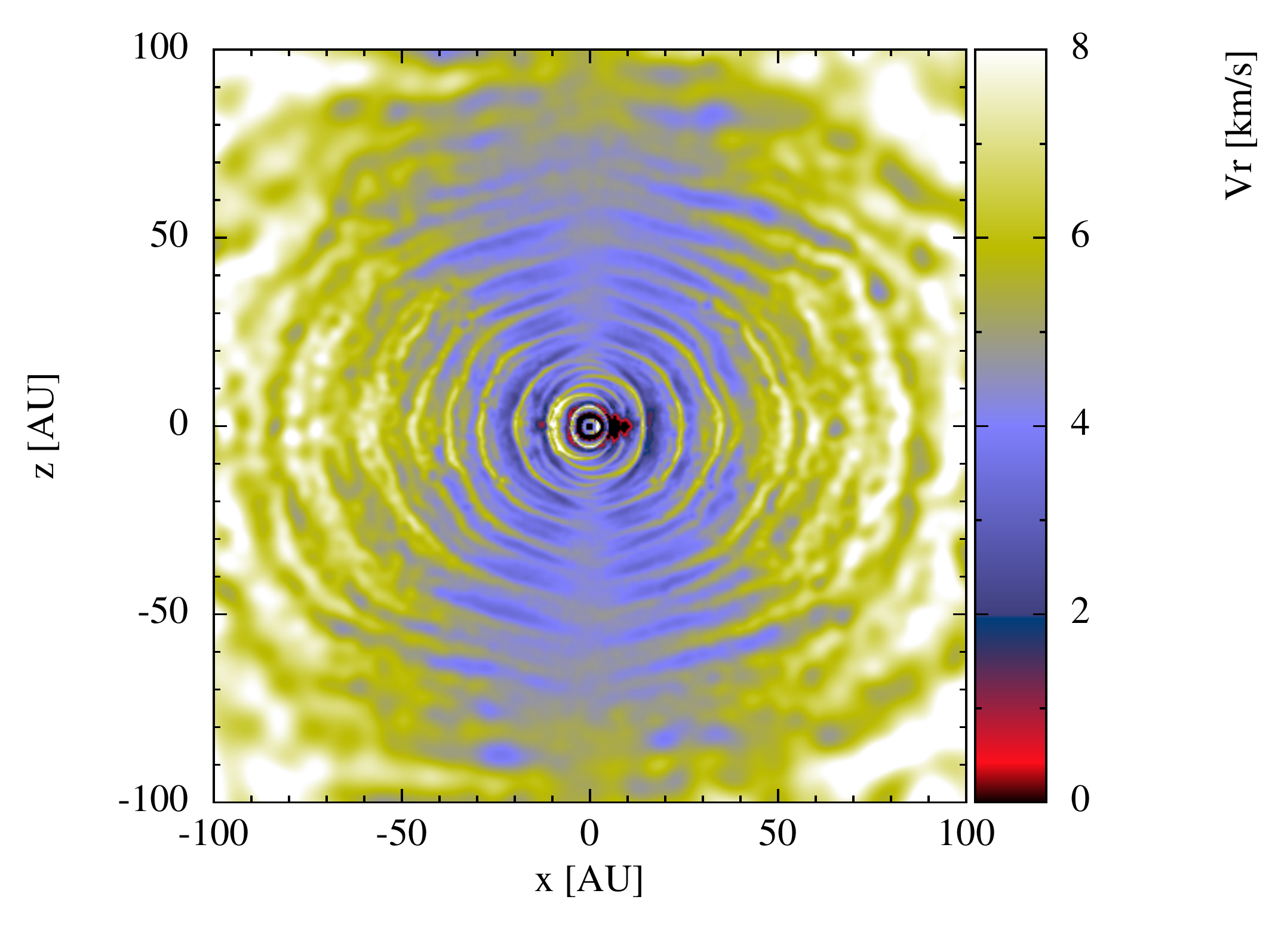}
\label{fig:right}
\end{subfigure}
\vspace{-0.5cm}
\caption{2D velocity cross-section slices through the x-z plane after $\sim$ 76 pulsation cycle for models 4 (\textit{Top left}), 9 (\textit{Top right}), 21 (\textit{Bottom left}), and 26 (\textit{Bottom right}).}
\label{Fig:xz}
\end{figure*}

\begin{figure*}
\begin{center}
\includegraphics[width=0.8\textwidth]{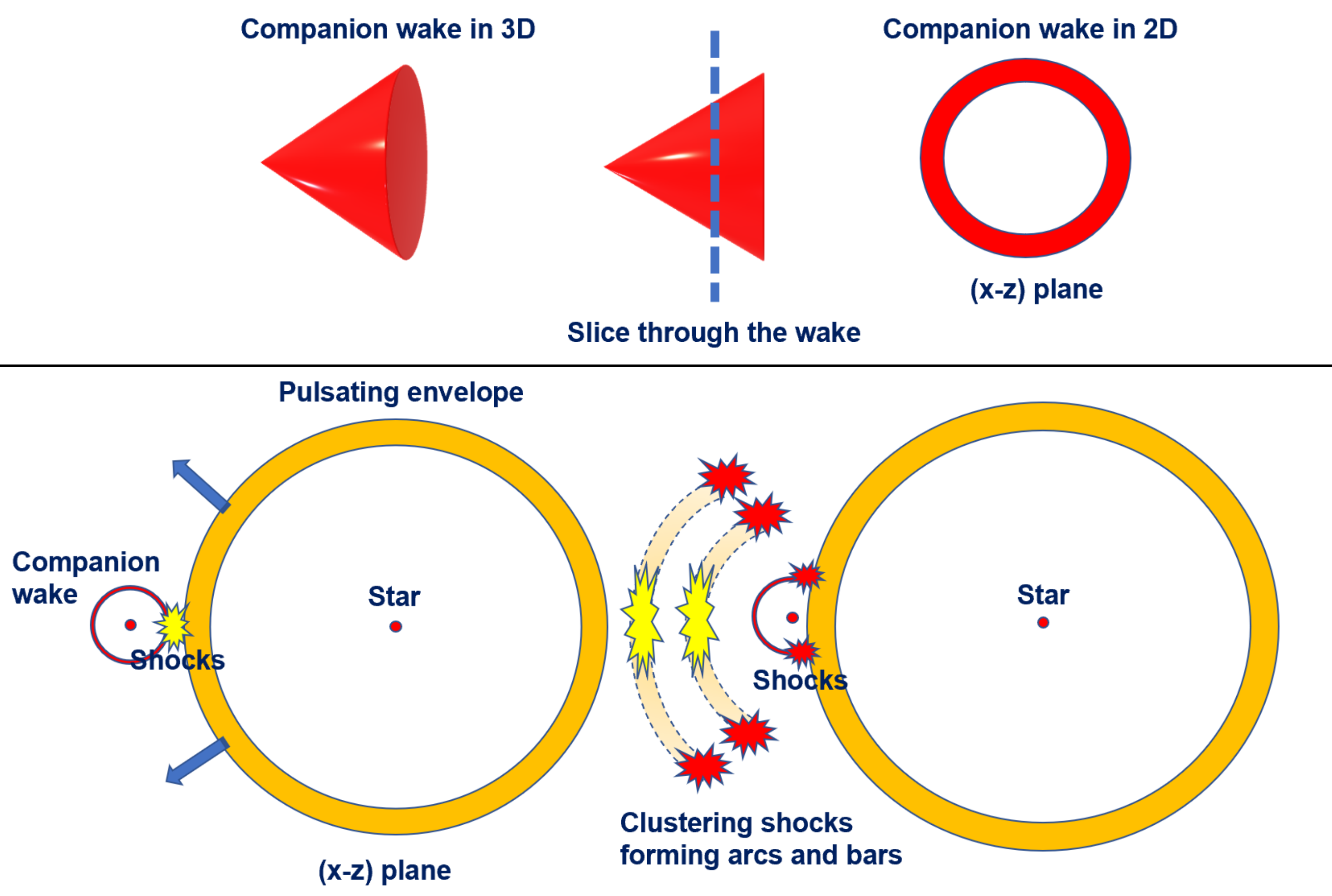}
\caption{\textit{Top:} Schematic illustration of the wake of the companion in 3D, showing a cone-like structure and a 2D cross-section slice through the wake forming a circle-shape in the x-z plane. \textit{Bottom:} Cartoon model (sizes not to scale) illustrating the formation of arcs and bars in the x-z plane due to the shocks between the stellar pulsations and the wake of the companion. The pulsation amplitude is highest when meeting the upper and lower edges of the wake creating stronger shocks that cluster into bars.}
\label{Fig:shocks_xz}
\end{center}
\end{figure*}

\begin{figure*}
\begin{center}
\includegraphics[width=0.8\textwidth]{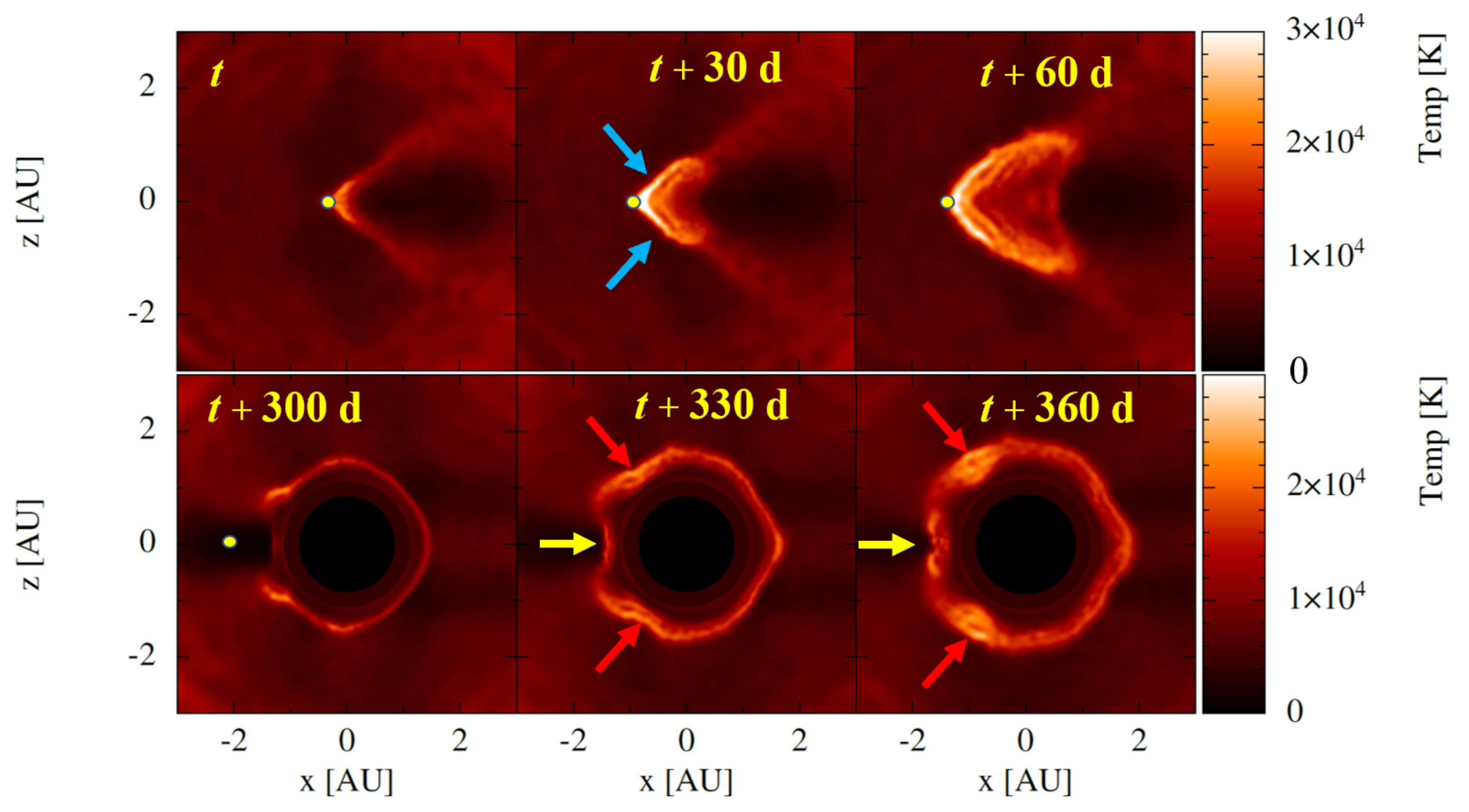}
\caption{\textit{Top:} 2D temperature cross-section through the x-z plane at y = 1.5 AU and  at different time-steps, showing the wake of the companion (blue arrows), forming a cone-like structure in 3D (V-shape in 2D). The yellow dots represent the location of the companion. The cross-section does not show the stellar atmosphere which is at y $\approx$ 0. \textit{Bottom:} 2D temperature cross-section through the the x-z plane at y = 0 and at different time-steps, showing the shocks between the wake of the companion (a circle in this case; see Figure~\ref{Fig:shocks_xz}) and the stellar pulsations. The shocks are highlighted in yellow and red arrows, and they eventually cluster into X-shaped arcs and bars. The companion is moving towards the inside (y$<$0) of the x-z  plane.}
\label{Fig:wake_xz}
\end{center}
\end{figure*}

\begin{figure*}
\begin{center}
\includegraphics[width=0.8\textwidth]{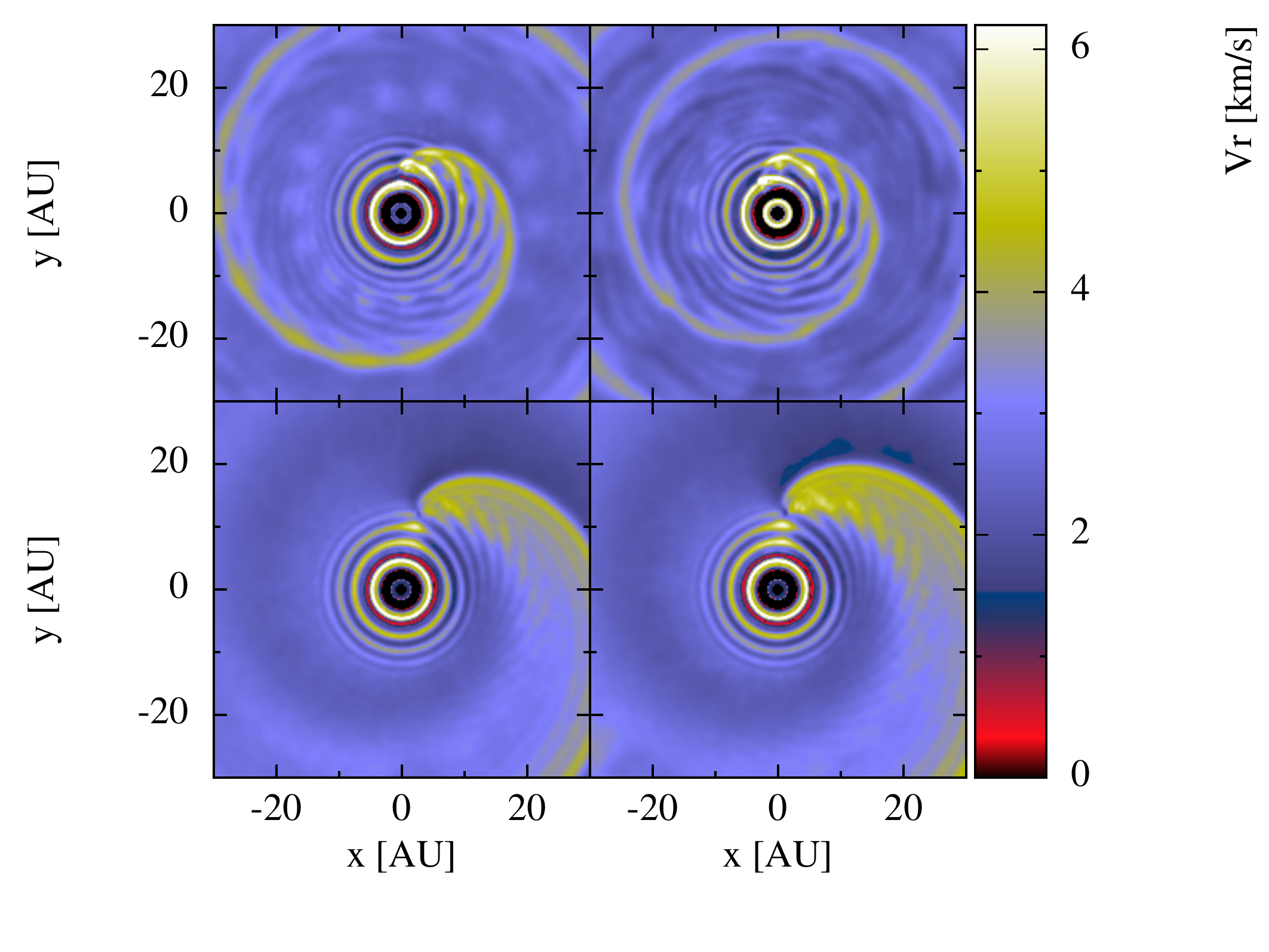}
\vspace{-1.cm}
\caption{Same as Figure~\ref{Fig:1spiral} but zoomed-in on the inner region of the spiral arms.}
\label{Fig:1spiral_zoom}
\end{center}
\end{figure*}

\begin{figure*}
\begin{center}
\includegraphics[width=0.8\textwidth]{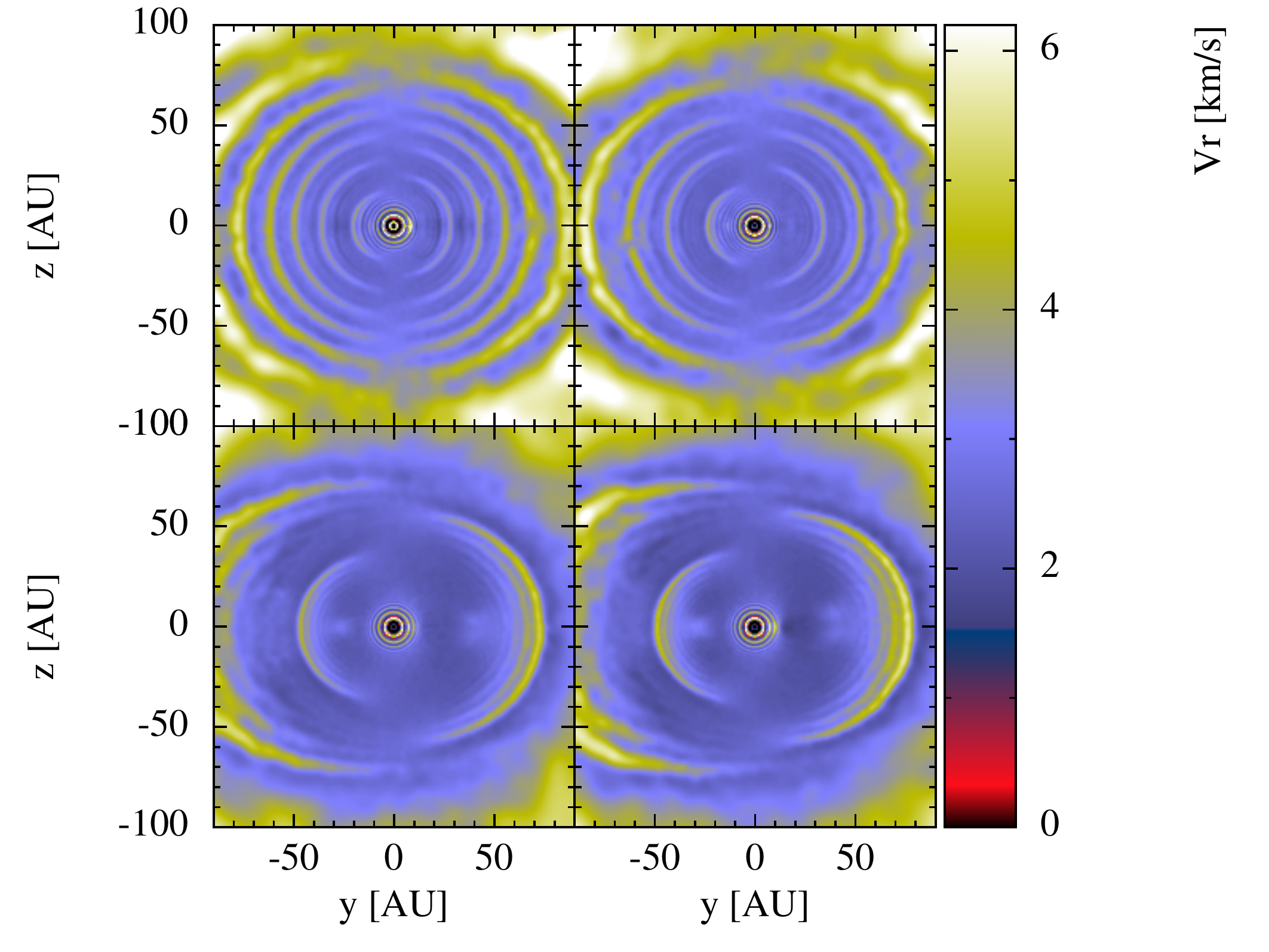}
\vspace{-0.5cm}
\caption{\textit{Top}: 2D radial velocity cross-section in the y-z plane from models 33 (\textit{left}) and 36 (\textit{right}), after $\approx$ 76 pulsation cycles. \textit{bottom}: same for models 39 (\textit{left}) and 40 (\textit{right}), after $\approx$ 89 pulsation cycles.}
\label{Fig:1spiral_yz}
\end{center}
\end{figure*}

\begin{figure*}
\begin{center}
\includegraphics[width=0.8\textwidth]{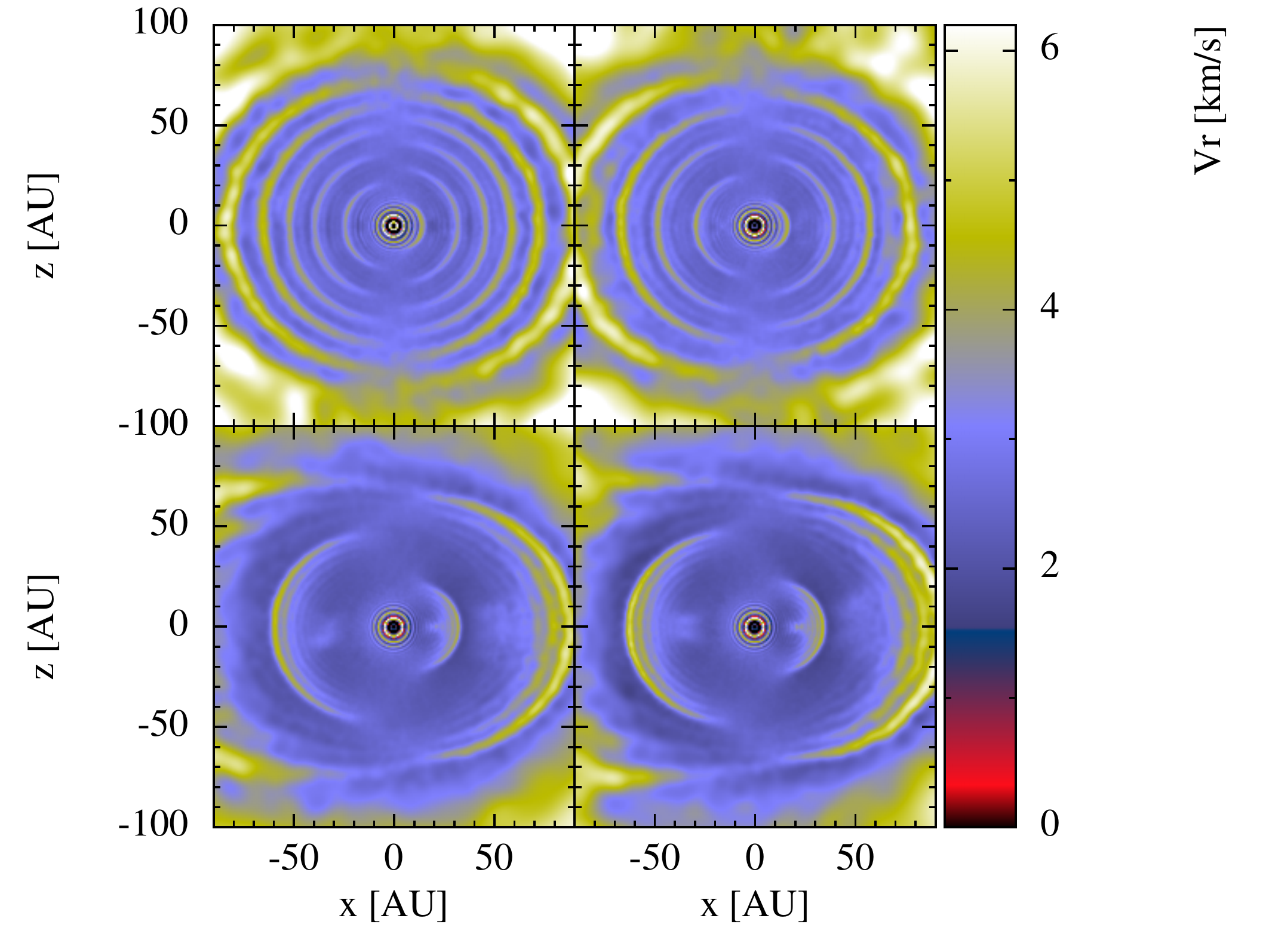}
\vspace{-0.5cm}
\caption{\textit{Top}: 2D radial velocity cross-section in the x-z plane from models 33 (\textit{left}) and 36 (\textit{right}), after $\approx$ 76 pulsation cycles. \textit{bottom}: same for models 39 (\textit{left}) and 40 (\textit{right}), after $\approx$ 89 pulsation cycles.}
\label{Fig:1spiral_xz}
\end{center}
\end{figure*}

\begin{figure*}
\begin{center}
\includegraphics[width=\textwidth]{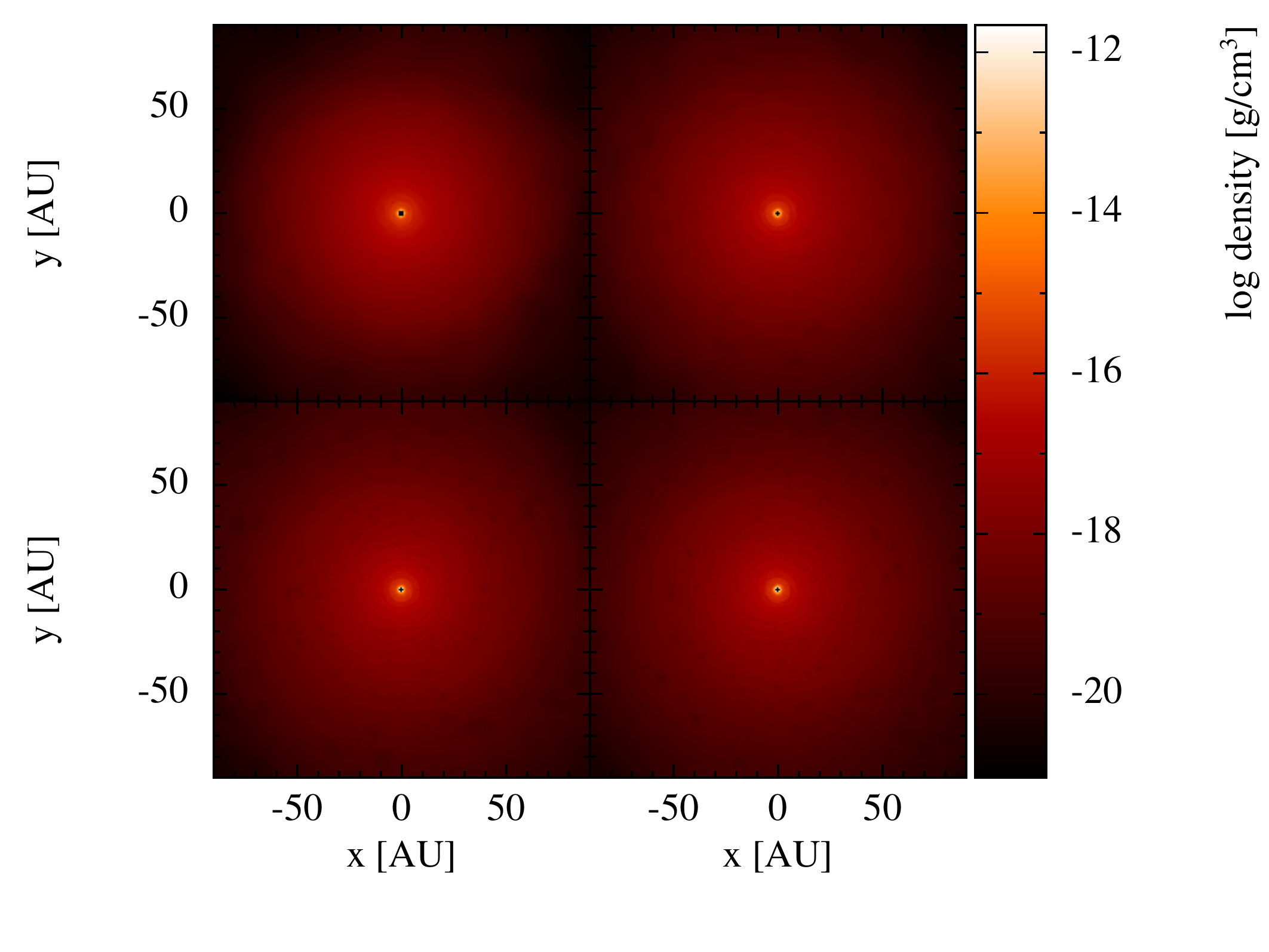}
\caption{2D density cross-section through the orbital plane from for model 9 after 76 pulsation cycles, but using different resolutions. The models from left to right, top to bottom have the following number of particles: $2.1\times10^5$, $1.1\times10^6$, $4.2\times10^6$, and $8.5\times10^6$, respectively.}
\label{Fig:resolution_density}
\end{center}
\end{figure*}

\label{lastpage}
\end{document}